\documentclass[longauth]{aa}
\usepackage{graphicx}
\usepackage{txfonts}
\usepackage{xcolor}
\usepackage{amssymb}
\usepackage[utf8]{inputenc}
\usepackage{float}
\usepackage{hyperref}
\usepackage{listings}
\usepackage{stfloats}

\newcommand\gdr[1]{\gaia~DR#1}
\newcommand{\orcit}[1]{\protect\href{https://orcid.org/#1}{\protect\includegraphics[width=8pt]{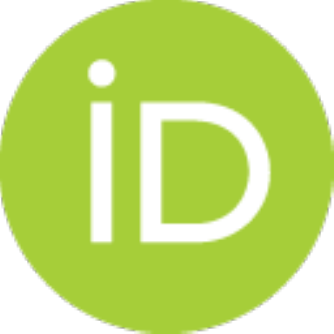}}}

\def\deg{\ensuremath{^\circ}}

\newcommand{\gaia}{\textit{Gaia}\xspace}
\providecommand{\gmag}{\ensuremath{G}}

\makeatletter
\renewcommand*\maketitle{%
  \thispagestyle{firstpage}
\begingroup
    \if@wideboxfn
    \setlength\bibindent{1.4\parindent}
    \else
    \setlength\bibindent{\parindent}
    \fi
    \renewcommand*\thefootnote{\@fnsymbol\c@footnote}%
    \renewcommand\@makefntext[1]{%
    \ifaa@longfn\hsize\textwidth\fi
    \noindent
    \hb@xt@\bibindent{\hss\@makefnmark\enspace}##1}
  \ifaa@twocolumn
  \begingroup
    \begin{aa@strip}
          \aa@maketitle
    \end{aa@strip}
    \@thanks            
  \endgroup
  \else
    \begingroup
      \let\thanks\footnote
      \aa@maketitle
    \endgroup
  \fi
\endgroup
  \setcounter{footnote}{0}%
}
\makeatother
\usepackage[normalem]{ulem}

\begin{document}

   \title{\gaia Data Release 3: Reflectance spectra of Solar System small bodies
   \thanks{This article is dedicated to the memory of Dimitri Pourbaix, 
                who laid the foundations and directed the Coordination Unit 4 (CU4) of the Data Processing and Analysis Consortium (DPAC) of the ESA mission Gaia.}}

\author{
{\it Gaia} Collaboration
\and         L.~                     Galluccio\orcit{0000-0002-8541-0476}\inst{\ref{inst:0001}}
\and         M.~                         Delbo\orcit{0000-0002-8963-2404}\inst{\ref{inst:0001}}
\and         F.~                     De Angeli\orcit{0000-0003-1879-0488}\inst{\ref{inst:0003}}
\and         T.~                       Pauwels\inst{\ref{inst:0004}}
\and         P.~                         Tanga\orcit{0000-0002-2718-997X}\inst{\ref{inst:0001}}
\and         F.~                       Mignard\inst{\ref{inst:0001}}
\and         A.~                       Cellino\orcit{0000-0002-6645-334X}\inst{\ref{inst:0007}}
\and     A.G.A.~                         Brown\orcit{0000-0002-7419-9679}\inst{\ref{inst:0008}}
\and         K.~                      Muinonen\orcit{0000-0001-8058-2642}\inst{\ref{inst:0119},\ref{inst:0120}}
\and         A.~                 Penttil\"{ a}\orcit{0000-0001-7403-1721}\inst{\ref{inst:0119}}
\and         S.~                        Jordan\orcit{0000-0001-6316-6831}\inst{\ref{inst:0015}}
\and  A.~                     Vallenari\orcit{0000-0003-0014-519X}\inst{\ref{inst:0009}}
\and         T.~                        Prusti\orcit{0000-0003-3120-7867}\inst{\ref{inst:0010}}
\and     J.H.J.~                    de Bruijne\orcit{0000-0001-6459-8599}\inst{\ref{inst:0010}}
\and         F.~                        Arenou\orcit{0000-0003-2837-3899}\inst{\ref{inst:0012}}
\and         C.~                     Babusiaux\orcit{0000-0002-7631-348X}\inst{\ref{inst:0013},\ref{inst:0012}}
\and         M.~                      Biermann\inst{\ref{inst:0015}}
\and       O.L.~                       Creevey\orcit{0000-0003-1853-6631}\inst{\ref{inst:0001}}
\and         C.~                     Ducourant\orcit{0000-0003-4843-8979}\inst{\ref{inst:0017}}
\and       D.W.~                         Evans\orcit{0000-0002-6685-5998}\inst{\ref{inst:0003}}
\and         L.~                          Eyer\orcit{0000-0002-0182-8040}\inst{\ref{inst:0019}}
\and         R.~                        Guerra\orcit{0000-0002-9850-8982}\inst{\ref{inst:0020}}
\and         A.~                        Hutton\inst{\ref{inst:0021}}
\and         C.~                         Jordi\orcit{0000-0001-5495-9602}\inst{\ref{inst:0022}}
\and       S.A.~                       Klioner\orcit{0000-0003-4682-7831}\inst{\ref{inst:0023}}
\and       U.L.~                       Lammers\orcit{0000-0001-8309-3801}\inst{\ref{inst:0020}}
\and         L.~                     Lindegren\orcit{0000-0002-5443-3026}\inst{\ref{inst:0025}}
\and         X.~                          Luri\orcit{0000-0001-5428-9397}\inst{\ref{inst:0022}}
\and         C.~                         Panem\inst{\ref{inst:0027}}
\and         D.~            Pourbaix$^\dagger$\orcit{0000-0002-3020-1837}\inst{\ref{inst:0028},\ref{inst:0029}}
\and         S.~                       Randich\orcit{0000-0003-2438-0899}\inst{\ref{inst:0030}}
\and         P.~                    Sartoretti\inst{\ref{inst:0012}}
\and         C.~                      Soubiran\orcit{0000-0003-3304-8134}\inst{\ref{inst:0017}}
\and       N.A.~                        Walton\orcit{0000-0003-3983-8778}\inst{\ref{inst:0003}}
\and     C.A.L.~                  Bailer-Jones\inst{\ref{inst:0034}}
\and         U.~                       Bastian\orcit{0000-0002-8667-1715}\inst{\ref{inst:0015}}
\and         R.~                       Drimmel\orcit{0000-0002-1777-5502}\inst{\ref{inst:0007}}
\and         F.~                        Jansen\inst{\ref{inst:0037}}
\and         D.~                          Katz\orcit{0000-0001-7986-3164}\inst{\ref{inst:0012}}
\and       M.G.~                      Lattanzi\orcit{0000-0003-0429-7748}\inst{\ref{inst:0007},\ref{inst:0040}}
\and         F.~                   van Leeuwen\inst{\ref{inst:0003}}
\and         J.~                        Bakker\inst{\ref{inst:0020}}
\and         C.~                      Cacciari\orcit{0000-0001-5174-3179}\inst{\ref{inst:0043}}
\and         J.~                 Casta\~{n}eda\orcit{0000-0001-7820-946X}\inst{\ref{inst:0044}}
\and         C.~                     Fabricius\orcit{0000-0003-2639-1372}\inst{\ref{inst:0022}}
\and         M.~                     Fouesneau\orcit{0000-0001-9256-5516}\inst{\ref{inst:0034}}
\and         Y.~                    Fr\'{e}mat\orcit{0000-0002-4645-6017}\inst{\ref{inst:0004}}
\and         A.~                      Guerrier\inst{\ref{inst:0027}}
\and         U.~                        Heiter\orcit{0000-0001-6825-1066}\inst{\ref{inst:0049}}
\and         E.~                        Masana\orcit{0000-0002-4819-329X}\inst{\ref{inst:0022}}
\and         R.~                      Messineo\inst{\ref{inst:0051}}
\and         N.~                       Mowlavi\orcit{0000-0003-1578-6993}\inst{\ref{inst:0019}}
\and         C.~                       Nicolas\inst{\ref{inst:0027}}
\and         K.~                  Nienartowicz\orcit{0000-0001-5415-0547}\inst{\ref{inst:0054},\ref{inst:0055}}
\and         F.~                       Pailler\orcit{0000-0002-4834-481X}\inst{\ref{inst:0027}}
\and         P.~                       Panuzzo\orcit{0000-0002-0016-8271}\inst{\ref{inst:0012}}
\and         F.~                        Riclet\inst{\ref{inst:0027}}
\and         W.~                          Roux\orcit{0000-0002-7816-1950}\inst{\ref{inst:0027}}
\and       G.M.~                      Seabroke\orcit{0000-0003-4072-9536}\inst{\ref{inst:0060}}
\and         R.~                         Sordo\orcit{0000-0003-4979-0659}\inst{\ref{inst:0009}}
\and         F.~                  Th\'{e}venin\inst{\ref{inst:0001}}
\and         G.~                  Gracia-Abril\inst{\ref{inst:0063},\ref{inst:0015}}
\and         J.~                       Portell\orcit{0000-0002-8886-8925}\inst{\ref{inst:0022}}
\and         D.~                      Teyssier\orcit{0000-0002-6261-5292}\inst{\ref{inst:0066}}
\and         M.~                       Altmann\orcit{0000-0002-0530-0913}\inst{\ref{inst:0015},\ref{inst:0068}}
\and         R.~                        Andrae\orcit{0000-0001-8006-6365}\inst{\ref{inst:0034}}
\and         M.~                        Audard\orcit{0000-0003-4721-034X}\inst{\ref{inst:0019},\ref{inst:0055}}
\and         I.~                Bellas-Velidis\inst{\ref{inst:0072}}
\and         K.~                        Benson\inst{\ref{inst:0060}}
\and         J.~                      Berthier\orcit{0000-0003-1846-6485}\inst{\ref{inst:0074}}
\and         R.~                        Blomme\orcit{0000-0002-2526-346X}\inst{\ref{inst:0004}}
\and       P.W.~                       Burgess\inst{\ref{inst:0003}}
\and         D.~                      Busonero\orcit{0000-0002-3903-7076}\inst{\ref{inst:0007}}
\and         G.~                         Busso\orcit{0000-0003-0937-9849}\inst{\ref{inst:0003}}
\and         H.~                   C\'{a}novas\orcit{0000-0001-7668-8022}\inst{\ref{inst:0066}}
\and         B.~                         Carry\orcit{0000-0001-5242-3089}\inst{\ref{inst:0001}}
\and         N.~                         Cheek\inst{\ref{inst:0081}}
\and         G.~                    Clementini\orcit{0000-0001-9206-9723}\inst{\ref{inst:0043}}
\and         Y.~                      Damerdji\orcit{0000-0002-3107-4024}\inst{\ref{inst:0083},\ref{inst:0084}}
\and         M.~                      Davidson\inst{\ref{inst:0085}}
\and         P.~                    de Teodoro\inst{\ref{inst:0020}}
\and         M.~              Nu\~{n}ez Campos\inst{\ref{inst:0021}}
\and         L.~                    Delchambre\orcit{0000-0003-2559-408X}\inst{\ref{inst:0083}}
\and         A.~                      Dell'Oro\orcit{0000-0003-1561-9685}\inst{\ref{inst:0030}}
\and         P.~                        Esquej\orcit{0000-0001-8195-628X}\inst{\ref{inst:0090}}
\and         J.~   Fern\'{a}ndez-Hern\'{a}ndez\inst{\ref{inst:0091}}
\and         E.~                        Fraile\inst{\ref{inst:0090}}
\and         D.~                      Garabato\orcit{0000-0002-7133-6623}\inst{\ref{inst:0093}}
\and         P.~              Garc\'{i}a-Lario\orcit{0000-0003-4039-8212}\inst{\ref{inst:0020}}
\and         E.~                        Gosset\inst{\ref{inst:0083},\ref{inst:0029}}
\and         R.~                       Haigron\inst{\ref{inst:0012}}
\and      J.-L.~                     Halbwachs\orcit{0000-0003-2968-6395}\inst{\ref{inst:0098}}
\and       N.C.~                        Hambly\orcit{0000-0002-9901-9064}\inst{\ref{inst:0085}}
\and       D.L.~                      Harrison\orcit{0000-0001-8687-6588}\inst{\ref{inst:0003},\ref{inst:0101}}
\and         J.~                 Hern\'{a}ndez\orcit{0000-0002-0361-4994}\inst{\ref{inst:0020}}
\and         D.~                    Hestroffer\orcit{0000-0003-0472-9459}\inst{\ref{inst:0074}}
\and       S.T.~                       Hodgkin\orcit{0000-0002-5470-3962}\inst{\ref{inst:0003}}
\and         B.~                          Holl\orcit{0000-0001-6220-3266}\inst{\ref{inst:0019},\ref{inst:0055}}
\and         K.~                    Jan{\ss}en\orcit{0000-0002-8163-2493}\inst{\ref{inst:0107}}
\and         G.~          Jevardat de Fombelle\inst{\ref{inst:0019}}
\and         A.~                 Krone-Martins\orcit{0000-0002-2308-6623}\inst{\ref{inst:0110},\ref{inst:0111}}
\and       A.C.~                     Lanzafame\orcit{0000-0002-2697-3607}\inst{\ref{inst:0112},\ref{inst:0113}}
\and         W.~                  L\"{ o}ffler\inst{\ref{inst:0015}}
\and         O.~                       Marchal\orcit{ 0000-0001-7461-892}\inst{\ref{inst:0098}}
\and       P.M.~                       Marrese\orcit{0000-0002-8162-3810}\inst{\ref{inst:0116},\ref{inst:0117}}
\and         A.~                      Moitinho\orcit{0000-0003-0822-5995}\inst{\ref{inst:0110}}
\and         P.~                       Osborne\inst{\ref{inst:0003}}
\and         E.~                       Pancino\orcit{0000-0003-0788-5879}\inst{\ref{inst:0030},\ref{inst:0117}}
\and         A.~                  Recio-Blanco\orcit{0000-0002-6550-7377}\inst{\ref{inst:0001}}
\and         C.~                     Reyl\'{e}\orcit{0000-0003-2258-2403}\inst{\ref{inst:0125}}
\and         M.~                        Riello\orcit{0000-0002-3134-0935}\inst{\ref{inst:0003}}
\and         L.~                     Rimoldini\orcit{0000-0002-0306-585X}\inst{\ref{inst:0055}}
\and         T.~                      Roegiers\orcit{0000-0002-1231-4440}\inst{\ref{inst:0128}}
\and         J.~                       Rybizki\orcit{0000-0002-0993-6089}\inst{\ref{inst:0034}}
\and       L.M.~                         Sarro\orcit{0000-0002-5622-5191}\inst{\ref{inst:0130}}
\and         C.~                        Siopis\orcit{0000-0002-6267-2924}\inst{\ref{inst:0028}}
\and         M.~                         Smith\inst{\ref{inst:0060}}
\and         A.~                      Sozzetti\orcit{0000-0002-7504-365X}\inst{\ref{inst:0007}}
\and         E.~                       Utrilla\inst{\ref{inst:0021}}
\and         M.~                   van Leeuwen\orcit{0000-0001-9698-2392}\inst{\ref{inst:0003}}
\and         U.~                         Abbas\orcit{0000-0002-5076-766X}\inst{\ref{inst:0007}}
\and         P.~               \'{A}brah\'{a}m\orcit{0000-0001-6015-646X}\inst{\ref{inst:0137},\ref{inst:0138}}
\and         A.~                Abreu Aramburu\inst{\ref{inst:0091}}
\and         C.~                         Aerts\orcit{0000-0003-1822-7126}\inst{\ref{inst:0140},\ref{inst:0141},\ref{inst:0034}}
\and       J.J.~                        Aguado\inst{\ref{inst:0130}}
\and         M.~                          Ajaj\inst{\ref{inst:0012}}
\and         F.~                 Aldea-Montero\inst{\ref{inst:0020}}
\and         G.~                     Altavilla\orcit{0000-0002-9934-1352}\inst{\ref{inst:0116},\ref{inst:0117}}
\and       M.A.~                   \'{A}lvarez\orcit{0000-0002-6786-2620}\inst{\ref{inst:0093}}
\and         J.~                         Alves\orcit{0000-0002-4355-0921}\inst{\ref{inst:0149}}
\and       R.I.~                      Anderson\orcit{0000-0001-8089-4419}\inst{\ref{inst:0150}}
\and         E.~                Anglada Varela\orcit{0000-0001-7563-0689}\inst{\ref{inst:0091}}
\and         T.~                        Antoja\orcit{0000-0003-2595-5148}\inst{\ref{inst:0022}}
\and         D.~                        Baines\orcit{0000-0002-6923-3756}\inst{\ref{inst:0066}}
\and       S.G.~                         Baker\orcit{0000-0002-6436-1257}\inst{\ref{inst:0060}}
\and         L.~        Balaguer-N\'{u}\~{n}ez\orcit{0000-0001-9789-7069}\inst{\ref{inst:0022}}
\and         E.~                      Balbinot\orcit{0000-0002-1322-3153}\inst{\ref{inst:0156}}
\and         Z.~                         Balog\orcit{0000-0003-1748-2926}\inst{\ref{inst:0015},\ref{inst:0034}}
\and         C.~                       Barache\inst{\ref{inst:0068}}
\and         D.~                       Barbato\inst{\ref{inst:0019},\ref{inst:0007}}
\and         M.~                        Barros\orcit{0000-0002-9728-9618}\inst{\ref{inst:0110}}
\and       M.A.~                       Barstow\orcit{0000-0002-7116-3259}\inst{\ref{inst:0163}}
\and         S.~                 Bartolom\'{e}\orcit{0000-0002-6290-6030}\inst{\ref{inst:0022}}
\and      J.-L.~                     Bassilana\inst{\ref{inst:0165}}
\and         N.~                       Bauchet\inst{\ref{inst:0012}}
\and         U.~                      Becciani\orcit{0000-0002-4389-8688}\inst{\ref{inst:0112}}
\and         M.~                    Bellazzini\orcit{0000-0001-8200-810X}\inst{\ref{inst:0043}}
\and         A.~                     Berihuete\orcit{0000-0002-8589-4423}\inst{\ref{inst:0169}}
\and         M.~                        Bernet\orcit{0000-0001-7503-1010}\inst{\ref{inst:0022}}
\and         S.~                       Bertone\orcit{0000-0001-9885-8440}\inst{\ref{inst:0171},\ref{inst:0172},\ref{inst:0007}}
\and         L.~                       Bianchi\orcit{0000-0002-7999-4372}\inst{\ref{inst:0174}}
\and         A.~                    Binnenfeld\orcit{0000-0002-9319-3838}\inst{\ref{inst:0175}}
\and         S.~               Blanco-Cuaresma\orcit{0000-0002-1584-0171}\inst{\ref{inst:0176}}
\and         T.~                          Boch\orcit{0000-0001-5818-2781}\inst{\ref{inst:0098}}
\and         A.~                       Bombrun\inst{\ref{inst:0178}}
\and         D.~                       Bossini\orcit{0000-0002-9480-8400}\inst{\ref{inst:0179}}
\and         S.~                    Bouquillon\inst{\ref{inst:0068},\ref{inst:0181}}
\and         A.~                     Bragaglia\orcit{0000-0002-0338-7883}\inst{\ref{inst:0043}}
\and         L.~                      Bramante\inst{\ref{inst:0051}}
\and         E.~                        Breedt\orcit{0000-0001-6180-3438}\inst{\ref{inst:0003}}
\and         A.~                       Bressan\orcit{0000-0002-7922-8440}\inst{\ref{inst:0185}}
\and         N.~                     Brouillet\orcit{0000-0002-3274-7024}\inst{\ref{inst:0017}}
\and         E.~                    Brugaletta\orcit{0000-0003-2598-6737}\inst{\ref{inst:0112}}
\and         B.~                   Bucciarelli\orcit{0000-0002-5303-0268}\inst{\ref{inst:0007},\ref{inst:0040}}
\and         A.~                       Burlacu\inst{\ref{inst:0190}}
\and       A.G.~                     Butkevich\orcit{0000-0002-4098-3588}\inst{\ref{inst:0007}}
\and         R.~                         Buzzi\orcit{0000-0001-9389-5701}\inst{\ref{inst:0007}}
\and         E.~                        Caffau\orcit{0000-0001-6011-6134}\inst{\ref{inst:0012}}
\and         R.~                   Cancelliere\orcit{0000-0002-9120-3799}\inst{\ref{inst:0194}}
\and         T.~                 Cantat-Gaudin\orcit{0000-0001-8726-2588}\inst{\ref{inst:0022},\ref{inst:0034}}
\and         R.~                      Carballo\orcit{0000-0001-7412-2498}\inst{\ref{inst:0197}}
\and         T.~                      Carlucci\inst{\ref{inst:0068}}
\and       M.I.~                     Carnerero\orcit{0000-0001-5843-5515}\inst{\ref{inst:0007}}
\and       J.M.~                      Carrasco\orcit{0000-0002-3029-5853}\inst{\ref{inst:0022}}
\and         L.~                   Casamiquela\orcit{0000-0001-5238-8674}\inst{\ref{inst:0017},\ref{inst:0012}}
\and         M.~                    Castellani\orcit{0000-0002-7650-7428}\inst{\ref{inst:0116}}
\and         A.~                 Castro-Ginard\orcit{0000-0002-9419-3725}\inst{\ref{inst:0008}}
\and         L.~                        Chaoul\inst{\ref{inst:0027}}
\and         P.~                       Charlot\orcit{0000-0002-9142-716X}\inst{\ref{inst:0017}}
\and         L.~                        Chemin\orcit{0000-0002-3834-7937}\inst{\ref{inst:0207}}
\and         V.~                    Chiaramida\inst{\ref{inst:0051}}
\and         A.~                     Chiavassa\orcit{0000-0003-3891-7554}\inst{\ref{inst:0001}}
\and         N.~                       Chornay\orcit{0000-0002-8767-3907}\inst{\ref{inst:0003}}
\and         G.~                     Comoretto\inst{\ref{inst:0066},\ref{inst:0212}}
\and         G.~                      Contursi\orcit{0000-0001-5370-1511}\inst{\ref{inst:0001}}
\and       W.J.~                        Cooper\orcit{0000-0003-3501-8967}\inst{\ref{inst:0214},\ref{inst:0007}}
\and         T.~                        Cornez\inst{\ref{inst:0165}}
\and         S.~                        Cowell\inst{\ref{inst:0003}}
\and         F.~                         Crifo\inst{\ref{inst:0012}}
\and         M.~                       Cropper\orcit{0000-0003-4571-9468}\inst{\ref{inst:0060}}
\and         M.~                        Crosta\orcit{0000-0003-4369-3786}\inst{\ref{inst:0007},\ref{inst:0221}}
\and         C.~                       Crowley\inst{\ref{inst:0178}}
\and         C.~                       Dafonte\orcit{0000-0003-4693-7555}\inst{\ref{inst:0093}}
\and         A.~                    Dapergolas\inst{\ref{inst:0072}}
\and         P.~                         David\inst{\ref{inst:0074}}
\and         P.~                    de Laverny\orcit{0000-0002-2817-4104}\inst{\ref{inst:0001}}
\and         F.~                      De Luise\orcit{0000-0002-6570-8208}\inst{\ref{inst:0227}}
\and         R.~                      De March\orcit{0000-0003-0567-842X}\inst{\ref{inst:0051}}
\and         J.~                     De Ridder\orcit{0000-0001-6726-2863}\inst{\ref{inst:0140}}
\and         R.~                      de Souza\inst{\ref{inst:0230}}
\and         A.~                     de Torres\inst{\ref{inst:0178}}
\and       E.F.~                    del Peloso\inst{\ref{inst:0015}}
\and         E.~                      del Pozo\inst{\ref{inst:0021}}
\and         A.~                       Delgado\inst{\ref{inst:0090}}
\and      J.-B.~                       Delisle\orcit{0000-0001-5844-9888}\inst{\ref{inst:0019}}
\and         C.~                      Demouchy\inst{\ref{inst:0236}}
\and       T.E.~                 Dharmawardena\orcit{0000-0002-9583-5216}\inst{\ref{inst:0034}}
\and         S.~                       Diakite\inst{\ref{inst:0238}}
\and         C.~                        Diener\inst{\ref{inst:0003}}
\and         E.~                     Distefano\orcit{0000-0002-2448-2513}\inst{\ref{inst:0112}}
\and         C.~                       Dolding\inst{\ref{inst:0060}}
\and         H.~                          Enke\orcit{0000-0002-2366-8316}\inst{\ref{inst:0107}}
\and         C.~                         Fabre\inst{\ref{inst:0243}}
\and         M.~                      Fabrizio\orcit{0000-0001-5829-111X}\inst{\ref{inst:0116},\ref{inst:0117}}
\and         S.~                       Faigler\orcit{0000-0002-8368-5724}\inst{\ref{inst:0246}}
\and         G.~                      Fedorets\orcit{0000-0002-8418-4809}\inst{\ref{inst:0119},\ref{inst:0248}}
\and         P.~                      Fernique\orcit{0000-0002-3304-2923}\inst{\ref{inst:0098},\ref{inst:0250}}
\and         F.~                      Figueras\orcit{0000-0002-3393-0007}\inst{\ref{inst:0022}}
\and         Y.~                      Fournier\orcit{0000-0002-6633-9088}\inst{\ref{inst:0107}}
\and         C.~                        Fouron\inst{\ref{inst:0190}}
\and         F.~                     Fragkoudi\orcit{0000-0002-0897-3013}\inst{\ref{inst:0254},\ref{inst:0255},\ref{inst:0256}}
\and         M.~                           Gai\orcit{0000-0001-9008-134X}\inst{\ref{inst:0007}}
\and         A.~              Garcia-Gutierrez\inst{\ref{inst:0022}}
\and         M.~              Garcia-Reinaldos\inst{\ref{inst:0020}}
\and         M.~             Garc\'{i}a-Torres\orcit{0000-0002-6867-7080}\inst{\ref{inst:0260}}
\and         A.~                      Garofalo\orcit{0000-0002-5907-0375}\inst{\ref{inst:0043}}
\and         A.~                         Gavel\orcit{0000-0002-2963-722X}\inst{\ref{inst:0049}}
\and         P.~                        Gavras\orcit{0000-0002-4383-4836}\inst{\ref{inst:0090}}
\and         E.~                       Gerlach\orcit{0000-0002-9533-2168}\inst{\ref{inst:0023}}
\and         R.~                         Geyer\orcit{0000-0001-6967-8707}\inst{\ref{inst:0023}}
\and         P.~                      Giacobbe\orcit{0000-0001-7034-7024}\inst{\ref{inst:0007}}
\and         G.~                       Gilmore\orcit{0000-0003-4632-0213}\inst{\ref{inst:0003}}
\and         S.~                        Girona\orcit{0000-0002-1975-1918}\inst{\ref{inst:0268}}
\and         G.~                     Giuffrida\inst{\ref{inst:0116}}
\and         R.~                         Gomel\inst{\ref{inst:0246}}
\and         A.~                         Gomez\orcit{0000-0002-3796-3690}\inst{\ref{inst:0093}}
\and         J.~    Gonz\'{a}lez-N\'{u}\~{n}ez\orcit{0000-0001-5311-5555}\inst{\ref{inst:0081},\ref{inst:0273}}
\and         I.~   Gonz\'{a}lez-Santamar\'{i}a\orcit{0000-0002-8537-9384}\inst{\ref{inst:0093}}
\and       J.J.~            Gonz\'{a}lez-Vidal\inst{\ref{inst:0022}}
\and         M.~                       Granvik\orcit{0000-0002-5624-1888}\inst{\ref{inst:0119},\ref{inst:0277}}
\and         P.~                      Guillout\inst{\ref{inst:0098}}
\and         J.~                       Guiraud\inst{\ref{inst:0027}}
\and         R.~     Guti\'{e}rrez-S\'{a}nchez\inst{\ref{inst:0066}}
\and       L.P.~                           Guy\orcit{0000-0003-0800-8755}\inst{\ref{inst:0055},\ref{inst:0282}}
\and         D.~                Hatzidimitriou\orcit{0000-0002-5415-0464}\inst{\ref{inst:0283},\ref{inst:0072}}
\and         M.~                        Hauser\inst{\ref{inst:0034},\ref{inst:0286}}
\and         M.~                       Haywood\orcit{0000-0003-0434-0400}\inst{\ref{inst:0012}}
\and         A.~                        Helmer\inst{\ref{inst:0165}}
\and         A.~                         Helmi\orcit{0000-0003-3937-7641}\inst{\ref{inst:0156}}
\and       M.H.~                     Sarmiento\orcit{0000-0003-4252-5115}\inst{\ref{inst:0021}}
\and       S.L.~                       Hidalgo\orcit{0000-0002-0002-9298}\inst{\ref{inst:0291},\ref{inst:0292}}
\and         N.~                   H\l{}adczuk\orcit{0000-0001-9163-4209}\inst{\ref{inst:0020},\ref{inst:0294}}
\and         D.~                         Hobbs\orcit{0000-0002-2696-1366}\inst{\ref{inst:0025}}
\and         G.~                       Holland\inst{\ref{inst:0003}}
\and       H.E.~                        Huckle\inst{\ref{inst:0060}}
\and         K.~                       Jardine\inst{\ref{inst:0298}}
\and         G.~                    Jasniewicz\inst{\ref{inst:0299}}
\and         A.~          Jean-Antoine Piccolo\orcit{0000-0001-8622-212X}\inst{\ref{inst:0027}}
\and     \'{O}.~            Jim\'{e}nez-Arranz\orcit{0000-0001-7434-5165}\inst{\ref{inst:0022}}
\and         J.~             Juaristi Campillo\inst{\ref{inst:0015}}
\and         F.~                         Julbe\inst{\ref{inst:0022}}
\and         L.~                     Karbevska\inst{\ref{inst:0055},\ref{inst:0305}}
\and         P.~                      Kervella\orcit{0000-0003-0626-1749}\inst{\ref{inst:0306}}
\and         S.~                        Khanna\orcit{0000-0002-2604-4277}\inst{\ref{inst:0156},\ref{inst:0007}}
\and         G.~                    Kordopatis\orcit{0000-0002-9035-3920}\inst{\ref{inst:0001}}
\and       A.J.~                          Korn\orcit{0000-0002-3881-6756}\inst{\ref{inst:0049}}
\and      \'{A}~                K\'{o}sp\'{a}l\orcit{\'{u}t 15-17, 1121 }\inst{\ref{inst:0137},\ref{inst:0034},\ref{inst:0138}}
\and         Z.~           Kostrzewa-Rutkowska\inst{\ref{inst:0008},\ref{inst:0315}}
\and         K.~                Kruszy\'{n}ska\orcit{0000-0002-2729-5369}\inst{\ref{inst:0316}}
\and         M.~                           Kun\orcit{0000-0002-7538-5166}\inst{\ref{inst:0137}}
\and         P.~                       Laizeau\inst{\ref{inst:0318}}
\and         S.~                       Lambert\orcit{0000-0001-6759-5502}\inst{\ref{inst:0068}}
\and       A.F.~                         Lanza\orcit{0000-0001-5928-7251}\inst{\ref{inst:0112}}
\and         Y.~                         Lasne\inst{\ref{inst:0165}}
\and      J.-F.~                    Le Campion\inst{\ref{inst:0017}}
\and         Y.~                      Lebreton\orcit{0000-0002-4834-2144}\inst{\ref{inst:0306},\ref{inst:0324}}
\and         T.~                     Lebzelter\orcit{0000-0002-0702-7551}\inst{\ref{inst:0149}}
\and         S.~                        Leccia\orcit{0000-0001-5685-6930}\inst{\ref{inst:0326}}
\and         N.~                       Leclerc\inst{\ref{inst:0012}}
\and         I.~                 Lecoeur-Taibi\orcit{0000-0003-0029-8575}\inst{\ref{inst:0055}}
\and         S.~                          Liao\orcit{0000-0002-9346-0211}\inst{\ref{inst:0329},\ref{inst:0007},\ref{inst:0331}}
\and       E.L.~                        Licata\orcit{0000-0002-5203-0135}\inst{\ref{inst:0007}}
\and     H.E.P.~                  Lindstr{\o}m\inst{\ref{inst:0007},\ref{inst:0334},\ref{inst:0335}}
\and       T.A.~                        Lister\orcit{0000-0002-3818-7769}\inst{\ref{inst:0336}}
\and         E.~                       Livanou\orcit{0000-0003-0628-2347}\inst{\ref{inst:0283}}
\and         A.~                         Lobel\orcit{0000-0001-5030-019X}\inst{\ref{inst:0004}}
\and         A.~                         Lorca\inst{\ref{inst:0021}}
\and         C.~                          Loup\inst{\ref{inst:0098}}
\and         P.~                 Madrero Pardo\inst{\ref{inst:0022}}
\and         A.~               Magdaleno Romeo\inst{\ref{inst:0190}}
\and         S.~                       Managau\inst{\ref{inst:0165}}
\and       R.G.~                          Mann\orcit{0000-0002-0194-325X}\inst{\ref{inst:0085}}
\and         M.~                      Manteiga\orcit{0000-0002-7711-5581}\inst{\ref{inst:0345}}
\and       J.M.~                      Marchant\orcit{0000-0002-3678-3145}\inst{\ref{inst:0346}}
\and         M.~                       Marconi\orcit{0000-0002-1330-2927}\inst{\ref{inst:0326}}
\and         J.~                        Marcos\inst{\ref{inst:0066}}
\and     M.M.S.~                 Marcos Santos\inst{\ref{inst:0081}}
\and         D.~                Mar\'{i}n Pina\orcit{0000-0001-6482-1842}\inst{\ref{inst:0022}}
\and         S.~                      Marinoni\orcit{0000-0001-7990-6849}\inst{\ref{inst:0116},\ref{inst:0117}}
\and         F.~                       Marocco\orcit{0000-0001-7519-1700}\inst{\ref{inst:0353}}
\and       D.J.~                      Marshall\orcit{0000-0003-3956-3524}\inst{\ref{inst:0354}}
\and         L.~                   Martin Polo\inst{\ref{inst:0081}}
\and       J.M.~            Mart\'{i}n-Fleitas\orcit{0000-0002-8594-569X}\inst{\ref{inst:0021}}
\and         G.~                        Marton\orcit{0000-0002-1326-1686}\inst{\ref{inst:0137}}
\and         N.~                          Mary\inst{\ref{inst:0165}}
\and         A.~                         Masip\orcit{0000-0003-1419-0020}\inst{\ref{inst:0022}}
\and         D.~                       Massari\orcit{0000-0001-8892-4301}\inst{\ref{inst:0043}}
\and         A.~          Mastrobuono-Battisti\orcit{0000-0002-2386-9142}\inst{\ref{inst:0012}}
\and         T.~                         Mazeh\orcit{0000-0002-3569-3391}\inst{\ref{inst:0246}}
\and       P.J.~                      McMillan\orcit{0000-0002-8861-2620}\inst{\ref{inst:0025}}
\and         S.~                       Messina\orcit{0000-0002-2851-2468}\inst{\ref{inst:0112}}
\and         D.~                      Michalik\orcit{0000-0002-7618-6556}\inst{\ref{inst:0010}}
\and       N.R.~                        Millar\inst{\ref{inst:0003}}
\and         A.~                         Mints\orcit{0000-0002-8440-1455}\inst{\ref{inst:0107}}
\and         D.~                        Molina\orcit{0000-0003-4814-0275}\inst{\ref{inst:0022}}
\and         R.~                      Molinaro\orcit{0000-0003-3055-6002}\inst{\ref{inst:0326}}
\and         L.~                    Moln\'{a}r\orcit{0000-0002-8159-1599}\inst{\ref{inst:0137},\ref{inst:0371},\ref{inst:0138}}
\and         G.~                        Monari\orcit{0000-0002-6863-0661}\inst{\ref{inst:0098}}
\and         M.~                   Mongui\'{o}\orcit{0000-0002-4519-6700}\inst{\ref{inst:0022}}
\and         P.~                   Montegriffo\orcit{0000-0001-5013-5948}\inst{\ref{inst:0043}}
\and         A.~                       Montero\inst{\ref{inst:0021}}
\and         R.~                           Mor\orcit{0000-0002-8179-6527}\inst{\ref{inst:0022}}
\and         A.~                          Mora\inst{\ref{inst:0021}}
\and         R.~                    Morbidelli\orcit{0000-0001-7627-4946}\inst{\ref{inst:0007}}
\and         T.~                         Morel\orcit{0000-0002-8176-4816}\inst{\ref{inst:0083}}
\and         D.~                        Morris\inst{\ref{inst:0085}}
\and         T.~                      Muraveva\orcit{0000-0002-0969-1915}\inst{\ref{inst:0043}}
\and       C.P.~                        Murphy\inst{\ref{inst:0020}}
\and         I.~                       Musella\orcit{0000-0001-5909-6615}\inst{\ref{inst:0326}}
\and         Z.~                          Nagy\orcit{0000-0002-3632-1194}\inst{\ref{inst:0137}}
\and         L.~                         Noval\inst{\ref{inst:0165}}
\and         F.~                     Oca\~{n}a\inst{\ref{inst:0066},\ref{inst:0387}}
\and         A.~                         Ogden\inst{\ref{inst:0003}}
\and         C.~                     Ordenovic\inst{\ref{inst:0001}}
\and       J.O.~                        Osinde\inst{\ref{inst:0090}}
\and         C.~                        Pagani\orcit{0000-0001-5477-4720}\inst{\ref{inst:0163}}
\and         I.~                        Pagano\orcit{0000-0001-9573-4928}\inst{\ref{inst:0112}}
\and         L.~                     Palaversa\orcit{0000-0003-3710-0331}\inst{\ref{inst:0393},\ref{inst:0003}}
\and       P.A.~                       Palicio\orcit{0000-0002-7432-8709}\inst{\ref{inst:0001}}
\and         L.~               Pallas-Quintela\orcit{0000-0001-9296-3100}\inst{\ref{inst:0093}}
\and         A.~                        Panahi\orcit{0000-0001-5850-4373}\inst{\ref{inst:0246}}
\and         S.~               Payne-Wardenaar\inst{\ref{inst:0015}}
\and         X.~         Pe\~{n}alosa Esteller\inst{\ref{inst:0022}}
\and      J.-M.~                         Petit\orcit{0000-0003-0407-2266}\inst{\ref{inst:0125}}
\and         B.~                        Pichon\orcit{0000 0000 0062 1449}\inst{\ref{inst:0001}}
\and       A.M.~                    Piersimoni\orcit{0000-0002-8019-3708}\inst{\ref{inst:0227}}
\and      F.-X.~                        Pineau\orcit{0000-0002-2335-4499}\inst{\ref{inst:0098}}
\and         E.~                        Plachy\orcit{0000-0002-5481-3352}\inst{\ref{inst:0137},\ref{inst:0371},\ref{inst:0138}}
\and         G.~                          Plum\inst{\ref{inst:0012}}
\and         E.~                        Poggio\orcit{0000-0003-3793-8505}\inst{\ref{inst:0001},\ref{inst:0007}}
\and         A.~                      Pr\v{s}a\orcit{0000-0002-1913-0281}\inst{\ref{inst:0410}}
\and         L.~                        Pulone\orcit{0000-0002-5285-998X}\inst{\ref{inst:0116}}
\and         E.~                        Racero\orcit{0000-0002-6101-9050}\inst{\ref{inst:0081},\ref{inst:0387}}
\and         S.~                       Ragaini\inst{\ref{inst:0043}}
\and         M.~                        Rainer\orcit{0000-0002-8786-2572}\inst{\ref{inst:0030},\ref{inst:0416}}
\and       C.M.~                       Raiteri\orcit{0000-0003-1784-2784}\inst{\ref{inst:0007}}
\and         P.~                         Ramos\orcit{0000-0002-5080-7027}\inst{\ref{inst:0022},\ref{inst:0098}}
\and         M.~                  Ramos-Lerate\inst{\ref{inst:0066}}
\and         P.~                  Re Fiorentin\orcit{0000-0002-4995-0475}\inst{\ref{inst:0007}}
\and         S.~                        Regibo\inst{\ref{inst:0140}}
\and       P.J.~                      Richards\inst{\ref{inst:0423}}
\and         C.~                     Rios Diaz\inst{\ref{inst:0090}}
\and         V.~                        Ripepi\orcit{0000-0003-1801-426X}\inst{\ref{inst:0326}}
\and         A.~                          Riva\orcit{0000-0002-6928-8589}\inst{\ref{inst:0007}}
\and      H.-W.~                           Rix\orcit{0000-0003-4996-9069}\inst{\ref{inst:0034}}
\and         G.~                         Rixon\orcit{0000-0003-4399-6568}\inst{\ref{inst:0003}}
\and         N.~                      Robichon\orcit{0000-0003-4545-7517}\inst{\ref{inst:0012}}
\and       A.C.~                         Robin\orcit{0000-0001-8654-9499}\inst{\ref{inst:0125}}
\and         C.~                         Robin\inst{\ref{inst:0165}}
\and         M.~                       Roelens\orcit{0000-0003-0876-4673}\inst{\ref{inst:0019}}
\and     H.R.O.~                        Rogues\inst{\ref{inst:0236}}
\and         L.~                    Rohrbasser\inst{\ref{inst:0055}}
\and         M.~              Romero-G\'{o}mez\orcit{0000-0003-3936-1025}\inst{\ref{inst:0022}}
\and         N.~                        Rowell\orcit{0000-0003-3809-1895}\inst{\ref{inst:0085}}
\and         F.~                         Royer\orcit{0000-0002-9374-8645}\inst{\ref{inst:0012}}
\and         D.~                    Ruz Mieres\orcit{0000-0002-9455-157X}\inst{\ref{inst:0003}}
\and       K.A.~                       Rybicki\orcit{0000-0002-9326-9329}\inst{\ref{inst:0316}}
\and         G.~                      Sadowski\orcit{0000-0002-3411-1003}\inst{\ref{inst:0028}}
\and         A.~        S\'{a}ez N\'{u}\~{n}ez\inst{\ref{inst:0022}}
\and         A.~       Sagrist\`{a} Sell\'{e}s\orcit{0000-0001-6191-2028}\inst{\ref{inst:0015}}
\and         J.~                      Sahlmann\orcit{0000-0001-9525-3673}\inst{\ref{inst:0090}}
\and         E.~                      Salguero\inst{\ref{inst:0091}}
\and         N.~                       Samaras\orcit{0000-0001-8375-6652}\inst{\ref{inst:0004},\ref{inst:0446}}
\and         V.~               Sanchez Gimenez\orcit{0000-0003-1797-3557}\inst{\ref{inst:0022}}
\and         N.~                         Sanna\orcit{0000-0001-9275-9492}\inst{\ref{inst:0030}}
\and         R.~                 Santove\~{n}a\orcit{0000-0002-9257-2131}\inst{\ref{inst:0093}}
\and         M.~                       Sarasso\orcit{0000-0001-5121-0727}\inst{\ref{inst:0007}}
\and       M.~                    Schultheis\orcit{0000-0002-6590-1657}\inst{\ref{inst:0001}}
\and         E.~                       Sciacca\orcit{0000-0002-5574-2787}\inst{\ref{inst:0112}}
\and         M.~                         Segol\inst{\ref{inst:0236}}
\and       J.C.~                       Segovia\inst{\ref{inst:0081}}
\and         D.~                 S\'{e}gransan\orcit{0000-0003-2355-8034}\inst{\ref{inst:0019}}
\and         D.~                        Semeux\inst{\ref{inst:0243}}
\and         S.~                        Shahaf\orcit{0000-0001-9298-8068}\inst{\ref{inst:0457}}
\and       H.I.~                      Siddiqui\orcit{0000-0003-1853-6033}\inst{\ref{inst:0458}}
\and         A.~                       Siebert\orcit{0000-0001-8059-2840}\inst{\ref{inst:0098},\ref{inst:0250}}
\and         L.~                       Siltala\orcit{0000-0002-6938-794X}\inst{\ref{inst:0119}}
\and         A.~                       Silvelo\orcit{0000-0002-5126-6365}\inst{\ref{inst:0093}}
\and         E.~                        Slezak\inst{\ref{inst:0001}}
\and         I.~                        Slezak\inst{\ref{inst:0001}}
\and       R.L.~                         Smart\orcit{0000-0002-4424-4766}\inst{\ref{inst:0007}}
\and       O.N.~                        Snaith\inst{\ref{inst:0012}}
\and         E.~                        Solano\inst{\ref{inst:0467}}
\and         F.~                       Solitro\inst{\ref{inst:0051}}
\and         D.~                        Souami\orcit{0000-0003-4058-0815}\inst{\ref{inst:0306},\ref{inst:0470}}
\and         J.~                       Souchay\inst{\ref{inst:0068}}
\and         A.~                        Spagna\orcit{0000-0003-1732-2412}\inst{\ref{inst:0007}}
\and         L.~                         Spina\orcit{0000-0002-9760-6249}\inst{\ref{inst:0009}}
\and         F.~                         Spoto\orcit{0000-0001-7319-5847}\inst{\ref{inst:0176}}
\and       I.A.~                        Steele\orcit{0000-0001-8397-5759}\inst{\ref{inst:0346}}
\and         H.~            Steidelm\"{ u}ller\inst{\ref{inst:0023}}
\and       C.A.~                    Stephenson\inst{\ref{inst:0066},\ref{inst:0478}}
\and         M.~                  S\"{ u}veges\orcit{0000-0003-3017-5322}\inst{\ref{inst:0479}}
\and         J.~                        Surdej\orcit{0000-0002-7005-1976}\inst{\ref{inst:0083},\ref{inst:0481}}
\and         L.~                      Szabados\orcit{0000-0002-2046-4131}\inst{\ref{inst:0137}}
\and         E.~                  Szegedi-Elek\orcit{0000-0001-7807-6644}\inst{\ref{inst:0137}}
\and         F.~                         Taris\inst{\ref{inst:0068}}
\and       M.B.~                        Taylor\orcit{0000-0002-4209-1479}\inst{\ref{inst:0485}}
\and         R.~                      Teixeira\orcit{0000-0002-6806-6626}\inst{\ref{inst:0230}}
\and         L.~                       Tolomei\orcit{0000-0002-3541-3230}\inst{\ref{inst:0051}}
\and         N.~                       Tonello\orcit{0000-0003-0550-1667}\inst{\ref{inst:0268}}
\and         F.~                         Torra\orcit{0000-0002-8429-299X}\inst{\ref{inst:0044}}
\and         J.~               Torra$^\dagger$\inst{\ref{inst:0022}}
\and         G.~                Torralba Elipe\orcit{0000-0001-8738-194X}\inst{\ref{inst:0093}}
\and         M.~                     Trabucchi\orcit{0000-0002-1429-2388}\inst{\ref{inst:0492},\ref{inst:0019}}
\and       A.T.~                       Tsounis\inst{\ref{inst:0494}}
\and         C.~                         Turon\orcit{0000-0003-1236-5157}\inst{\ref{inst:0012}}
\and         A.~                          Ulla\orcit{0000-0001-6424-5005}\inst{\ref{inst:0496}}
\and         N.~                         Unger\orcit{0000-0003-3993-7127}\inst{\ref{inst:0019}}
\and       M.V.~                      Vaillant\inst{\ref{inst:0165}}
\and         E.~                    van Dillen\inst{\ref{inst:0236}}
\and         W.~                    van Reeven\inst{\ref{inst:0500}}
\and         O.~                         Vanel\orcit{0000-0002-7898-0454}\inst{\ref{inst:0012}}
\and         A.~                     Vecchiato\orcit{0000-0003-1399-5556}\inst{\ref{inst:0007}}
\and         Y.~                         Viala\inst{\ref{inst:0012}}
\and         D.~                       Vicente\orcit{0000-0002-1584-1182}\inst{\ref{inst:0268}}
\and         S.~                     Voutsinas\inst{\ref{inst:0085}}
\and         M.~                        Weiler\inst{\ref{inst:0022}}
\and         T.~                        Wevers\orcit{0000-0002-4043-9400}\inst{\ref{inst:0003},\ref{inst:0508}}
\and      \L{}.~                   Wyrzykowski\orcit{0000-0002-9658-6151}\inst{\ref{inst:0316}}
\and         A.~                        Yoldas\inst{\ref{inst:0003}}
\and         P.~                         Yvard\inst{\ref{inst:0236}}
\and         H.~                          Zhao\orcit{0000-0003-2645-6869}\inst{\ref{inst:0001}}
\and         J.~                         Zorec\inst{\ref{inst:0513}}
\and         S.~                        Zucker\orcit{0000-0003-3173-3138}\inst{\ref{inst:0175}}
\and         T.~                       Zwitter\orcit{0000-0002-2325-8763}\inst{\ref{inst:0515}}
}
\institute{
     Universit\'{e} C\^{o}te d'Azur, Observatoire de la C\^{o}te d'Azur, CNRS, Laboratoire Lagrange, Bd de l'Observatoire, CS 34229, 06304 Nice Cedex 4, France\relax                                                                                                                                                                                              \label{inst:0001}
\and Institute of Astronomy, University of Cambridge, Madingley Road, Cambridge CB3 0HA, United Kingdom\relax                                                                                                                                                                                                                                                      \label{inst:0003}\vfill
\and Royal Observatory of Belgium, Ringlaan 3, 1180 Brussels, Belgium\relax                                                                                                                                                                                                                                                                                        \label{inst:0004}\vfill
\and INAF - Osservatorio Astrofisico di Torino, via Osservatorio 20, 10025 Pino Torinese (TO), Italy\relax                                                                                                                                                                                                                                                         \label{inst:0007}\vfill
\and Leiden Observatory, Leiden University, Niels Bohrweg 2, 2333 CA Leiden, The Netherlands\relax                                                                                                                                                                                                                                                                 \label{inst:0008}\vfill
\and Department of Physics, University of Helsinki, P.O. Box 64, 00014 Helsinki, Finland\relax                                                                                                                                                                                                                                                                     \label{inst:0119}\vfill
\and Finnish Geospatial Research Institute FGI, Geodeetinrinne 2, 02430 Masala, Finland\relax                                                                                                                                                                                                                                                                      \label{inst:0120}\vfill
\and Astronomisches Rechen-Institut, Zentrum f\"{ u}r Astronomie der Universit\"{ a}t Heidelberg, M\"{ o}nchhofstr. 12-14, 69120 Heidelberg, Germany\relax                                                                                                                                                                                                         \label{inst:0015}\vfill
\and INAF - Osservatorio astronomico di Padova, Vicolo Osservatorio 5, 35122 Padova, Italy\relax                                                                                                                                                                                                                                                                   \label{inst:0009}\vfill
\and European Space Agency (ESA), European Space Research and Technology Centre (ESTEC), Keplerlaan 1, 2201AZ, Noordwijk, The Netherlands\relax                                                                                                                                                                                                                    \label{inst:0010}\vfill
\and GEPI, Observatoire de Paris, Universit\'{e} PSL, CNRS, 5 Place Jules Janssen, 92190 Meudon, France\relax                                                                                                                                                                                                                                                      \label{inst:0012}\vfill
\and Univ. Grenoble Alpes, CNRS, IPAG, 38000 Grenoble, France\relax                                                                                                                                                                                                                                                                                                \label{inst:0013}\vfill
\and Laboratoire d'astrophysique de Bordeaux, Univ. Bordeaux, CNRS, B18N, all{\'e}e Geoffroy Saint-Hilaire, 33615 Pessac, France\relax                                                                                                                                                                                                                             \label{inst:0017}\vfill
\and Department of Astronomy, University of Geneva, Chemin Pegasi 51, 1290 Versoix, Switzerland\relax                                                                                                                                                                                                                                                              \label{inst:0019}\vfill
\and European Space Agency (ESA), European Space Astronomy Centre (ESAC), Camino bajo del Castillo, s/n, Urbanizacion Villafranca del Castillo, Villanueva de la Ca\~{n}ada, 28692 Madrid, Spain\relax                                                                                                                                                             \label{inst:0020}\vfill
\and Aurora Technology for European Space Agency (ESA), Camino bajo del Castillo, s/n, Urbanizacion Villafranca del Castillo, Villanueva de la Ca\~{n}ada, 28692 Madrid, Spain\relax                                                                                                                                                                               \label{inst:0021}\vfill
\and Institut de Ci\`{e}ncies del Cosmos (ICCUB), Universitat  de  Barcelona  (IEEC-UB), Mart\'{i} i  Franqu\`{e}s  1, 08028 Barcelona, Spain\relax                                                                                                                                                                                                                \label{inst:0022}\vfill
\and Lohrmann Observatory, Technische Universit\"{ a}t Dresden, Mommsenstra{\ss}e 13, 01062 Dresden, Germany\relax                                                                                                                                                                                                                                                 \label{inst:0023}\vfill
\and Lund Observatory, Department of Astronomy and Theoretical Physics, Lund University, Box 43, 22100 Lund, Sweden\relax                                                                                                                                                                                                                                          \label{inst:0025}\vfill
\and CNES Centre Spatial de Toulouse, 18 avenue Edouard Belin, 31401 Toulouse Cedex 9, France\relax                                                                                                                                                                                                                                                                \label{inst:0027}\vfill
\and Institut d'Astronomie et d'Astrophysique, Universit\'{e} Libre de Bruxelles CP 226, Boulevard du Triomphe, 1050 Brussels, Belgium\relax                                                                                                                                                                                                                       \label{inst:0028}\vfill
\and F.R.S.-FNRS, Rue d'Egmont 5, 1000 Brussels, Belgium\relax                                                                                                                                                                                                                                                                                                     \label{inst:0029}\vfill
\and INAF - Osservatorio Astrofisico di Arcetri, Largo Enrico Fermi 5, 50125 Firenze, Italy\relax                                                                                                                                                                                                                                                                  \label{inst:0030}\vfill
\and Max Planck Institute for Astronomy, K\"{ o}nigstuhl 17, 69117 Heidelberg, Germany\relax                                                                                                                                                                                                                                                                       \label{inst:0034}\vfill
\and European Space Agency (ESA, retired)\relax                                                                                                                                                                                                                                                                                                                    \label{inst:0037}\vfill
\and University of Turin, Department of Physics, Via Pietro Giuria 1, 10125 Torino, Italy\relax                                                                                                                                                                                                                                                                    \label{inst:0040}\vfill
\and INAF - Osservatorio di Astrofisica e Scienza dello Spazio di Bologna, via Piero Gobetti 93/3, 40129 Bologna, Italy\relax                                                                                                                                                                                                                                      \label{inst:0043}\vfill
\and DAPCOM for Institut de Ci\`{e}ncies del Cosmos (ICCUB), Universitat  de  Barcelona  (IEEC-UB), Mart\'{i} i  Franqu\`{e}s  1, 08028 Barcelona, Spain\relax                                                                                                                                                                                                     \label{inst:0044}\vfill
\and Observational Astrophysics, Division of Astronomy and Space Physics, Department of Physics and Astronomy, Uppsala University, Box 516, 751 20 Uppsala, Sweden\relax                                                                                                                                                                                           \label{inst:0049}\vfill
\and ALTEC S.p.a, Corso Marche, 79,10146 Torino, Italy\relax                                                                                                                                                                                                                                                                                                       \label{inst:0051}\vfill
\and S\`{a}rl, Geneva, Switzerland\relax                                                                                                                                                                                                                                                                                                                           \label{inst:0054}\vfill
\and Department of Astronomy, University of Geneva, Chemin d'Ecogia 16, 1290 Versoix, Switzerland\relax                                                                                                                                                                                                                                                            \label{inst:0055}\vfill
\and Mullard Space Science Laboratory, University College London, Holmbury St Mary, Dorking, Surrey RH5 6NT, United Kingdom\relax                                                                                                                                                                                                                                  \label{inst:0060}\vfill
\and Gaia DPAC Project Office, ESAC, Camino bajo del Castillo, s/n, Urbanizacion Villafranca del Castillo, Villanueva de la Ca\~{n}ada, 28692 Madrid, Spain\relax                                                                                                                                                                                                  \label{inst:0063}\vfill
\and Telespazio UK S.L. for European Space Agency (ESA), Camino bajo del Castillo, s/n, Urbanizacion Villafranca del Castillo, Villanueva de la Ca\~{n}ada, 28692 Madrid, Spain\relax                                                                                                                                                                              \label{inst:0066}\vfill
\and SYRTE, Observatoire de Paris, Universit\'{e} PSL, CNRS,  Sorbonne Universit\'{e}, LNE, 61 avenue de l'Observatoire 75014 Paris, France\relax                                                                                                                                                                                                                  \label{inst:0068}\vfill
\and National Observatory of Athens, I. Metaxa and Vas. Pavlou, Palaia Penteli, 15236 Athens, Greece\relax                                                                                                                                                                                                                                                         \label{inst:0072}\vfill
\and IMCCE, Observatoire de Paris, Universit\'{e} PSL, CNRS, Sorbonne Universit{\'e}, Univ. Lille, 77 av. Denfert-Rochereau, 75014 Paris, France\relax                                                                                                                                                                                                             \label{inst:0074}\vfill
\and Serco Gesti\'{o}n de Negocios for European Space Agency (ESA), Camino bajo del Castillo, s/n, Urbanizacion Villafranca del Castillo, Villanueva de la Ca\~{n}ada, 28692 Madrid, Spain\relax                                                                                                                                                                   \label{inst:0081}\vfill
\and Institut d'Astrophysique et de G\'{e}ophysique, Universit\'{e} de Li\`{e}ge, 19c, All\'{e}e du 6 Ao\^{u}t, B-4000 Li\`{e}ge, Belgium\relax                                                                                                                                                                                                                    \label{inst:0083}\vfill
\and CRAAG - Centre de Recherche en Astronomie, Astrophysique et G\'{e}ophysique, Route de l'Observatoire Bp 63 Bouzareah 16340 Algiers, Algeria\relax                                                                                                                                                                                                             \label{inst:0084}\vfill
\and Institute for Astronomy, University of Edinburgh, Royal Observatory, Blackford Hill, Edinburgh EH9 3HJ, United Kingdom\relax                                                                                                                                                                                                                                  \label{inst:0085}\vfill
\and RHEA for European Space Agency (ESA), Camino bajo del Castillo, s/n, Urbanizacion Villafranca del Castillo, Villanueva de la Ca\~{n}ada, 28692 Madrid, Spain\relax                                                                                                                                                                                            \label{inst:0090}\vfill
\and ATG Europe for European Space Agency (ESA), Camino bajo del Castillo, s/n, Urbanizacion Villafranca del Castillo, Villanueva de la Ca\~{n}ada, 28692 Madrid, Spain\relax                                                                                                                                                                                      \label{inst:0091}\vfill
\and CIGUS CITIC - Department of Computer Science and Information Technologies, University of A Coru\~{n}a, Campus de Elvi\~{n}a s/n, A Coru\~{n}a, 15071, Spain\relax                                                                                                                                                                                             \label{inst:0093}\vfill
\and Universit\'{e} de Strasbourg, CNRS, Observatoire astronomique de Strasbourg, UMR 7550,  11 rue de l'Universit\'{e}, 67000 Strasbourg, France\relax                                                                                                                                                                                                            \label{inst:0098}\vfill
\and Kavli Institute for Cosmology Cambridge, Institute of Astronomy, Madingley Road, Cambridge, CB3 0HA\relax                                                                                                                                                                                                                                                     \label{inst:0101}\vfill
\and Leibniz Institute for Astrophysics Potsdam (AIP), An der Sternwarte 16, 14482 Potsdam, Germany\relax                                                                                                                                                                                                                                                          \label{inst:0107}\vfill
\and CENTRA, Faculdade de Ci\^{e}ncias, Universidade de Lisboa, Edif. C8, Campo Grande, 1749-016 Lisboa, Portugal\relax                                                                                                                                                                                                                                            \label{inst:0110}\vfill
\and Department of Informatics, Donald Bren School of Information and Computer Sciences, University of California, Irvine, 5226 Donald Bren Hall, 92697-3440 CA Irvine, United States\relax                                                                                                                                                                        \label{inst:0111}\vfill
\and INAF - Osservatorio Astrofisico di Catania, via S. Sofia 78, 95123 Catania, Italy\relax                                                                                                                                                                                                                                                                       \label{inst:0112}\vfill
\and Dipartimento di Fisica e Astronomia ""Ettore Majorana"", Universit\`{a} di Catania, Via S. Sofia 64, 95123 Catania, Italy\relax                                                                                                                                                                                                                               \label{inst:0113}\vfill
\and INAF - Osservatorio Astronomico di Roma, Via Frascati 33, 00078 Monte Porzio Catone (Roma), Italy\relax                                                                                                                                                                                                                                                       \label{inst:0116}\vfill
\and Space Science Data Center - ASI, Via del Politecnico SNC, 00133 Roma, Italy\relax                                                                                                                                                                                                                                                                             \label{inst:0117}\vfill
\and Institut UTINAM CNRS UMR6213, Universit\'{e} Bourgogne Franche-Comt\'{e}, OSU THETA Franche-Comt\'{e} Bourgogne, Observatoire de Besan\c{c}on, BP1615, 25010 Besan\c{c}on Cedex, France\relax                                                                                                                                                                 \label{inst:0125}\vfill
\and HE Space Operations BV for European Space Agency (ESA), Keplerlaan 1, 2201AZ, Noordwijk, The Netherlands\relax                                                                                                                                                                                                                                                \label{inst:0128}\vfill
\and Dpto. de Inteligencia Artificial, UNED, c/ Juan del Rosal 16, 28040 Madrid, Spain\relax                                                                                                                                                                                                                                                                       \label{inst:0130}\vfill
\and Konkoly Observatory, Research Centre for Astronomy and Earth Sciences, E\"{ o}tv\"{ o}s Lor{\'a}nd Research Network (ELKH), MTA Centre of Excellence, Konkoly Thege Mikl\'{o}s \'{u}t 15-17, 1121 Budapest, Hungary\relax                                                                                                                                     \label{inst:0137}\vfill
\and ELTE E\"{ o}tv\"{ o}s Lor\'{a}nd University, Institute of Physics, 1117, P\'{a}zm\'{a}ny P\'{e}ter s\'{e}t\'{a}ny 1A, Budapest, Hungary\relax                                                                                                                                                                                                                 \label{inst:0138}\vfill
\and Instituut voor Sterrenkunde, KU Leuven, Celestijnenlaan 200D, 3001 Leuven, Belgium\relax                                                                                                                                                                                                                                                                      \label{inst:0140}\vfill
\and Department of Astrophysics/IMAPP, Radboud University, P.O.Box 9010, 6500 GL Nijmegen, The Netherlands\relax                                                                                                                                                                                                                                                   \label{inst:0141}\vfill
\and University of Vienna, Department of Astrophysics, T\"{ u}rkenschanzstra{\ss}e 17, A1180 Vienna, Austria\relax                                                                                                                                                                                                                                                 \label{inst:0149}\vfill
\and Institute of Physics, Laboratory of Astrophysics, Ecole Polytechnique F\'ed\'erale de Lausanne (EPFL), Observatoire de Sauverny, 1290 Versoix, Switzerland\relax                                                                                                                                                                                              \label{inst:0150}\vfill
\and Kapteyn Astronomical Institute, University of Groningen, Landleven 12, 9747 AD Groningen, The Netherlands\relax                                                                                                                                                                                                                                               \label{inst:0156}\vfill
\and School of Physics and Astronomy / Space Park Leicester, University of Leicester, University Road, Leicester LE1 7RH, United Kingdom\relax                                                                                                                                                                                                                     \label{inst:0163}\vfill
\and Thales Services for CNES Centre Spatial de Toulouse, 18 avenue Edouard Belin, 31401 Toulouse Cedex 9, France\relax                                                                                                                                                                                                                                            \label{inst:0165}\vfill
\and Depto. Estad\'istica e Investigaci\'on Operativa. Universidad de C\'adiz, Avda. Rep\'ublica Saharaui s/n, 11510 Puerto Real, C\'adiz, Spain\relax                                                                                                                                                                                                             \label{inst:0169}\vfill
\and Center for Research and Exploration in Space Science and Technology, University of Maryland Baltimore County, 1000 Hilltop Circle, Baltimore MD, USA\relax                                                                                                                                                                                                    \label{inst:0171}\vfill
\and GSFC - Goddard Space Flight Center, Code 698, 8800 Greenbelt Rd, 20771 MD Greenbelt, United States\relax                                                                                                                                                                                                                                                      \label{inst:0172}\vfill
\and EURIX S.r.l., Corso Vittorio Emanuele II 61, 10128, Torino, Italy\relax                                                                                                                                                                                                                                                                                       \label{inst:0174}\vfill
\and Porter School of the Environment and Earth Sciences, Tel Aviv University, Tel Aviv 6997801, Israel\relax                                                                                                                                                                                                                                                      \label{inst:0175}\vfill
\and Harvard-Smithsonian Center for Astrophysics, 60 Garden St., MS 15, Cambridge, MA 02138, USA\relax                                                                                                                                                                                                                                                             \label{inst:0176}\vfill
\and HE Space Operations BV for European Space Agency (ESA), Camino bajo del Castillo, s/n, Urbanizacion Villafranca del Castillo, Villanueva de la Ca\~{n}ada, 28692 Madrid, Spain\relax                                                                                                                                                                          \label{inst:0178}\vfill
\and Instituto de Astrof\'{i}sica e Ci\^{e}ncias do Espa\c{c}o, Universidade do Porto, CAUP, Rua das Estrelas, PT4150-762 Porto, Portugal\relax                                                                                                                                                                                                                    \label{inst:0179}\vfill
\and LFCA/DAS,Universidad de Chile,CNRS,Casilla 36-D, Santiago, Chile\relax                                                                                                                                                                                                                                                                                        \label{inst:0181}\vfill
\and SISSA - Scuola Internazionale Superiore di Studi Avanzati, via Bonomea 265, 34136 Trieste, Italy\relax                                                                                                                                                                                                                                                        \label{inst:0185}\vfill
\and Telespazio for CNES Centre Spatial de Toulouse, 18 avenue Edouard Belin, 31401 Toulouse Cedex 9, France\relax                                                                                                                                                                                                                                                 \label{inst:0190}\vfill
\and University of Turin, Department of Computer Sciences, Corso Svizzera 185, 10149 Torino, Italy\relax                                                                                                                                                                                                                                                           \label{inst:0194}\vfill
\and Dpto. de Matem\'{a}tica Aplicada y Ciencias de la Computaci\'{o}n, Univ. de Cantabria, ETS Ingenieros de Caminos, Canales y Puertos, Avda. de los Castros s/n, 39005 Santander, Spain\relax                                                                                                                                                                   \label{inst:0197}\vfill
\and Centro de Astronom\'{i}a - CITEVA, Universidad de Antofagasta, Avenida Angamos 601, Antofagasta 1270300, Chile\relax                                                                                                                                                                                                                                          \label{inst:0207}\vfill
\and DLR Gesellschaft f\"{ u}r Raumfahrtanwendungen (GfR) mbH M\"{ u}nchener Stra{\ss}e 20 , 82234 We{\ss}ling\relax                                                                                                                                                                                                                                               \label{inst:0212}\vfill
\and Centre for Astrophysics Research, University of Hertfordshire, College Lane, AL10 9AB, Hatfield, United Kingdom\relax                                                                                                                                                                                                                                         \label{inst:0214}\vfill
\and University of Turin, Mathematical Department ""G.Peano"", Via Carlo Alberto 10, 10123 Torino, Italy\relax                                                                                                                                                                                                                                                     \label{inst:0221}\vfill
\and INAF - Osservatorio Astronomico d'Abruzzo, Via Mentore Maggini, 64100 Teramo, Italy\relax                                                                                                                                                                                                                                                                     \label{inst:0227}\vfill
\and Instituto de Astronomia, Geof\`{i}sica e Ci\^{e}ncias Atmosf\'{e}ricas, Universidade de S\~{a}o Paulo, Rua do Mat\~{a}o, 1226, Cidade Universitaria, 05508-900 S\~{a}o Paulo, SP, Brazil\relax                                                                                                                                                                \label{inst:0230}\vfill
\and APAVE SUDEUROPE SAS for CNES Centre Spatial de Toulouse, 18 avenue Edouard Belin, 31401 Toulouse Cedex 9, France\relax                                                                                                                                                                                                                                        \label{inst:0236}\vfill
\and M\'{e}socentre de calcul de Franche-Comt\'{e}, Universit\'{e} de Franche-Comt\'{e}, 16 route de Gray, 25030 Besan\c{c}on Cedex, France\relax                                                                                                                                                                                                                  \label{inst:0238}\vfill
\and ATOS for CNES Centre Spatial de Toulouse, 18 avenue Edouard Belin, 31401 Toulouse Cedex 9, France\relax                                                                                                                                                                                                                                                       \label{inst:0243}\vfill
\and School of Physics and Astronomy, Tel Aviv University, Tel Aviv 6997801, Israel\relax                                                                                                                                                                                                                                                                          \label{inst:0246}\vfill
\and Astrophysics Research Centre, School of Mathematics and Physics, Queen's University Belfast, Belfast BT7 1NN, UK\relax                                                                                                                                                                                                                                        \label{inst:0248}\vfill
\and Centre de Donn\'{e}es Astronomique de Strasbourg, Strasbourg, France\relax                                                                                                                                                                                                                                                                                    \label{inst:0250}\vfill
\and Institute for Computational Cosmology, Department of Physics, Durham University, Durham DH1 3LE, UK\relax                                                                                                                                                                                                                                                     \label{inst:0254}\vfill
\and European Southern Observatory, Karl-Schwarzschild-Str. 2, 85748 Garching, Germany\relax                                                                                                                                                                                                                                                                       \label{inst:0255}\vfill
\and Max-Planck-Institut f\"{ u}r Astrophysik, Karl-Schwarzschild-Stra{\ss}e 1, 85748 Garching, Germany\relax                                                                                                                                                                                                                                                      \label{inst:0256}\vfill
\and Data Science and Big Data Lab, Pablo de Olavide University, 41013, Seville, Spain\relax                                                                                                                                                                                                                                                                       \label{inst:0260}\vfill
\and Barcelona Supercomputing Center (BSC), Pla\c{c}a Eusebi G\"{ u}ell 1-3, 08034-Barcelona, Spain\relax                                                                                                                                                                                                                                                          \label{inst:0268}\vfill
\and ETSE Telecomunicaci\'{o}n, Universidade de Vigo, Campus Lagoas-Marcosende, 36310 Vigo, Galicia, Spain\relax                                                                                                                                                                                                                                                   \label{inst:0273}\vfill
\and Asteroid Engineering Laboratory, Space Systems, Lule\aa{} University of Technology, Box 848, S-981 28 Kiruna, Sweden\relax                                                                                                                                                                                                                                    \label{inst:0277}\vfill
\and Vera C Rubin Observatory,  950 N. Cherry Avenue, Tucson, AZ 85719, USA\relax                                                                                                                                                                                                                                                                                  \label{inst:0282}\vfill
\and Department of Astrophysics, Astronomy and Mechanics, National and Kapodistrian University of Athens, Panepistimiopolis, Zografos, 15783 Athens, Greece\relax                                                                                                                                                                                                  \label{inst:0283}\vfill
\and TRUMPF Photonic Components GmbH, Lise-Meitner-Stra{\ss}e 13,  89081 Ulm, Germany\relax                                                                                                                                                                                                                                                                        \label{inst:0286}\vfill
\and IAC - Instituto de Astrofisica de Canarias, Via L\'{a}ctea s/n, 38200 La Laguna S.C., Tenerife, Spain\relax                                                                                                                                                                                                                                                   \label{inst:0291}\vfill
\and Department of Astrophysics, University of La Laguna, Via L\'{a}ctea s/n, 38200 La Laguna S.C., Tenerife, Spain\relax                                                                                                                                                                                                                                          \label{inst:0292}\vfill
\and Faculty of Aerospace Engineering, Delft University of Technology, Kluyverweg 1, 2629 HS Delft, The Netherlands\relax                                                                                                                                                                                                                                          \label{inst:0294}\vfill
\and Radagast Solutions\relax                                                                                                                                                                                                                                                                                                                                      \label{inst:0298}\vfill
\and Laboratoire Univers et Particules de Montpellier, CNRS Universit\'{e} Montpellier, Place Eug\`{e}ne Bataillon, CC72, 34095 Montpellier Cedex 05, France\relax                                                                                                                                                                                                 \label{inst:0299}\vfill
\and Universit\'{e} de Caen Normandie, C\^{o}te de Nacre Boulevard Mar\'{e}chal Juin, 14032 Caen, France\relax                                                                                                                                                                                                                                                     \label{inst:0305}\vfill
\and LESIA, Observatoire de Paris, Universit\'{e} PSL, CNRS, Sorbonne Universit\'{e}, Universit\'{e} de Paris, 5 Place Jules Janssen, 92190 Meudon, France\relax                                                                                                                                                                                                   \label{inst:0306}\vfill
\and SRON Netherlands Institute for Space Research, Niels Bohrweg 4, 2333 CA Leiden, The Netherlands\relax                                                                                                                                                                                                                                                         \label{inst:0315}\vfill
\and Astronomical Observatory, University of Warsaw,  Al. Ujazdowskie 4, 00-478 Warszawa, Poland\relax                                                                                                                                                                                                                                                             \label{inst:0316}\vfill
\and Scalian for CNES Centre Spatial de Toulouse, 18 avenue Edouard Belin, 31401 Toulouse Cedex 9, France\relax                                                                                                                                                                                                                                                    \label{inst:0318}\vfill
\and Universit\'{e} Rennes, CNRS, IPR (Institut de Physique de Rennes) - UMR 6251, 35000 Rennes, France\relax                                                                                                                                                                                                                                                      \label{inst:0324}\vfill
\and INAF - Osservatorio Astronomico di Capodimonte, Via Moiariello 16, 80131, Napoli, Italy\relax                                                                                                                                                                                                                                                                 \label{inst:0326}\vfill
\and Shanghai Astronomical Observatory, Chinese Academy of Sciences, 80 Nandan Road, Shanghai 200030, People's Republic of China\relax                                                                                                                                                                                                                             \label{inst:0329}\vfill
\and University of Chinese Academy of Sciences, No.19(A) Yuquan Road, Shijingshan District, Beijing 100049, People's Republic of China\relax                                                                                                                                                                                                                       \label{inst:0331}\vfill
\and Niels Bohr Institute, University of Copenhagen, Juliane Maries Vej 30, 2100 Copenhagen {\O}, Denmark\relax                                                                                                                                                                                                                                                    \label{inst:0334}\vfill
\and DXC Technology, Retortvej 8, 2500 Valby, Denmark\relax                                                                                                                                                                                                                                                                                                        \label{inst:0335}\vfill
\and Las Cumbres Observatory, 6740 Cortona Drive Suite 102, Goleta, CA 93117, USA\relax                                                                                                                                                                                                                                                                            \label{inst:0336}\vfill
\and CIGUS CITIC, Department of Nautical Sciences and Marine Engineering, University of A Coru\~{n}a, Paseo de Ronda 51, 15071, A Coru\~{n}a, Spain\relax                                                                                                                                                                                                          \label{inst:0345}\vfill
\and Astrophysics Research Institute, Liverpool John Moores University, 146 Brownlow Hill, Liverpool L3 5RF, United Kingdom\relax                                                                                                                                                                                                                                  \label{inst:0346}\vfill
\and IPAC, Mail Code 100-22, California Institute of Technology, 1200 E. California Blvd., Pasadena, CA 91125, USA\relax                                                                                                                                                                                                                                           \label{inst:0353}\vfill
\and IRAP, Universit\'{e} de Toulouse, CNRS, UPS, CNES, 9 Av. colonel Roche, BP 44346, 31028 Toulouse Cedex 4, France\relax                                                                                                                                                                                                                                        \label{inst:0354}\vfill
\and MTA CSFK Lend\"{ u}let Near-Field Cosmology Research Group, Konkoly Observatory, MTA Research Centre for Astronomy and Earth Sciences, Konkoly Thege Mikl\'{o}s \'{u}t 15-17, 1121 Budapest, Hungary\relax                                                                                                                                                    \label{inst:0371}\vfill
\and Departmento de F\'{i}sica de la Tierra y Astrof\'{i}sica, Universidad Complutense de Madrid, 28040 Madrid, Spain\relax                                                                                                                                                                                                                                        \label{inst:0387}\vfill
\and Ru{\dj}er Bo\v{s}kovi\'{c} Institute, Bijeni\v{c}ka cesta 54, 10000 Zagreb, Croatia\relax                                                                                                                                                                                                                                                                     \label{inst:0393}\vfill
\and Villanova University, Department of Astrophysics and Planetary Science, 800 E Lancaster Avenue, Villanova PA 19085, USA\relax                                                                                                                                                                                                                                 \label{inst:0410}\vfill
\and INAF - Osservatorio Astronomico di Brera, via E. Bianchi, 46, 23807 Merate (LC), Italy\relax                                                                                                                                                                                                                                                                  \label{inst:0416}\vfill
\and STFC, Rutherford Appleton Laboratory, Harwell, Didcot, OX11 0QX, United Kingdom\relax                                                                                                                                                                                                                                                                         \label{inst:0423}\vfill
\and Charles University, Faculty of Mathematics and Physics, Astronomical Institute of Charles University, V Holesovickach 2, 18000 Prague, Czech Republic\relax                                                                                                                                                                                                   \label{inst:0446}\vfill
\and Department of Particle Physics and Astrophysics, Weizmann Institute of Science, Rehovot 7610001, Israel\relax                                                                                                                                                                                                                                                 \label{inst:0457}\vfill
\and Department of Astrophysical Sciences, 4 Ivy Lane, Princeton University, Princeton NJ 08544, USA\relax                                                                                                                                                                                                                                                         \label{inst:0458}\vfill
\and Departamento de Astrof\'{i}sica, Centro de Astrobiolog\'{i}a (CSIC-INTA), ESA-ESAC. Camino Bajo del Castillo s/n. 28692 Villanueva de la Ca\~{n}ada, Madrid, Spain\relax                                                                                                                                                                                      \label{inst:0467}\vfill
\and naXys, University of Namur, Rempart de la Vierge, 5000 Namur, Belgium\relax                                                                                                                                                                                                                                                                                   \label{inst:0470}\vfill
\and CGI Deutschland B.V. \& Co. KG, Mornewegstr. 30, 64293 Darmstadt, Germany\relax                                                                                                                                                                                                                                                                               \label{inst:0478}\vfill
\and Institute of Global Health, University of Geneva\relax                                                                                                                                                                                                                                                                                                        \label{inst:0479}\vfill
\and Astronomical Observatory Institute, Faculty of Physics, Adam Mickiewicz University, Pozna\'{n}, Poland\relax                                                                                                                                                                                                                                                  \label{inst:0481}\vfill
\and H H Wills Physics Laboratory, University of Bristol, Tyndall Avenue, Bristol BS8 1TL, United Kingdom\relax                                                                                                                                                                                                                                                    \label{inst:0485}\vfill
\and Department of Physics and Astronomy G. Galilei, University of Padova, Vicolo dell'Osservatorio 3, 35122, Padova, Italy\relax                                                                                                                                                                                                                                  \label{inst:0492}\vfill
\and CERN, Geneva, Switzerland\relax                                                                                                                                                                                                                                                                                                                               \label{inst:0494}\vfill
\and Applied Physics Department, Universidade de Vigo, 36310 Vigo, Spain\relax                                                                                                                                                                                                                                                                                     \label{inst:0496}\vfill
\and Association of Universities for Research in Astronomy, 1331 Pennsylvania Ave. NW, Washington, DC 20004, USA\relax                                                                                                                                                                                                                                             \label{inst:0500}\vfill
\and European Southern Observatory, Alonso de C\'ordova 3107, Casilla 19, Santiago, Chile\relax                                                                                                                                                                                                                                                                    \label{inst:0508}\vfill
\and Sorbonne Universit\'{e}, CNRS, UMR7095, Institut d'Astrophysique de Paris, 98bis bd. Arago, 75014 Paris, France\relax                                                                                                                                                                                                                                         \label{inst:0513}\vfill
\and Faculty of Mathematics and Physics, University of Ljubljana, Jadranska ulica 19, 1000 Ljubljana, Slovenia\relax                                                                                                                                                                                                                                               \label{inst:0515}\vfill
}

    

             
\authorrunning{Gaia Collaboration}

   \date{}

 
  \abstract
   {The \gaia mission of the European Space Agency (ESA) has been routinely observing Solar System objects (SSOs) since the beginning of its operations in August 2014. The \gaia data release three (DR3) includes, for the first time, the mean reflectance spectra of a selected sample of 60\,518 SSOs, primarily asteroids, observed between August 5, 2014, and May 28, 2017. Each reflectance spectrum was derived from measurements obtained by means of the Blue and Red photometers (BP/RP), which were binned in 16 discrete wavelength bands. For every spectrum, the DR3 also contains additional information about the data quality for each band. }
   {We describe the processing of the \gaia spectral data of SSOs, explaining both the criteria used to select the subset of asteroid spectra published in \gaia DR3, and the different steps of our internal validation procedures. In order to further assess the quality of \gaia
SSO reflectance spectra, we carried out external
validation against SSO reflectance spectra obtained
from ground-based and space-borne telescopes and available in the literature; we present our validation approach.}
   {For each selected SSO, an epoch reflectance was computed by dividing the calibrated spectrum observed by the BP/RP at each transit on the focal plane by the  mean spectrum of a solar analogue. The latter was obtained by averaging the \gaia spectral measurements of a selected sample of stars known to have very similar spectra to that of the Sun. Finally, a mean of the epoch reflectance spectra was calculated in 16 spectral bands for each SSO.}
   {\gaia SSO reflectance spectra are in general agreement with those obtained from a ground-based spectroscopic campaign specifically designed to cover the same spectral interval as \gaia and mimic the illumination and observing geometry characterising \gaia SSO observations.
   In addition, the agreement between \gaia mean reflectance spectra and those available in the literature is good for bright SSOs, regardless of their taxonomic spectral class. We identify an increase in the spectral slope of S-type SSOs with increasing phase angle. Moreover, we show that the spectral slope increases and the depth of the 1 $\mu$m absorption band decreases for increasing ages of S-type asteroid families.
   The latter can be interpreted as proof of progressive ageing of S-type asteroid surfaces due to their exposure to space weathering effects. }
  {}


   \maketitle
%


\section{Introduction}
A major breakthrough of the last several decades in astrophysics has been the discovery of the great diversity of planetary systems in our Galaxy and their marked differences with respect to our Solar System \citep{Winn2015ARA&A..53..409W}. This observational progress has boosted research in one of the oldest subjects of planetary science, namely understanding the formation of planets and their evolution \citep{Morbidelli2016JGRE..121.1962M,Raymond2020plas.book..287R}. How discs of dust and gas around similar stars evolved and eventually led to the great planetary diversity that we observe and why our own Solar System took a path that is uncommon amongst others are fundamental questions of planetary science and astrophysics. 

Studying the Solar System objects (SSOs) is key to answering the above questions. For instance, the current orbital structure of asteroids informs us about dynamical events that the planets of our Solar System have undergone during their formation and evolution \citep{Minton2009Natur.457.1109M,Walsh2011Natur.475..206W,Raymond2017SciA....3E1138R,Nesvorny2018ARA&A..56..137N,Raymond2020arXiv201207932R}. One of these events was a brief and violent phase of orbital instability of the giant planets \citep{Tsiganis2005Natur.435..459T,Nesvorny2012AJ....144..117N}. 
In addition, asteroids contain material that is the most pristine of all the material dating back to the formation of our Solar System 4.5 billion of years ago \citep[see the review of][and references therein]{Libourel2017Icar..282..375L}. Moreover, some asteroids are the parent bodies of the meteorites, which are a major source of information about the evolution of the material in the protoplanetary disc \citep{Zolensky2006mess.book..869Z}. 

The largest repository of asteroids is the main belt, which comprises bodies with stable orbits between Mars and Jupiter. However, over time, collisions have shattered some of these asteroids, creating families of fragments; these have drifted along the orbital semi-major axis \citep[due to a non-gravitational force known as the Yarkovsky effect;][]{Vokrouhlicky2000Natur.407..606V,Bottke2000Icar..145..301B,Rubincam2000Icar..148....2R,Bottke2006AREPS..34..157B} until reaching orbital instability zones capable of increasing orbital eccentricity, causing these  asteroid fragments to cross the orbits of the inner planets \citep{Morbidelli2003Icar..163..120M,Granvik2017A&A...598A..52G,Granvik2018Icar..312..181G}. Close encounters with planets can fully change the orbits of these bodies to be in the terrestrial planet region. Because these near-Earth asteroids can impact our planet, substantial effort has been devoted to the studying their population \citep{Mainzer2011ApJ...743..156M,Mainzer2015aste.book...89M,Morbidelli2020Icar..34013631M}, in some cases in order to assess impact hazard \citep{Michel2013AcAau..90....6M}.

More recently, asteroids have been targeted by space missions of Solar System exploration. Several missions flew by or rendezvoused with asteroids, such as 
(951) Gaspra \citep{Belton1992Sci...257.1647B}, 
(243) Ida \citep{Belton1995Natur.374..785B}, 
(253) Mathilde \citep{Veverka1996Icar..120...66V}, 
(433) Eros \citep{2000Sci...289.2088V}, 
(25143) Itokawa \citep{Abe2006Sci...312.1334A},
(2867) \u{S}teins \citep{2010Sci...327..190K},
(21) Lutetia \citep{2011Sci...334..487S},
(4) Vesta \citep{2012Sci...336..700R},
(4179) Toutatis \citep{Huang2013NatSR...3E3411H},
(1) Ceres \citep{2004P&SS...52..465R},
(162173) Ryugu \citep{Sugita2019Sci...364..252S}, and
(101955) Bennu \citep{Lauretta2019Natur.568...55L}. These visits revealed a great variety in the composition and nature of the surfaces of these objects. 
More missions are flying towards asteroids, such as  NASA's Lucy, which is bound to explore the Jupiter Trojan asteroids \citep{Olkin2021PSJ.....2..172O}, and the NASA Double Asteroid Redirection Test (DART), which plans to impact the natural satellite of the double asteroid (65803) Didymos \citep{Rivkin2021PSJ.....2..173R}. In August 2022, NASA will also launch the Psyche mission to explore the main-belt asteroid (16) Psyche, which is thought to be a remnant of the metallic core of a disrupted planetesimal \citep{Elkins-Tanton2016LPI....47.1631E}. Furthermore, in 2024, ESA will launch Hera to investigate the Didymos binary asteroid, including the very first assessment of its internal properties, and to measure the outcome of the DART mission kinetic impactor test \citep{Michel2021EPSC...15...71M}.

Given all of the above, determining the composition of asteroids is of utmost importance. Most of the data collected so far  have been collected from the ground using different techniques, including spectroscopy, photometry, polarimetry, radar experiments, and adaptive optics imaging. Spectroscopy is the preferred method to estimate asteroid surface composition from the wavelength dependent reflectance of the surface \citep[see][for a review]{Bus2002aste.conf..169B}. The observed diversity of the reflectance spectra of asteroids has been traditionally used to develop taxonomic classifications \citep[see][for a review]{DeMeo2015aste.book...13D}. These taxonomic classes express the relative abundances of asteroids across the Solar System \citep{Gradie1982Sci...216.1405G,Gradie1989aste.conf..316G} and their mixing \citep{DeMeo2014Natur.505..629D}. Reflectance spectra are also typically used to link meteorites to their parent asteroids \citep[e.g.][]{Popescu2016A&A...591A.115P,DeMeo2022Icar..38014971D}. These links are extremely useful for relating detailed laboratory measurements to the orbital distribution and classes of small bodies. 

The existence of different classes of asteroids is interpreted by many authors in terms of a variety of surface compositions likely resulting from differences in origin and evolution. The different orbital distributions of distinct taxonomic classes are believed to be diagnostic of phenomena of early mixing 
of different classes of planetesimals across the Solar System \citep{Gradie1982Sci...216.1405G,Gradie1989aste.conf..316G,DeMeo2014Natur.505..629D}, and provide an important input for theoretical models of the early phases of evolution of our planetary system \citep{Gomes2005Natur.435..466G,Pierens2014ApJ...795L..11P,Walsh2012M&PS...47.1941W}.

In this context, it is very important to be able to disentangle properties due to the early history of the Solar System from those resulting from long-term evolution beginning from when the current structure of the Solar System was attained.
The knowledge accumulated over decades of investigations suggests that there are essentially three physical processes that play a major role in the evolution of the asteroid population. 
The first is collisional evolution, which progressively affects the inventory and size distribution of the main belt asteroids, and their surfaces; for example by producing craters \citep{Davis1979aste.book..528D,Farinella1981Icar...46..114F,Farinella1992A&A...253..604F,Davis1985Icar...62...30D,Davis2002aste.book..545D,Morbidelli2009Icar..204..558M,Bottke2015aste.book..701B}.
The second is space weathering. This is due to the exposure of asteroid surfaces to irradiation from cosmic rays, solar wind, and collisions with micro-meteorites. 
For decades, we have known that space weathering progressively modifies the reflectance spectra of asteroids, the most important effects having been found to affect the class of asteroids belonging to the so-called S-complex, which includes objects believed to be the parent bodies of the most common class of meteorites, the ordinary chondrites \citep{Brunetto2006Icar..184..327B}. The third is the realisation that the simple cycle of thermal expansion and contraction of the material constituting the outer layer of surface regolith, which is due to rotation of the body, leads to progressive evolution of the regolith structural and thermal properties \citep{Delbo2014Natur.508..233D,Molaro2017Icar..294..247M,Molaro2020NatCo..11.2913M}.  
Of course, there is interplay between the above-mentioned evolution mechanisms. For instance, energetic collisions not only generate families, producing the exposure of the internal layers of their parent bodies, but also restart the space weathering clock and trigger a Yarkovsky-driven dynamical evolution of the smallest fragments. 

Spectrophotometry has been a very important tool for understanding the compositional big picture of the asteroid population. Initiated in the 1980s with the Eight Color Asteroid Survey \citep[ECAS,][]{Zellner1985Icar...61..355Z}, spectrophotometric asteroid surveys evolved with the  
24-colour asteroid survey \citep{Chapman2005PDSS...27.....C},
the 52-colour survey \citep{Bell1988LPI....19...57B}, 
the Seven Colour Asteroid Survey in the infrared \citep{Clark1993LPI....24..299C},
the moving object component of the Sloan digital sky survey (SDSS) \citep{Ivezic2019ApJ...873..111I}, which in its latest analysis provided measurements for 379\,714 known asteroids \citep{Sergeyev2021A&A...652A..59S}. Moreover, we recall
the near-infrared (NIR) colours of asteroids recovered from the Visible and Infrared Survey Telescope for Astronomy - VISTA Hemisphere Survey (VISTA-VHS)  and the Moving Objects from VISTA survey \citep[MOVIS;][]{Popescu2016A&A...591A.115P}, the Korea Microlensing Telescope Network-South African Astronomical Observatory (KMTNET-SAAO) Multiband Photometry survey \citep{Erasmus2019ApJS..242...15E}, the moving object observations from the Javalambre Photometric Local Universe Survey (J-PLUS) \citep{Morate2021A&A...655A..47M}, and the multi-filter photometry of Solar System objects from the SkyMapper Southern Survey \citep{Sergeyev2022A&A...658A.109S}. In total, over 1.5 million spectrophotometric observations of asteroids exist.

Among several spectroscopic surveys of small bodies carried out by different authors, we mention the SMall Asteroid Spectroscopic Survey of the MIT in the visible light (SMASS) phase I \citep{Xu1995Icar..115....1X}, II  \citep{Bus2002Icar..158..146B,Bus2002Icar..158..106B}, and in the NIR \citep{Burbine2002Icar..159..468B}; the MIT-Hawaii Near-Earth Object Spectroscopic Survey \cite[MITHNEOS,][]{Binzel2019Icar..324...41B,Marsset2022AJ....163..165M}; the Small Solar System Objects Spectroscopic Survey (S$^3$OS$^2$) \citep{Lazzaro2004Icar..172..179L}; the Mission Accessible Near-Earth Objects Survey (MANOS) of the Lowell Observatory \citep{Devogele2019AJ....158..196D}; the PRIMitive Asteroids Spectroscopic Survey \citep[PRIMASS,][]{deLeon2018DPS....5031005D}; and efforts devoted to the characterisation of small near-Earth objects \citep{Perna2018P&SS..157...82P}. In the literature, more than 7600 asteroid spectra are available today. We note that the most modern asteroid spectroscopic surveys covered preferentially the visible and NIR spectral regions, whereas the blue region has been lost downward of about 450-500 nm in many cases. This makes an interesting difference with the reflectance spectra  obtained by \gaia, as explained below.

It is in this framework that here we present the survey of reflectance spectra of 60\,518 Solar System small bodies contained in the Data Release 3 (DR3) of the ESA mission \gaia. Successfully launched from Kourou spaceport, French Guiana, on 19 December 2013, \gaia\ started its nominal mission on 25 July 2014; it continuously observed celestial bodies, including SSOs, with magnitudes $\lesssim$21 entering the field of view according to a predefined (so-called nominal) sky scanning law \citep{2016Prusti}. The detectors on the focal plane of \gaia, which is optimised for achieving unprecedented astrometric accuracy, include two low-resolution slit-less spectrographs capable of providing SSO spectroscopy. One of them is optimised for observations in the blue region of the visible light and is called Blue-Photometer (BP), while the other is optimised for the red region and is called Red-Photometer (RP). Both spectrographs are sometimes collectively referred to herein as XP. 

The \gdr{3} is the largest space-based survey of asteroid reflectance spectrophotometry in the visible range to date. \gdr{3} contains averaged spectra of main belt asteroids (MBAs), near-Earth asteroids (NEAs), Centaurs, Jupiter Trojans, and a few transneptunian objects (TNOs) (see Table~\ref{tab:sso_dynamic_class}). For each SSO, one reflectance spectrum sampled in 16 wavelength bands is provided. This is the result of averaging several epoch reflectance spectra. While the DR3 also contains astrometry and photometry of 158\,152 SSOs \citep{DR3-DPACP-150}, it does not contain epoch spectra, nor spectra of natural satellites or comets. The publication of these are foreseen for later releases.

Gaia SSO reflectance spectra will be complementary to spectrophotometric data that are expected from the ESA Euclid mission, which will observe several tens of thousand of asteroids in three wide wavelength bands covering the near-infrared region of the electromagnetic spectrum \citep{Carry2018A&A...609A.113C}.
In addition, at the end of 2022, the Large Synoptic Survey Telescope (LSST) of the Vera Rubin Observatory will be commissioned and will begin operations. Approximately two years later, the LSST teams will start publishing the fully calibrated spectrophotometric data. In a single visit, LSST is expected to be able to detect up to 5000 Solar System objects. Over its ten-year nominal lifespan, LSST could catalogue over 5 million MBAs, almost 300,000 Jupiter Trojans, over 100,000 NEAs, and over 40,000 TNOs. Many of these objects will be observed hundreds of times in six broad bands from 0.35 to 1.1 microns \citep{lsst2009arXiv0912.0201L,lsst2019ApJ...873..111I,lsst2020arXiv200907653V}. Gaia spectroscopic data of SSOs will offer a key comparison against LSST spectrophotomtery, allowing us to study the biases of both surveys and also the potential time spectral variability of asteroids.

This article is organised as follows:  in section \ref{S:Observations}, we present \gaia observations, in section \ref{S:dataProcessing}, we describe the methods used to create the SSO reflectance spectrophotometry, and in section \ref{S:validation}, we present our validation of SSO reflectances. In section \ref{sec:discussion}, we discuss our main results.

\section{Observations}
\label{S:Observations}

\begin{figure}[!ht]
   \centering
   \includegraphics[width=\columnwidth]{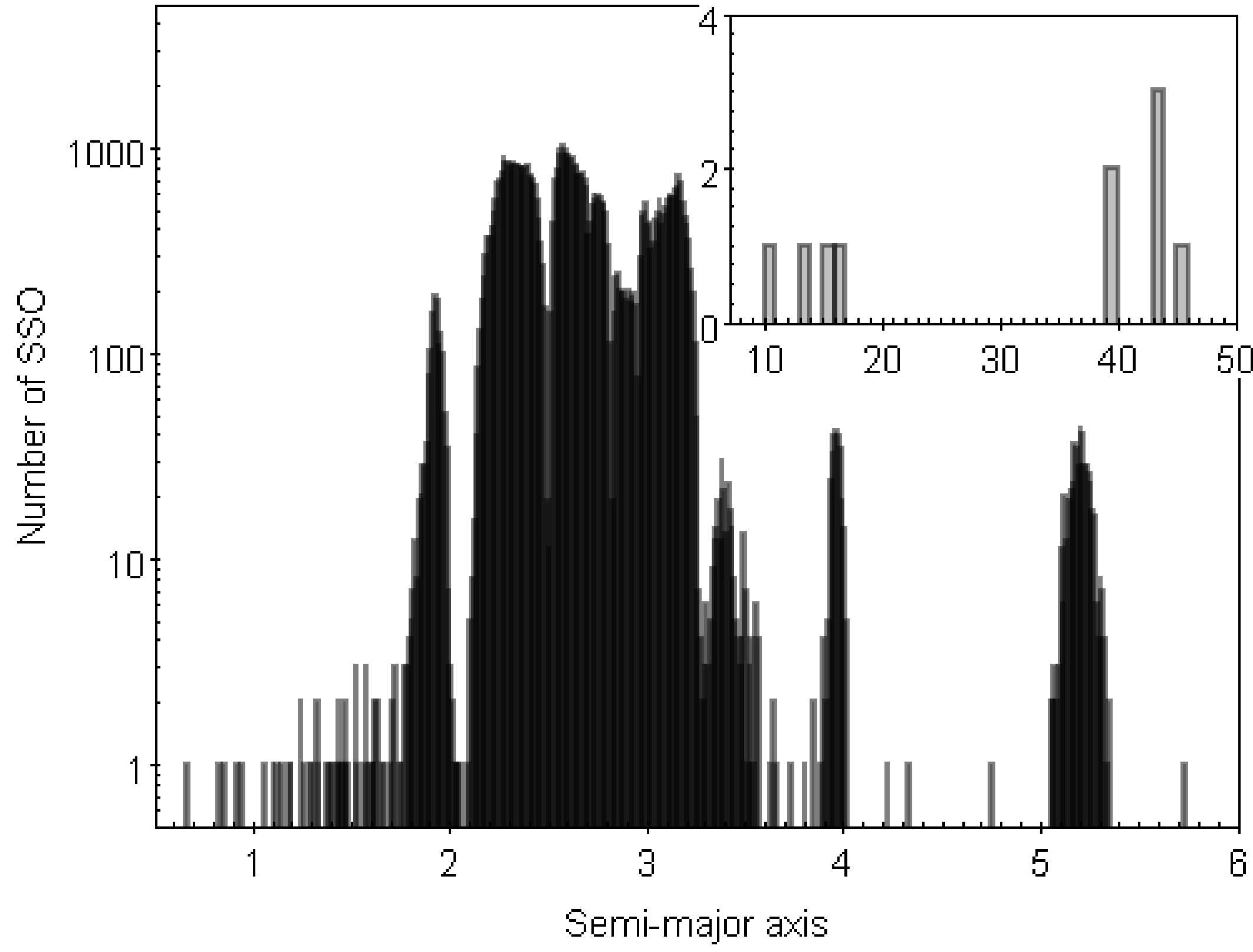}
   \caption{Orbital distribution of SSOs with reflectance spectra in Gaia DR3.}
   \label{Fig:hist_a}%
\end{figure}

\begin{figure}[!h]
   \centering
   \includegraphics[width=\columnwidth]{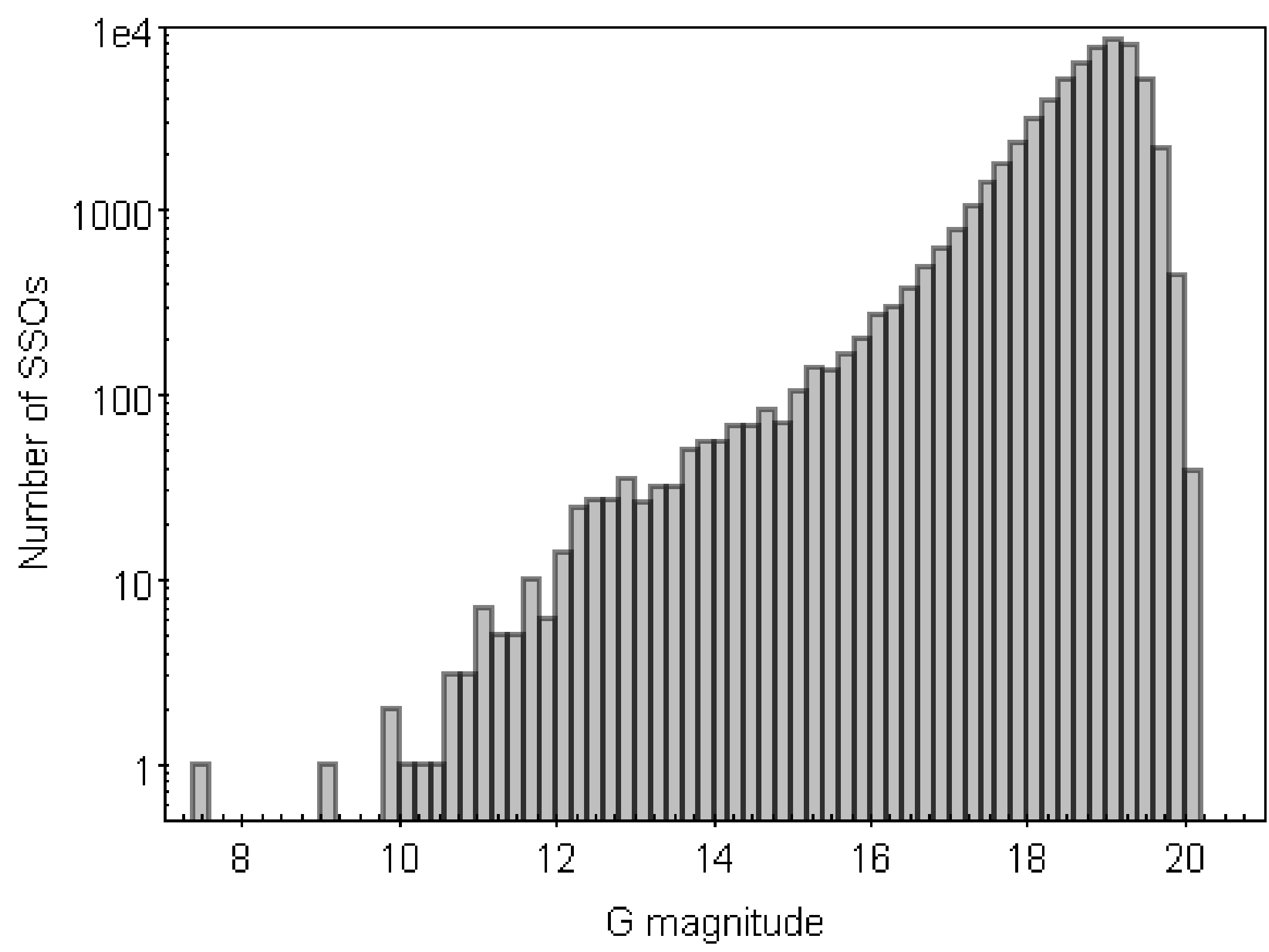}
   \caption{G-magnitude distribution of SSOs with reflectance spectra.}
   \label{Fig:hist_gMag}%
\end{figure}

Observations that resulted in the DR3 data were collected by \gaia\ during the nominal mission operations from Earth’s Lagrangian Point L2 between 5 August 2014 and 28 May 2017. We processed 
158 152 SSOs \citep[see][for a complete description on their selection and their complete processing]{DR3-DPACP-150}. However, not all these observations resulted in usable reflectance spectra. 
Figures~\ref{Fig:hist_a} and \ref{Fig:hist_gMag} show the orbital distribution and the mean value of the G magnitude, respectively, of the 60\,518 SSOs that have a valid mean reflectance spectrum in the DR3. This shows that the majority of these SSOs have been observed with magnitudes of between $\sim$18 and 20. The faintest SSO with \gdr{3} spectrophotometry is asteroid 2004 RH$_{319}$, which was  observed with a mean G magnitude of 20.19. Table~\ref{tab:sso_dynamic_class} presents the number of SSOs with DR3 reflectance spectra for each dynamical class. These dynamical classes are listed at the NASA Jet Propulsion Laboratory web site \footnote{\url{https://ssd.jpl.nasa.gov/tools/sbdb\_query.html}}.

%

The \gaia satellite is equipped with two telescopes that collect the light of astronomical sources on a shared focal plane composed of 106 charge-coupled device (CCD) detectors. Due to the rotation of \gaia, the sources move on the focal plane and encounter a series of different instruments. The first are the SkyMapper (SM) instruments that are used by the on-board electronics to detect the sources. The light is then measured by each of the nine astrometric field (AF) CCDs. Next, it is dispersed by the two slit-less prisms and collected by the XP instruments \citep[for the instrument layout, see][ in particular their Figure 2]{Jordi2010A&A...523A..48J}. 
On each XP, the on-board electronics consider only a small window of 60 $\times$ 12 pixels (along scan, AL, $\times$ across scan, AC; the angular size of an AL pixel is 58.933 milli-arcsec) centred on each spectrum. This window is also binned in the AC direction in order to compose a 1D spectrum of 60 samples. Among these, in general, only the 40 central pixels contain exploitable signal. The edges of the windows are mainly dominated by the background and the extended wings of the line spread function \citep[LSF;][]{2021A&A...652A..86C}. One spectrum per XP is produced.
The BP operates in the wavelength range between 330 and 680~nm, while the RP in the range between 640 and 1050~nm (see Appendix \ref{appendix:xpresolution} for the spectral resolution of each spectrophotometer).

\begin{figure}[!h]
   \centering
   \includegraphics[width=\columnwidth]{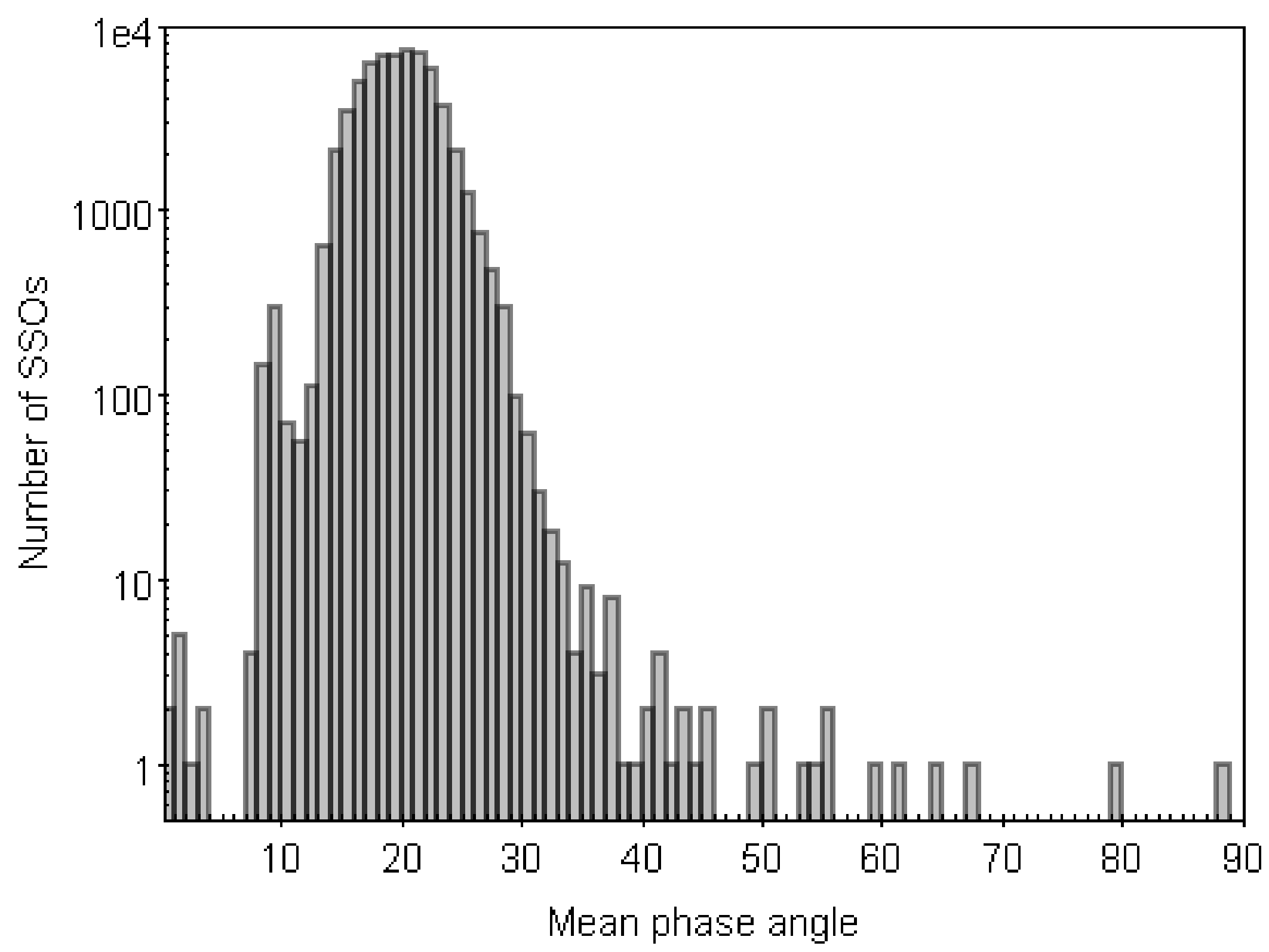}
   \caption{Phase angle distribution of SSOs with reflectance spectra. For each SSO, only the phase angles of the transits used to compute the mean reflectance spectrum are averaged and contribute to the plot.}
   \label{Fig:hist_phase_Angle}%

\end{figure}

%
Due to the \gaia scanning law of the sky, SSOs are never observed at opposition. Figure~\ref{Fig:hist_phase_Angle} presents the histogram of the average phase angles (the Sun-SSO-\gaia angle) of SSO spectroscopic observations. For each SSO, this is calculated as the straight arithmetic average of the phase angles of the observations producing valid epoch reflectance spectra. Most of the MBAs are observed by \gaia with a phase angle of around 20 degrees. However, Jupiter Trojans, Centaurs, and TNOs are in general observed with lower phase angles than MBAs.

\begin{table}
    \centering
\caption{Number (No.) of SSOs with \gdr{3} spectra for each of the dynamical classes listed on the NASA JPL website. The classes are defined according to criteria based on the orbital semimajor axis, $a$, the perihelion distance, $q$, and the aphelion distance, $Q$.}
    \label{tab:sso_dynamic_class}
    \begin{tabular}{l|r|l}
    Dynamical class & No. of SSO & Criterion (values in au)\\
    \hline
    \hline
    NEA Aten        & 6       & $a <$ 1.0 \& $Q >$ 0.983\\
    NEA Apollo      & 52      & $a >$ 1.0 \& $q <$ 1.017\\
    NEA Amor        & 47      & 1.017 $< q <$ 1.3 \\
    Mars-Crosser    & 729     & 1.3 $< q <$ 1.666 \\
    Inner Main Belt & 1\,221  & $a <$ 2.0 \& $q >$ 1.666\\ 
    Main Belt       & 55\,976 & 2.0 $< a <$ 3.2 \& $q >$ 1.666\\
    Outer Main Belt &1\,995   & 3.2 $< a <$ 4.6 \\
    Jupiter Trojan  & 477     & 4.6 $< a <$ 5.5\\
    Centaur         & 5       & 5.5 $< a <$ 30.1\\
    TNO             & 7       & $a >$ 30.1\\
    Other           & 2       & none of the above\\
    \hline
    \end{tabular}
\end{table}

Each SSO is typically observed at multiple epochs, typically 60 times during the nominal five-year duration of the mission \citep[][however, the \gaia\ mission has been extended]{2012P&SS...73....5T}. Each epoch corresponds to a transit of the SSO on the \gaia focal plane, and receives a unique identifier called \textit{transit\_id}. However, not all observations obtained by \gaia\ eventually produce an exploitable epoch spectrum. Several factors can affect the production of spectra, as in the case where a window is affected by multiple peaks, a charge injection, a gate release, a satellite event, and so on. Some issues can also happen during the calibration of the spectrum; for example, a poor prediction of the source position might lower the quality of the spectrum. In the case of \gdr{3}, we found that almost 80\% of the epoch spectra for each SSO produced by the spectrophotometric pipeline PhotPipe (see below) could be used to compute the mean reflectance spectra. 

\section{Data processing}\label{S:dataProcessing}
The process that leads to the generation of internally calibrated BP/RP spectra \citep{2021A&A...652A..86C,DPACP-118} converts the observed raw pixel data into an internal system that is homogeneous across all different instrumental configurations. This is obtained by calibrating and removing a number of instrumental and astrophysical effects such as the CCD bias, background, geometry, differential dispersion, variations in response, and variation in the LSF across the focal plane. The main product of this process is a set of internally calibrated epoch spectra for each source and each epoch (i.e. transit on the focal plane). Internally calibrated epoch spectra are represented by arrays of 60$\times$1 internal flux values and flux uncertainties, corresponding to the 60 pixel-long CCD windows on each XP. The 60 samples are given as a function of the so-called pseudo-wavelengths. 
Pseudo-wavelengths correspond to the wavelengths measured in units of sample in a reference location in the focal plane, before they are transformed into physical wavelength units (such as nanometers). This internal pseudo-wavelength scale provides homogeneity across all observations, while being close to the actual wavelength sampling of each XP instrument, which allows the alignment of observed spectra. For each source that is not a SSO (mainly stars), several internally calibrated epoch spectra were aligned thanks to this pseudo-wavelength scale and then averaged over the epochs (transits) to produce so-called mean spectra. \gaia DR3 includes mean spectra for about 220 million sources. The mean spectra of a set of stars known to have a spectrum that is analogous to that of our Sun were extracted and averaged to produce a single solar analogue reference spectrum (see Appendix \ref{a:solar_analog}). 

However, due to the intrinsic variability and proper motion of SSOs, the calculation of their mean spectra was not performed. Instead, for each SSO and each epoch, the nominal, pre-launch dispersion function was used to convert pseudo-wavelengths to physical wavelengths. This procedure was complicated by the dispersed image formation in the time-delay integration (TDI) mode that is used by \gaia, whereby the dispersion function can only be defined in terms of a relation between wavelengths and the offset in the data space with respect to some reference point in the dispersed image.
%
%
%
%

%
This reference point is given by the prediction of the location of a given reference wavelength via the knowledge of the source position in the sky, the satellite attitude, and the geometry of the XP CCDs. For SSOs, data from the AF and predicted sky-plane motion of the SSO from the ephemerides were instead used to determine the field angles ---namely the angular coordinates of the source on the focal plane--- for each transit as a function of time \citep[see][in particular their Figure 2]{Lindegren2012A&A...538A..78L}.
The use of the ephemerides, rather than the estimate of the sky-plane motion that can be obtained from a single AF transit, allows us to obtain higher accuracy in the predictions of the aforementioned reference point. Three different epochs were used, namely 45, 50, and 55 seconds after the epoch of the read-out of the AF1 window. The latter is measured by \gaia on-board electronics and coded in the \textit{transit\_id}. These three epochs bracket the times of XP observations. The field angles at the exact observation time for the BP and the RP were then determined for each SSO transit by interpolation. The wavelength reference point was therefore determined, and the nominal dispersion function applied. This resulted in epoch spectra whose 60 flux and flux uncertainty samples were expressed in terms of physical wavelengths. 



Subsequently, an epoch reflectance $R(\lambda_i)_t$ was determined by dividing the flux $f_t(\lambda_i)$ of each SSO epoch spectrum with \textit{transit\_id} $t$ by the reference solar analogue spectrum $F(\lambda)$. 
The index $i$ refers to the discrete samples of the epoch spectrum and can therefore take integer values between 0 and 59; $\lambda_i$ is the corresponding physical wavelength. We note that $(\lambda_i)_t \ne (\lambda_i)_{t'}$ for $t \ne t'$ due to the different sampling of epoch spectra:
\begin{equation}
R(\lambda_i)_t = \frac{1}{\xi_t} \frac{f(\lambda_i)_t}{F(\lambda_i)}.
\label{E:cu4_sso_reflectance_function}
\end{equation}
The $\xi_t$ factor was defined to allow for normalisation of the epoch reflectance to 1 at 550~nm and was calculated as the mean reflectance value of samples with wavelength between 525~nm~$ \leq \lambda_i \leq$~575~nm:
\begin{equation}
\xi_t = \frac{1}{N} \sum_{i\geq i_{525}}^{i\leq i_{575}} \frac{f(\lambda_i)_t}{F(\lambda_i)},
\label{E:cu4_sso_reflectance_normalization_function}
\end{equation}
where $i_{525}$ and $i_{575}$ denote the index range corresponding to the wavelength range of interest and N is the number of reflectance values being measured between 525 and 575 nm.

The uncertainty of $\xi_t$ was computed as the standard error of the mean. We defined the sample standard deviation $\sigma_{\xi_t}$ as
\begin{equation}
   \sigma_{\xi_t} = \frac{1}{\sqrt{N-1}}  \sqrt{\sum_{i=1}^N (R(\lambda_i)_t - \xi_t)^2},
\label{E:cu4_sso_std_normalization}
\end{equation}
from which the standard error on the mean was computed as
\begin{equation}
    \sigma_{\overline{\xi_t}} = \frac{\sigma_{\xi_t}}{\sqrt{N}}.
\label{E:cu4_sso_se_normalization}
\end{equation}

The uncertainty $\sigma(\lambda_i)_t$ on $R(\lambda_i)_t$ was calculated by propagating the uncertainties on $f(\lambda_i)_t$ and $F(\lambda)$ assuming them to be independent. 
The $R(\lambda_i)_t$ values were calculated separately for BP and RP, but the $\xi_t$ factor was assumed to be  the same for both BP and RP. Samples for which $R(\lambda_i)_t\leq$ 0 or $\sigma(\lambda_i)_t >$ 1 were rejected. 
Only BP and RP epoch reflectance samples corresponding to wavelengths in the ranges $[325, 650]$ nm and $[650, 1125]$ nm, respectively, were used to generate the mean reflectance spectra. 

\begin{figure}[!ht]
   \centering
   \includegraphics[width=\columnwidth]{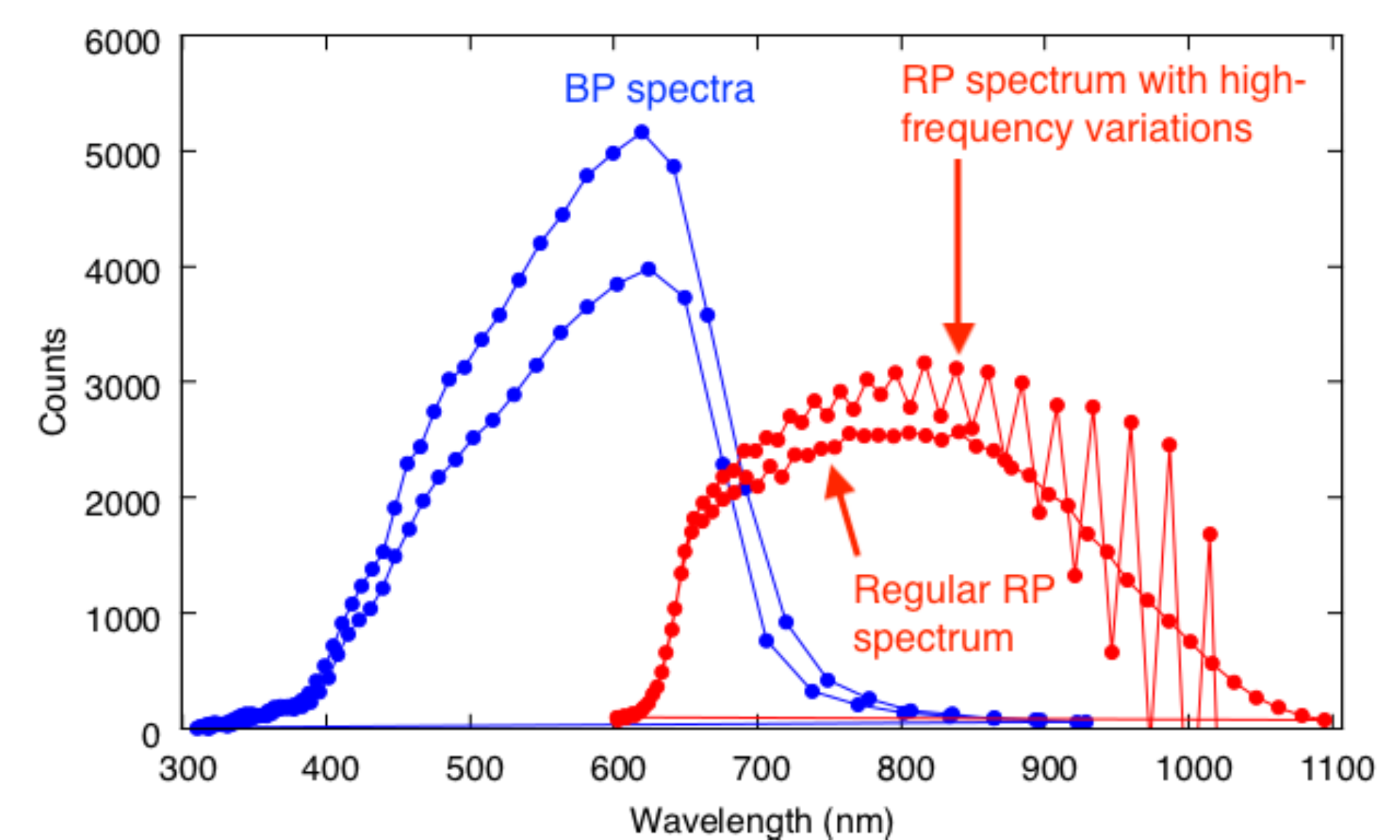}
   \caption{Example of a regular epoch spectrum and of a spectrum whose RP part shows anomalous high-frequency variations. Epoch spectra are from the same main belt asteroid, (90) Antiope.}
    \label{Fig:XP_Raw_Data}%
\end{figure}

Visual inspection of the resulting epoch reflectances showed well-behaved continuous curves, with the exception of a few rare cases for some of the brightest SSOs, which display high spectral frequency variability. 
An extreme example of this rare problem is shown in Fig.~\ref{Fig:XP_Raw_Data}.\\
In the large majority of cases, the BP and RP epoch reflectances overlap in the wavelength range 650 $\lesssim \lambda \lesssim$ 680~nm covered by both instruments. However, in some rare cases, the superposition did not occur and the RP reflectance is lower, or higher, than the BP one (see Fig.~\ref{Fig:mean_refl_computation}). In order to avoid the degradation of the mean reflectance spectra because of this discrepancy, some filters described below were put in place.

\begin{figure}[!ht]
   \centering
   \includegraphics[width=\columnwidth]{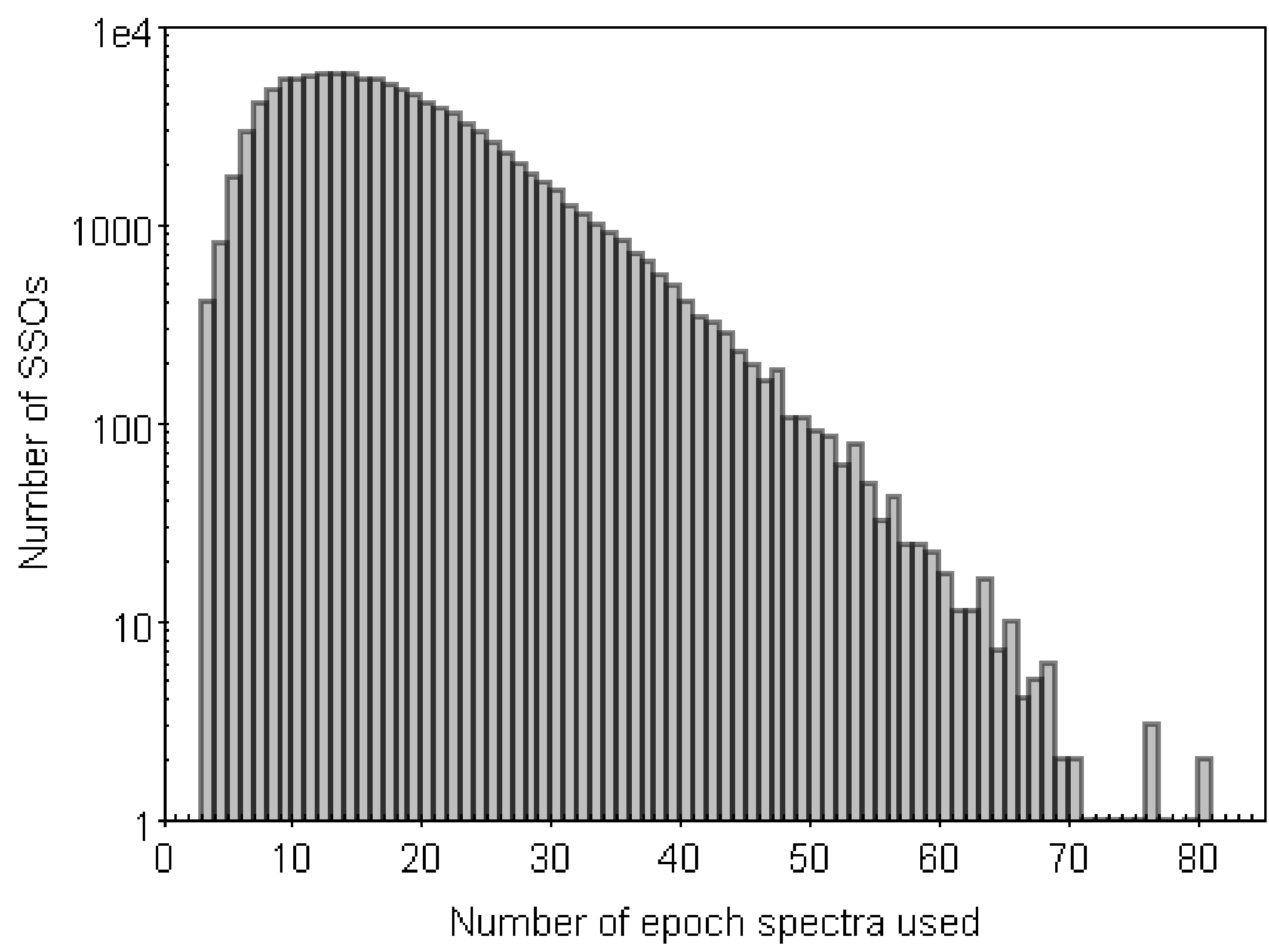}
   \caption{Distribution of the number of epoch spectra used for the calculation of SSO mean reflectance spectra.}
   \label{Fig:hist_epoch_spectra}%
\end{figure}

\begin{figure}[!ht]
   \centering
   \includegraphics[width=\columnwidth]{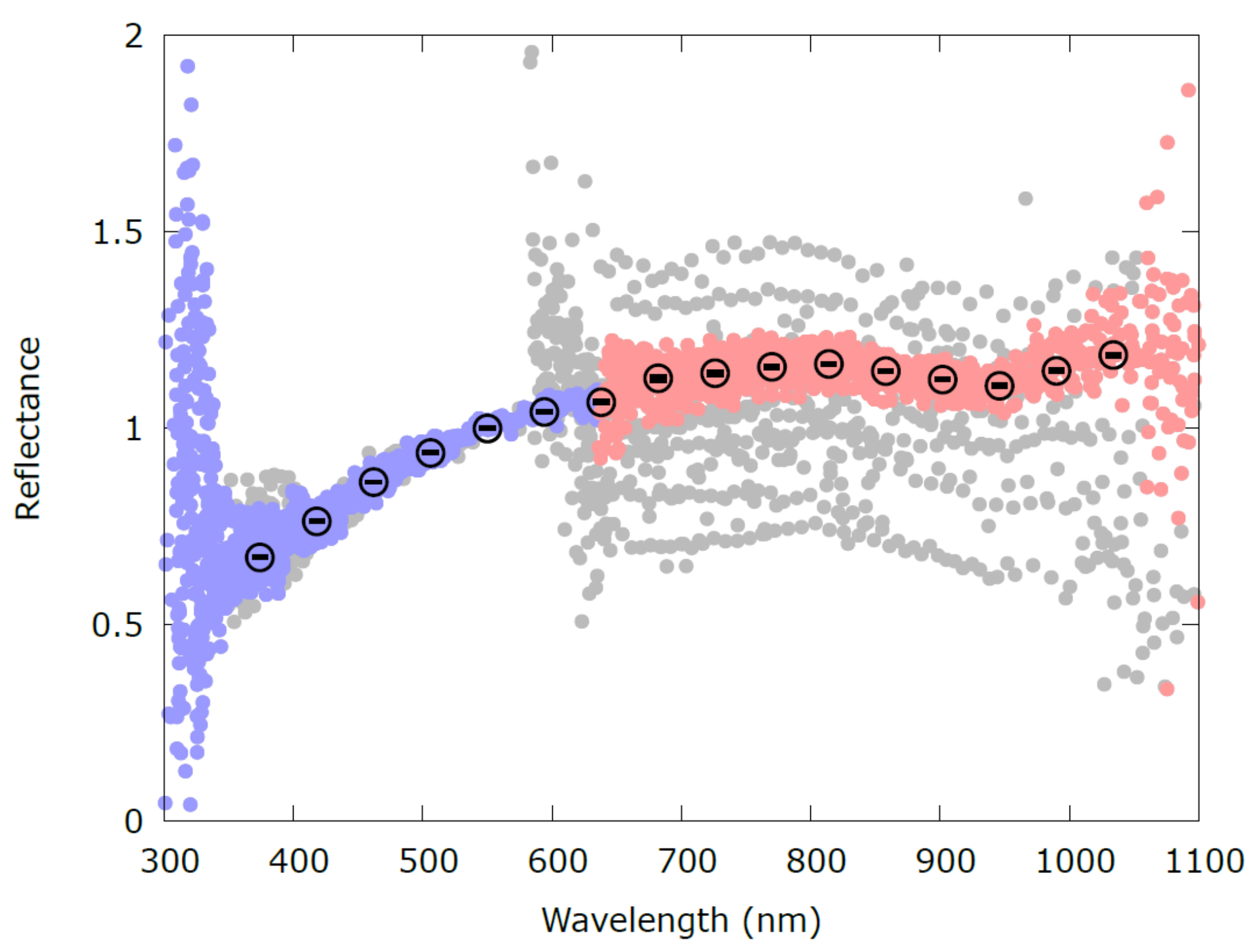}
   \caption{Example of the mean computed reflectance $\overline{R}(\lambda_i)$ (black circles) for the asteroid (61) Danae. The grey points are the $R(\lambda_i)_t$ corresponding to each epoch reflectance. The $R(\lambda_i)_t$ values that are accepted by our filtering method are coloured in light blue (BP) and light red (RP).}
\label{Fig:mean_refl_computation}%
\end{figure}

In \gaia DR3, we present one mean reflectance spectrum, $\overline{R}$, per SSO. This mean reflectance is calculated by averaging $R(\lambda_i)_t$ over the set of epochs (or transits) $t$ for each SSO. Figure~\ref{Fig:hist_epoch_spectra} reports the number of epoch spectra per SSO used to compute the mean reflectance spectrum. The minimum number of epoch spectra per asteroid considered to allow the computation of the mean reflectance is three. The peak of the distribution is around 15. We expect an important increase in the number of epoch spectra per SSO for Gaia DR4 because of the increase in the number of observations by a factor of two and the improvements in the calibration process, which will lead to a decrease in the number of transits that we had to filter out for the current DR3.

The averaging of SSO epoch reflectances was performed as follows: firstly, we defined a set of fixed wavelengths $\lambda_j$ every 44~nm in the range between 374 and 1034~nm. Next, we defined a set of wavelength bins 44~nm wide centred on each $\lambda_j$. Inside each wavelength bin, we calculated the weighted mean of the values of $R(\lambda_i)_t$ using $1 / \sigma_{R(\lambda_i)_t}^2$ as weight. For each band, the median and the median absolute deviation (MAD) values were computed. A $\sigma$-clipping approach was used for filtering out all values in each band that were outside the range (median$_{\lambda_i} - 2.5$~MAD, median$_{\lambda_i} + 2.5$~MAD); this was repeated twice in order to remove outliers. 

A weighted average of each band was computed using the surviving epoch reflectance values:
\begin{equation}
    \overline{R}(\lambda_i) = \frac{1}{\sum w_{i_t}}  \sum  w_{i_t} R(\lambda_i)_t.
\end{equation}
Finally, all reflectances were normalised by the value at $\lambda = 550$~nm.
Figure~\ref{Fig:mean_refl_computation} shows the final result for one particular case. 

The choice of the positioning of the 16 bands was dictated by two criteria: First, we required a whole band on a wavelength interval between 352 and 396 nm corresponding to an acceptable response of the BP. This has been verified on spectra with S/N $\gtrsim$ 100. Second, we needed to preserve a band centred at 550 nm to facilitate normalisation and comparison with literature data. 

\begin{figure*}[!ht]
   \centering
   \includegraphics[width=2\columnwidth]{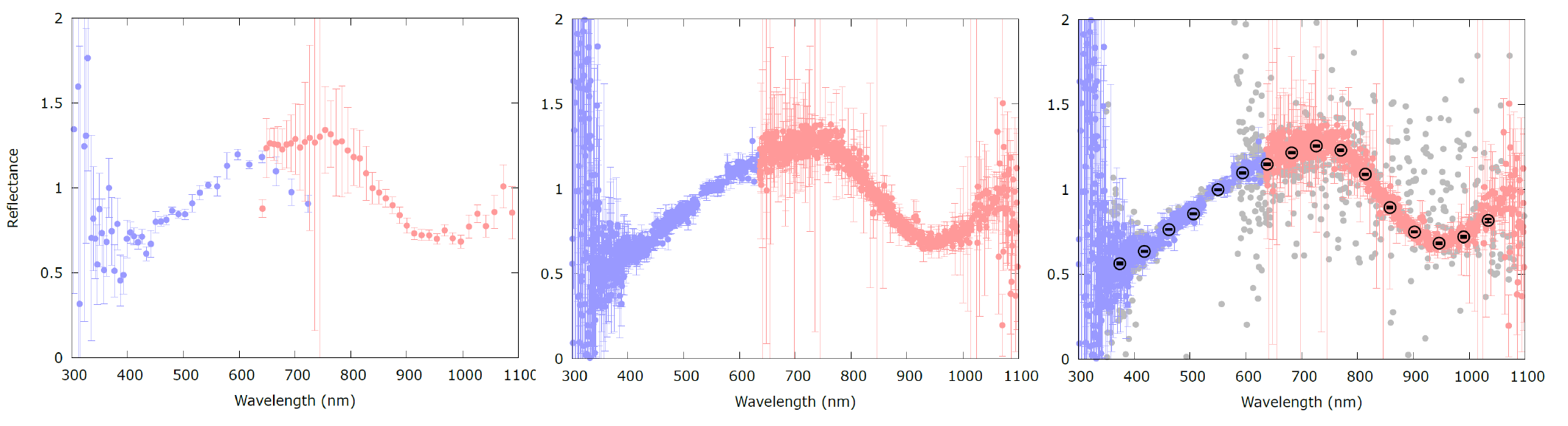}
   \caption{Illustration of the procedure adopted to compute mean reflectance spectra. The SSO chosen as an example here is (1459) Magnya, which is a basaltic asteroid presenting a deep 950~nm-centred absorption band and a red-sloped spectrum. In the left panel, one epoch reflectance spectrum is plotted. BP and RP data are blue and red respectively. 
   In the middle panel, all epoch reflectance spectra of Magnya are plotted. 
   In the right panel,  data filtered out by our sigma-clipping procedure are shown in grey; the over-plotted black dots correspond to the final average reflectance spectrum sampled in the 16 bands.}
\label{Fig:computation_procedure}%
\end{figure*}

In order to clean up our catalogue of mean reflectance spectra from anomalous cases, we put in place a filtering procedure: we determined the best-fit straight line to  $\overline{R}(\lambda_i)_t$ in the range 450 $\leq (\lambda_i)_t \leq$ 600~nm; we used the straight line equation to calculate a value $R^N_t$ at 550~nm and an extrapolated value of the $R^E_t$ at 726~nm; we calculated a value of the mean reflectance $R^M_t$ and of its standard deviation $\sigma^M_t$ by averaging those values of $\overline{R}(\lambda_i)_t$ with 680 $\leq (\lambda_i)_t \leq$ 780~nm; we measured the discrepancy $\delta_t$ = $ | R^M_t - R^E_t |$; we rejected all values of $\overline{R}(\lambda_i)_t$ where $\delta_t \geq$~0.30 or $\sigma^M_t \geq$~0.2 or $\sum_i^\text{BP} \overline{R}(\lambda_i)_t\leq$~0 or $\sum_i^\text{RP} \overline{R}(\lambda_i)_t\leq$~0 or $|R^N_t-1|\geq$~0.15.
Figure~\ref{Fig:computation_procedure} presents epoch and mean reflectance spectra computed for one typical SSO.

Apart from the bright asteroids, most of which have known spectra already, epoch spectra have quite low $\text{S/N}$. For \gdr{3}, the DPAC decided to provide the most reliable data, hence the mean reflectances. Epoch spectra will be published in future releases. This will allow us to (1) use improved mission calibrations (calibrations covering the entire mission are updated and improved for each release) and (2) have a larger number of epoch spectra per SSO, thus permitting a more reliable detection of outliers compared to the DR3.

\section{Validation}
\label{S:validation}
Having produced the mean reflectance spectra, the next step was to assess their quality and compare them with external data. We performed this validation in three steps, which are presented in the subsections below. 

\subsection{Internal consistency and S/N threshold for publication}\label{ssec:snr}
In Section~\ref{S:dataProcessing}, we described how the 16-band mean reflectance spectra were initially produced for 111\,818 asteroids. However, visual inspection of some (thousands) of these spectra clearly showed that objects of different magnitude classes displayed spectra of different quality, with the lowest quality spectra obviously associated with the objects observed at the faintest magnitudes.
The average $\text{S/N}$ was considered as an initial parameter to assess the quality of the spectra:
\begin{equation}
   \left< \text{S/N} \right> = \frac{1}{12} \sum_{i=3}^{14}\frac{\overline{R}(\lambda_i)}{\sigma_{\overline{R}(\lambda_i)}}.
 \label{eq:cu4sso_validationspectra_snr}
\end{equation}
Data at the wavelengths of 374, 418, 990, and 1034~nm were omitted from the computation, as they are often affected by large random and systematic errors (see Fig. \ref{Fig:computation_procedure}). On the other hand, the data point at 632 nm is included in the S/N calculation. This point can be problematic for bright objects, when in the BP-RP overlapping region the epoch reflectance values of the two spectrometers differs much more than the standard errors of the data. However, this is not a problem for the majority of SSOs, and, in particular, for those with S/N < 50.

\begin{figure*}[!ht]
\centering
\includegraphics[width=2\columnwidth]{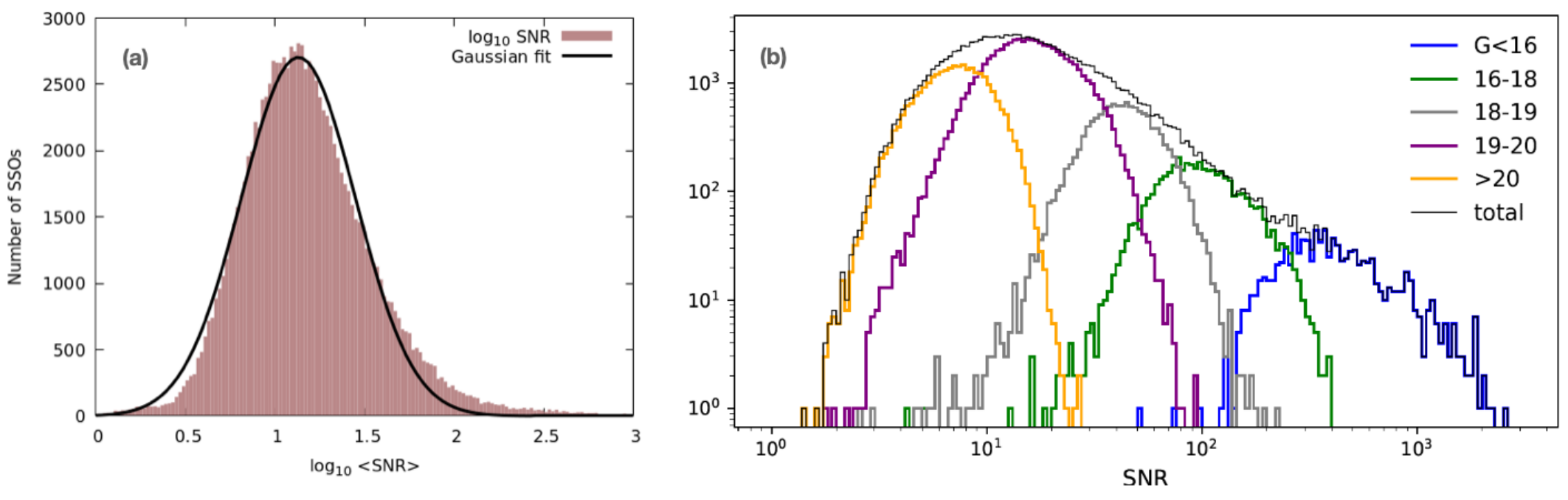}
\caption[Mean signal-to-noise ratio of SSO spectra]{(a) Mean $\left< \text{S/N} \right>$ for the 111\-818 SSOs for which the pipeline produced mean reflectance spectra. (b) $\left< \text{S/N} \right>$ for the 111\,818 SSOs of different magnitude classes. Each G-band magnitude class is represented by a separate curve (from right to left): SSOs brighter than 16~mag, between 16~mag and 18~mag, between 18~mag and 19~mag, between 19~mag and 20~mag, and fainter than 20~mag. 
The dark grey enclosing curve is the global histogram of $\left< \text{S/N} \right>$, the same shown in panel (a). }
\label{fig:cu4sso_validationspectra_SNRdistribution}
\end{figure*}

The histogram of the distribution of the $\left< \text{S/N} \right>$ value amongst the 111\,818 SSOs shows a quasi-lognormal distribution (Fig.~\ref{fig:cu4sso_validationspectra_SNRdistribution}) with a clear peak at $\left< \text{S/N} \right>$ = 13. The same figure also shows that reflectance spectra with $\left< \text{S/N} \right>$ values smaller than the peak value (13) are essentially due to the SSOs observed with magnitudes $>$19.

Visual inspection of randomly selected spectra with $\left< \text{S/N} \right> >$ 13 and with $\left< \text{S/N} \right> <$ 13 showed that the latter class is usually characterised by noisy spectra and the former by more accurate ones. The publication of spectra was limited to $\left< \text{S/N} \right> > $ 13 in \gdr{3}. The remaining spectra are waiting to be published in \gdr{4} based on 66 months of \gaia observations (cf. the 34 months of \gdr{3} observations). More transits will therefore be averaged for \gdr{4}. By applying the condition $\left< \text{S/N} \right> >$ 13, a total of 60\,518 SSOs was obtained. 

However, the condition of having $\left< \text{S/N} \right> >$ 13 does not necessarily guarantee that the reflectance spectrum of a single asteroid will be scientifically exploitable, whereas it could still be important for population studies. Therefore, it was decided not to reject additional reflectance spectra but to flag them on a wavelength-by-wavelength basis. Namely, an array of 16 integers, one for each wavelength of the spectral bands was created with the name of {\texttt{reflectance\_spectrum\_flag}} (\texttt{RSF}). A value equal to 0, 1, or 2 was assigned depending on whether the data at that band were validated, suspected to be of poorer quality, or deemed to be compromised, respectively. 

The procedure used to assign values to the RSF array consisted of three steps:
(1) all the elements of the RSF array were initially set to zero.
(2) The values of the mean reflectances and their uncertainties were explored for unreliable or non-numerical values, namely: The value of the RSF was set to 2 if the corresponding mean reflectance or its uncertainty were not numbers (NaN). The value of the RSF was set to 2 if the corresponding mean reflectance was larger than 2.5 or smaller than 0.2. This is to signal unrealistically high or low reflectance values. The value of the RSF was set to 2 if the corresponding uncertainty of the mean reflectance was larger than 0.5.
(3) The values of the mean reflectances and their uncertainties were explored in order to identify large discrepancies from an average continuous curve that would fit the discrete data. Specifically, a smoothing natural spline $S(\lambda)$ was fitted to the data points for which the corresponding RSF values were still zero after the previous step (see below for a description of how the spline was defined and implemented). The values of the RSF array were then set to 1 or 2 at those wavelengths where the mean reflectance has a distance from the smoothing spline larger than one or three times the corresponding uncertainty: if $\left | \overline{R}(\lambda_i) - S(\lambda_i) \right | > \sigma_{\overline{R}(\lambda_i)}$ then RSF$[i]$ = 1; if  $\left | \overline{R}(\lambda_i) - S(\lambda_i) \right | > 3 \times  \sigma_{\overline{R}(\lambda_i)}$ then RSF$[i]$ = 2. A value of spectral reflectance slope $\alpha_{\rm R}$ was then calculated by fitting a straight line to those data with RSF = 0 and 450~nm $\leq \lambda \leq$ 750~nm using weights equal to the inverse of the uncertainty squared. A smoothing natural spline $S'(\lambda)$ was fitted to the data points for which their corresponding RSF was still zero after the previous step and the value $z-i=2.5 \log_{10} (S'(\lambda_z) / S'(\lambda_i))$ where $\lambda_z$ = 893.2~nm and  $\lambda_i$=748.0~nm was calculated. 
Finally, only asteroids with -10~\%/100~nm  $<\alpha_{\rm R}<$ 40~\%/100~nm and $-0.75$ $< z-i < $ $0.5$ were validated. These latter conditions rejected only four asteroids with anomalously blue reflectance spectra, which we attributed to a flux loss of RP compared to BP.  

We used the cubic spline approximation (smoothing) CSAPS\footnote{\url{https://pypi.org/project/csaps/}} Python3 module to implement the smoothing spline with a smoothing coefficient equal to $5 \times 10^{-7}$.

\subsection{Comparison against ground-based spectrophotometry and spectra}\label{ssec:gbcomparison}
The consistency of \gdr{3} SSO mean spectra against literature data was estimated by comparing spectral parameters against the same parameters from ground-based surveys and comparing spectra against literature ones. Here, we calculated spectral parameters such as the spectral slope and the equivalent of the SDSS z-i colour using a well-established method \citep{Demeo2013Icar..226..723D}. Specifically, the spectral slope was determined from the angular coefficient of a straight line fitted to the mean reflectance values with a wavelength of between 450 and 760~nm and {\texttt {reflectance\_spectrum\_flag}} = 0. The z-i colour determination was obtained by fitting a natural smoothing spline (using Python package CSAPS; smoothing coefficient = $5 \times 10^{-7}$) to all the mean reflectance values with {\texttt {reflectance\_spectrum\_flag}} = 0 and then by calculating the z-i colour as $z-i = 2.5 \log_{10} R_z / R_i$, where z and i represent the values of the reflectances interpolated with the spline at 748 and 893~nm, respectively. 

We also downloaded literature (unsmoothed) spectra from the SMASSII website \footnote{\url{http://smass.mit.edu}}, selected those asteroids in common with the \gdr{3}, and applied the same aforementioned procedure to calculate their slopes and the z-i colours. We also obtained the taxonomic classification of the SMASS spectra from the SMASSII website \footnote{\url{http://smass.mit.edu/data/smass/Bus.Taxonomy.txt}}. Following known prescriptions \citep{DeMeo2009Icar..202..160D}, we grouped 
the classes S, Sa, Sk, Sl, Sq, and Sv into the S-complex, 
the classes X, Xc, Xe, Xk into the X-complex, 
and the classes C, Cb, Cg, Ch, Cgh into the C-complex.

\begin{figure}[h]
   \centering
   \includegraphics[width=\columnwidth]{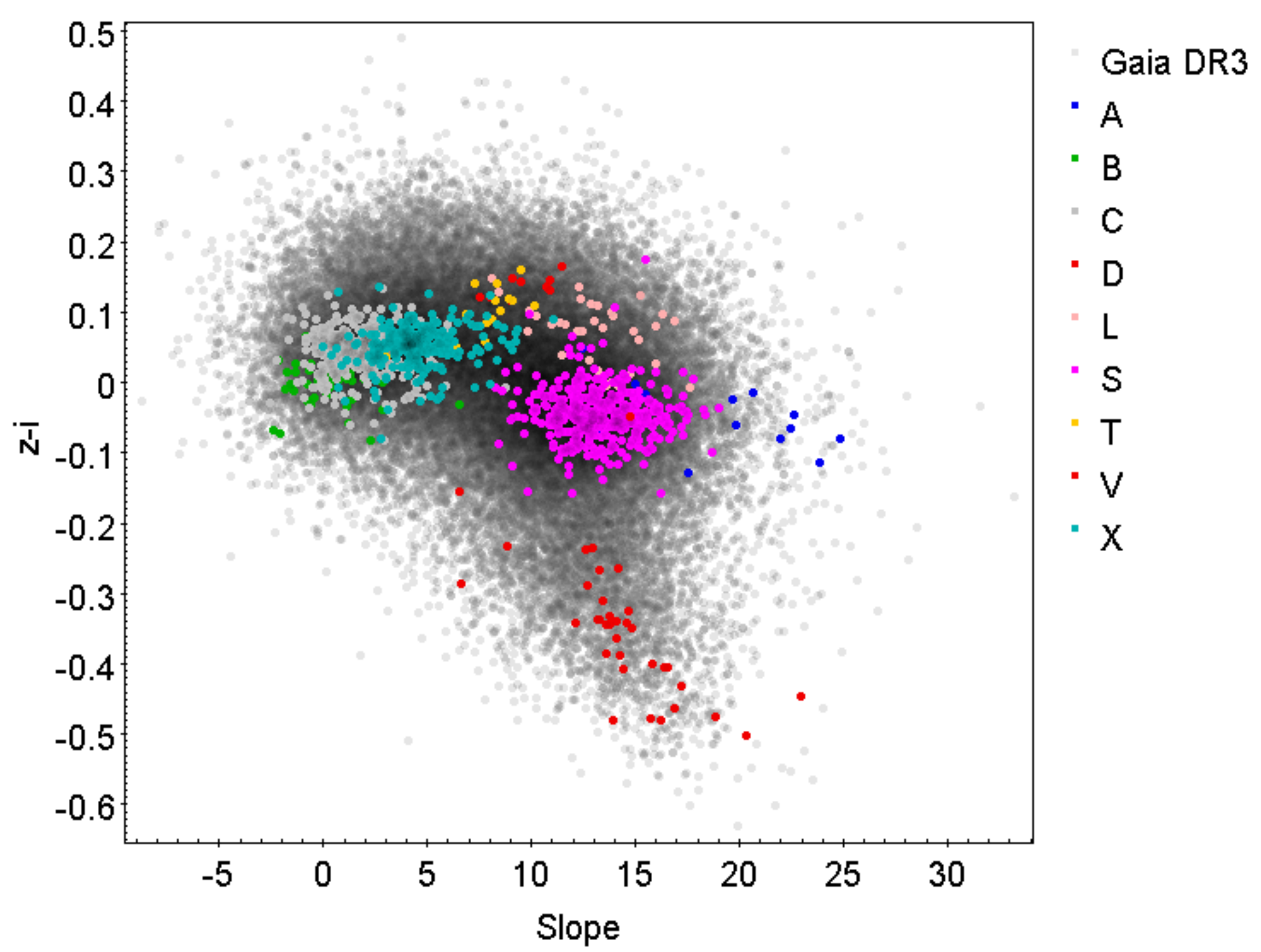}
   \caption{z-i colour vs. spectral slope of the asteroids of DR3 (grey dots). Over-plotted with circles of different colours are the same spectral parameters calculated by us (see Section~\ref{ssec:gbcomparison}) for the asteroids of SMASSII. The letters C, S, and X represent complexes, the other letters spectral classes. }
   \label{Fig:slope_vs_zi}%
\end{figure}

\begin{figure*}[h]
   \centering
   \includegraphics[width=2\columnwidth]{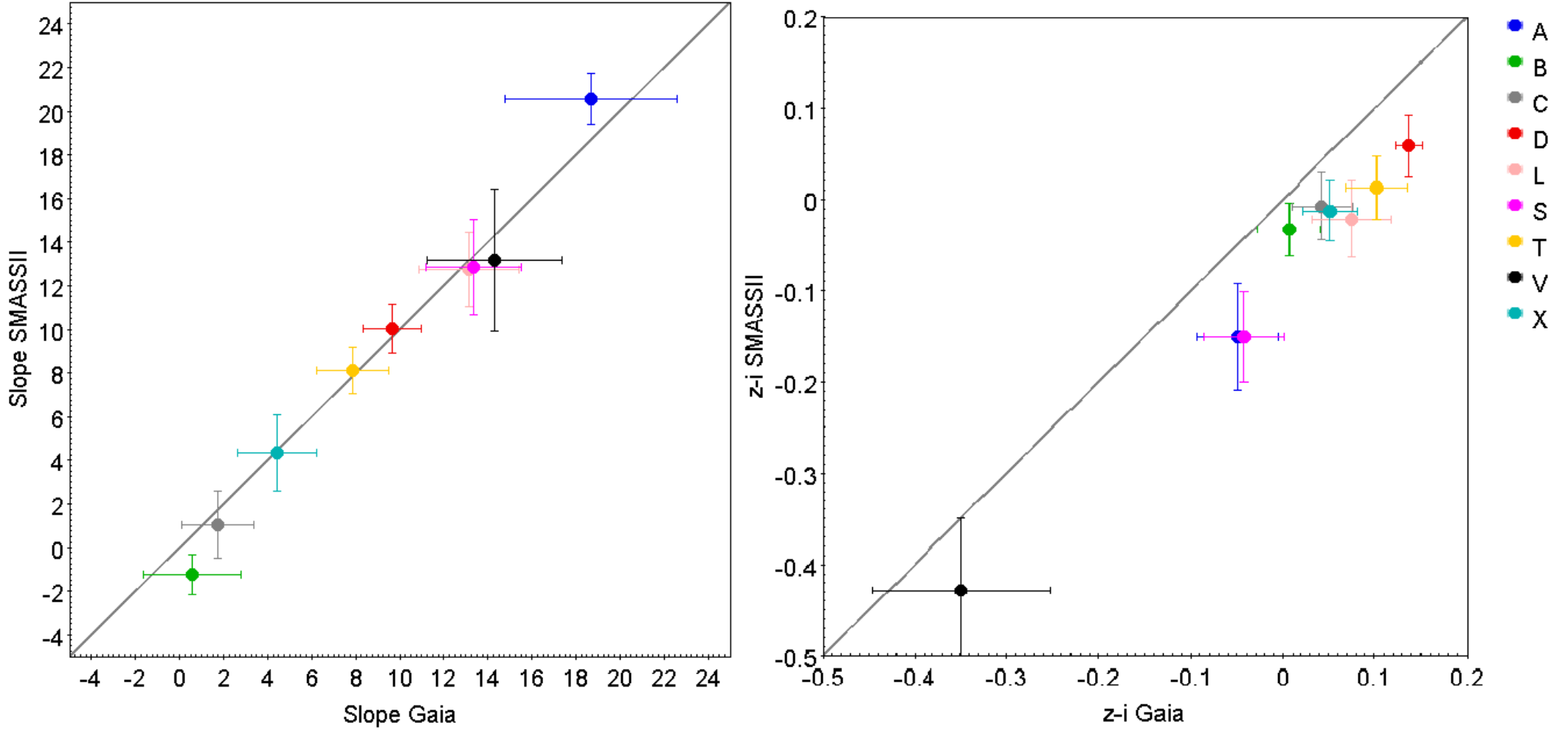}
   \caption{Comparison of mean spectral slopes and mean z-i colours of different taxonomic classes calculated on asteroids in common between \gdr{3} and the SMASSII survey.}
   \label{Fig:comp_slope_zi}%
\end{figure*}

Figure~\ref{Fig:slope_vs_zi} shows that the distribution of the spectral parameters for the \gdr{3} is qualitatively similar to that presented by other surveys \citep[e.g.][]{Parker2008Icar..198..138P,Demeo2013Icar..226..723D} in visible light and that SMASSII taxonomic classes and complexes overlap with the \gaia results as one would expect.
In order to highlight more subtle differences between asteroid reflectance spectra of the \gdr{3} and those of the SMASSII, we calculated the average spectral slope, the standard deviation of the spectral slope, the average z-i colour, and the standard deviation of the z-i colour for each complex and end-member spectral class for those asteroids that are in common between \gdr{3} and the SMASSII (Fig.~\ref{Fig:comp_slope_zi}); next we compared the average values and the dispersion of the aforementioned parameters between \gaia and SMASSII data (Fig.~\ref{Fig:comp_slope_zi}). We found a general good agreement between \gaia and SMASSII spectral slopes for all taxonomic classes and complexes. On the other hand, Fig.~\ref{Fig:comp_slope_zi} shows that \gdr{3} reflectance spectra have, in general, a smaller z-i colour index compared to those of SMASSII. The average difference between the z-i colours of the SMASS and Gaia is -0.08. This corresponds to a difference in the depth of the $\sim$1-$\mu$m band for those spectral classes with this feature. We investigate this difference in section~\ref{sec:discussion}.

\begin{figure}[h]
   \centering
   \includegraphics[width=\columnwidth]{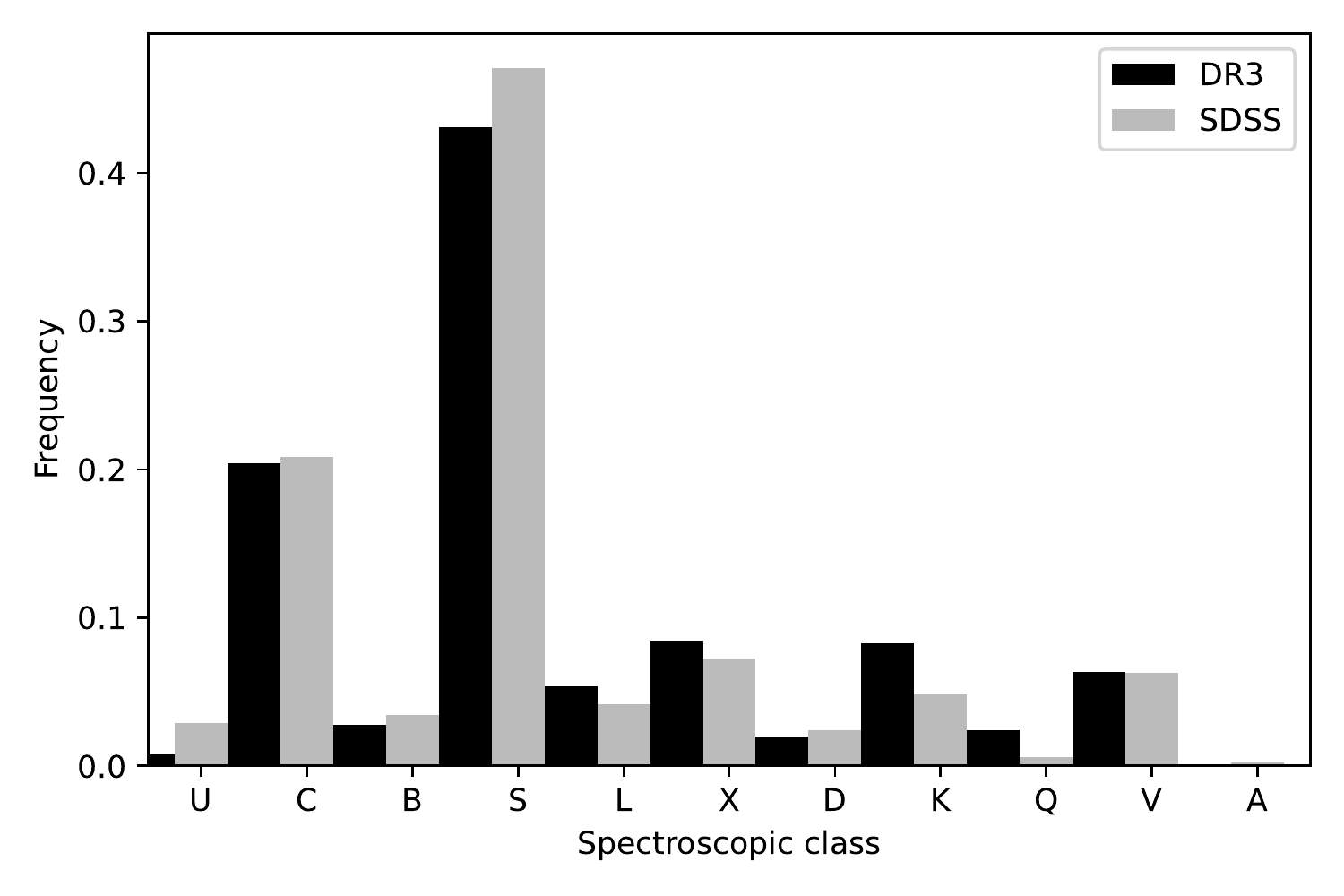}
   \caption{Histogram of the amount of asteroids for each spectral class in the \gdr{3} in comparison with the SDSS classification of \citet{Demeo2013Icar..226..723D}.}
   \label{Fig:class_histo}%
\end{figure}

An accurate classification of \gdr{3} SSO spectral reflectances is expected for future works. However, it is possible to divide the \gdr{3} SSO data set into broad taxonomic groups using slope and z-i values \citep{Demeo2013Icar..226..723D}. To this aim, the z-i versus slope plane is divided into the rectangular areas defined in Table 3 of \cite{Demeo2013Icar..226..723D}; we added -0.08 to the z-i values of the \gdr{3} in order to account for the offset found at the previous step of the analysis (as shown in Fig.~\ref{Fig:comp_slope_zi}, right panel); we classified asteroids following the same decision tree as that used by \citet{Demeo2013Icar..226..723D}, where the slope and z-i values of the SSO are compared with each region in
the following order: C, B, S, L, X, D, K, Q, V, A. If an object fell in more than one class, it was designated to the last class in which it resides. If an object did not obtain a class, that is, it was outside the boundaries of the previous classes, it was designated  `U', which is historically used to mark unusual objects. Next, we counted the number of asteroids in each class and divided each number by the total number of \gdr{3} SSOs to obtain the frequency per class, which we then displayed in Fig. \ref{Fig:class_histo}. In the same figure, we also reported the frequency of asteroids in each class from the SDSS as classified by \citet{Demeo2013Icar..226..723D} -- from their file alluniq\_adr4.dat. In general, Fig. \ref{Fig:class_histo} shows qualitative agreement between \gdr{3} and SDSS spectral classes.

%

Next, we calculated an RGB colour palette from the values of the slope and the z-i colour, following an approach similar to that of \cite{Parker2008Icar..198..138P} (an example code is given in Appendix~\ref{sec:appendix:spec2rgb}). According to this palette, asteroids that are spectrally blue or spectrally neutral are given a blue colour and those that are spectrally red are given a red colour, whereas the amount of green increases with decreasing value of the z-i magnitude; for example, with increasing depth of the $\sim$1-$\mu$m band. Hence, S-complex asteroids tend to have pale red to brownish colour, C-complex asteroids are blue, D- and L- types are red in colour, X-complex and K-types are magenta, and V-types are green. 

\begin{figure*}[!ht]
   \centering
   \includegraphics[width=\columnwidth]{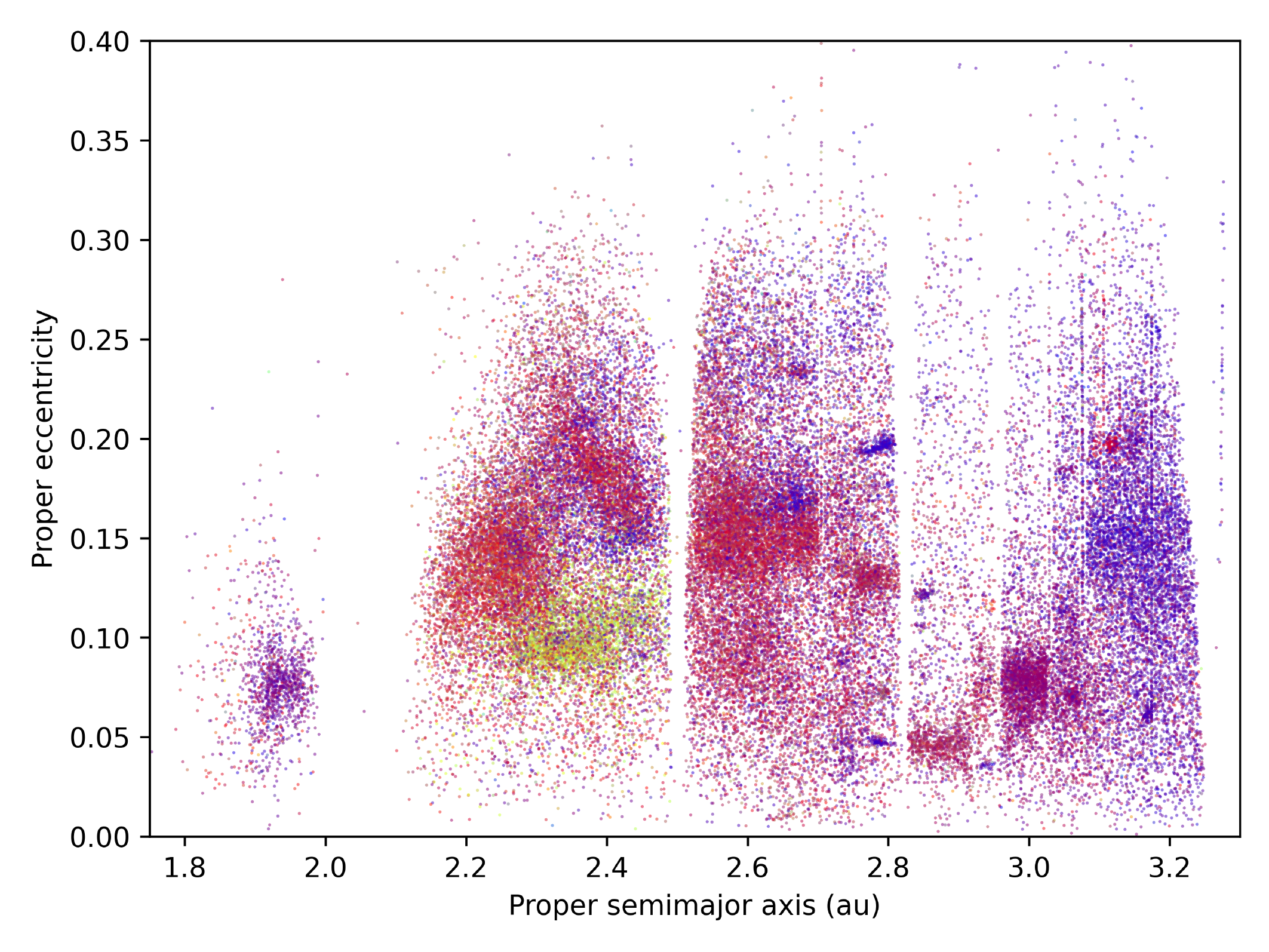}
   \includegraphics[width=\columnwidth]{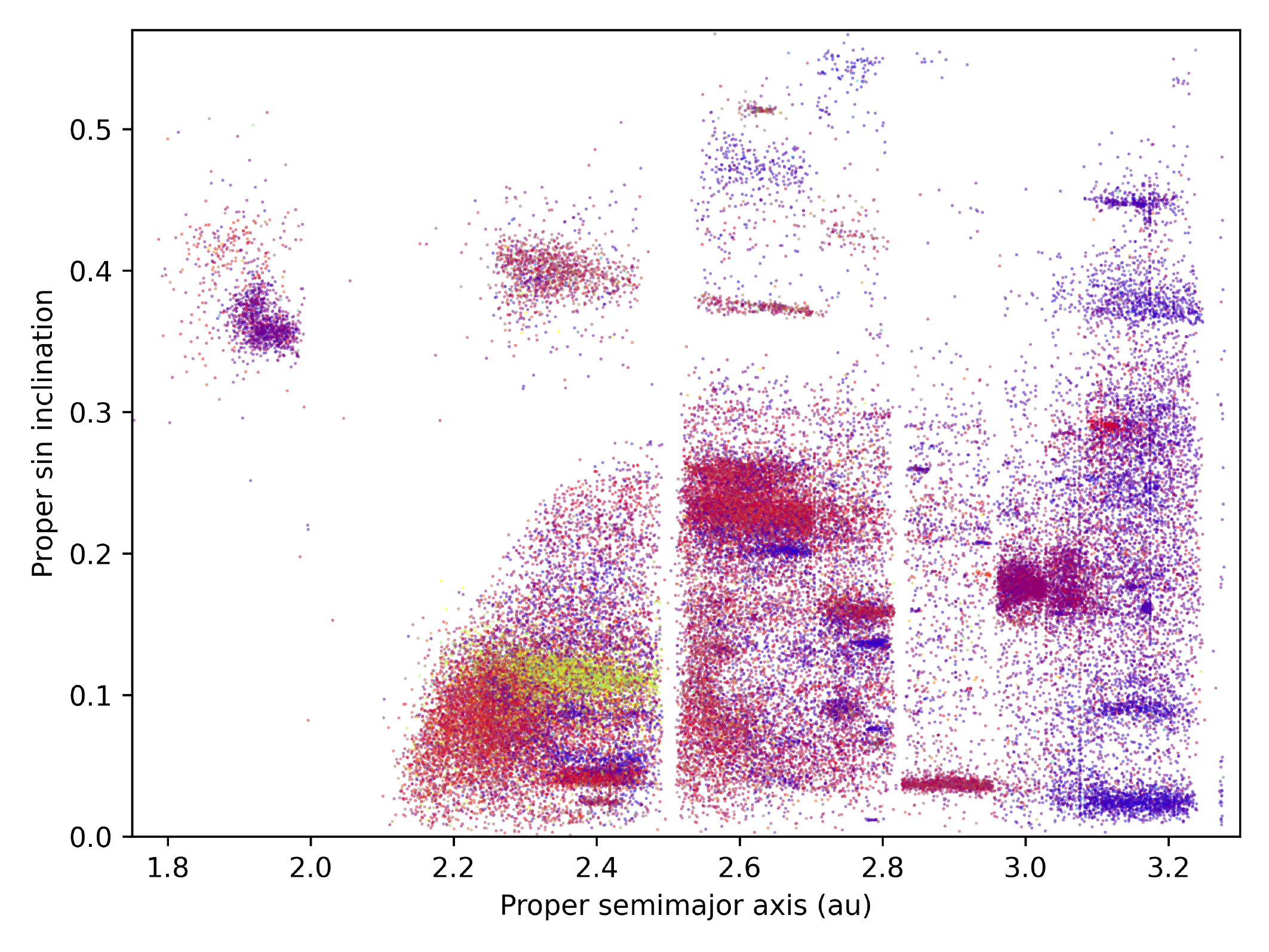}
   \caption{Plot of the proper semi-major axis vs. proper eccentricity and sin of proper inclination for \gdr{3} SSOs of the main-belt and Hungaria region. The colour of each dot is representative of the object’s colour measured by \gaia according to the colour scheme defined in Section~\ref{ssec:gbcomparison}.}
   \label{Fig:propa_ei_map}%
\end{figure*}

Having assigned to those SSOs of the main belt their proper orbital elements from \citet{Novakovic2009SerAJ.179...75N}, we then produced colour plots of the orbital distribution of asteroids (Fig.~\ref{Fig:propa_ei_map}). 
These figures show a global gradient of colours of asteroids consistent with previous findings \citep{Parker2008Icar..198..138P, Sergeyev2021A&A...652A..59S}. Asteroid collisional families are clearly distinguishable by the naked eye as groups of points with generally homogeneous colours. 

The average phase angle of \gaia's SSO observations is around 20 - 25$^\circ$, with considerable dispersion (Fig.~\ref{Fig:hist_phase_Angle}). It is known that phase angle has an effect on spectral reddening and the depth of the absorption bands \citep[e.g.][]{1971AJ.....76..141T, 1976Icar...28...53M, 2002Icar..155..189C, 2015aste.book...43R, 2020A&A...644A.142F}. In particular, an increase of the spectral slope and decrease in the depth of the absorption bands are observed for increasing phase angle  \citep{2012Icar..220...36S, 2015A&A...580A..98C}. However, there are also physical processes that affect
spectral slope and band depth, such as space weathering \citep{2010Icar..209..564G,Vernazza2016AJ....152...54V, 2018Icar..302...10L,Hasegawa2019PASJ...71..103H} and grain size. It is therefore important to be able to disentangle the effects of such physical processes from those due to the geometry of the observations on asteroid spectra. 
To investigate this issue, \cite{2020A&A...642A..80C} carried out spectroscopic observations of SSOs at phase angles and in the wavelength range similar to those obtained by \gaia. These authors selected objects in order to cover several taxonomic classes. In Appendix \ref{appendix:spectraPlots}, we compare \gdr{3} mean reflectance spectra with those of \cite{2020A&A...642A..80C} along with other literature spectra for reference. 

In general, we find satisfying agreement with literature spectra. This agreement is particularly good for asteroids with  reddish, featureless spectra (e.g. D-types, such as (269) Justitia and (624) Hektor). 
(269) Justitia seems to be a very peculiar and interesting object based on some recent works  \citep{2020A&A...642A..80C,Hasegawa2021ApJ...916L...6H}, since this asteroid turned out to be unique and well distinct from other spectrally reddish asteroids in that sample.
Good agreement between the \gaia reflectance spectra and those of the literature is visible for asteroids with moderate reddish spectra but mostly located in the inner main belt \citep[e.g. for (96) Aegle we can observe that its \gaia reflectance spectrum goes deeper in the blue, in accordance with spectra of troilite-dominated objects;][]{1992Metic..27Q.207B}. 
The \gaia reflectance spectra of asteroids (1904) Massevitch and (1929) Kollaa, representatives of the V-type taxonomic class in the sample of \cite{2020A&A...642A..80C}, show slopes that are coherent with literature ones, except for a small wavelength shift of the centre of the 1-$\mu$m absorption band. A potential caveat is that the observations of these two asteroids were made at very low phase angles for the telescope-based studies.
Concerning olivine-dominated asteroids, such as the A-types, with a red spectrum and an absorption band characteristic of the olivine at 1-$\mu$m, we observe marked differences between Gaia reflectance spectra of (246) Asporina and that of \cite{2020A&A...642A..80C}. However, the reflectance spectrum of Asporina of \cite{2020A&A...642A..80C} is also very different from that of \cite{Bus2002Icar..158..106B}. 
%
\gaia reflectance spectra of C-type asteroids (175), (207), (261), and (3451) are in very good agreement with the ground-based ones. The absorption present in the wavelength range 400--500~nm, the ultraviolet (UV) downturn, is clearly visible. The \gaia spectrum of asteroid (8424) has some issues at the edge of the spectrum.
Stony asteroids typical of the S-type are also included in this comparison set. These asteroids present a moderate slope between 400 and 700~nm and a tiny absorption band around 1 $\mu$m, representative of the presence of silicates. We can observe that the \gaia reflectance spectra of asteroids (39), (82), (179), (720), (808), (1662), and (2715) are very compatible with S-type reflectances. With the exception of the case of asteroid (39) Laetitia, the \gaia reflectance spectra of all the other aforementioned SSOs have  slopes similar to those of the corresponding reflectance spectra measured from ground-based telescopes. We note that for some cases, namely asteroids (720), (808), (1662), and (2715), the absorption band is weaker in \gaia data compared to the literature. This is as expected from the basis of results presented in Fig.~\ref{Fig:comp_slope_zi}.
The \gaia reflectance spectrum of asteroid (43962), despite being noisy with an average $\text{S/N}$ value of $\sim$13.7, thus very close to the rejection threshold, still shows that \gaia data show a nice match when superimposed on the ground-based reflectance spectra.

\subsection{Comparison with literature reflectance spectra from space observations}
The large majority of asteroid reflectance spectra available in the literature were obtained using ground-based telescopes. However, the reflectance spectra of some SSOs have also been obtained from space missions, and they can be compared with those derived from Gaia observations. 

\begin{figure}[!ht]
   \centering
   \includegraphics[width=\columnwidth]{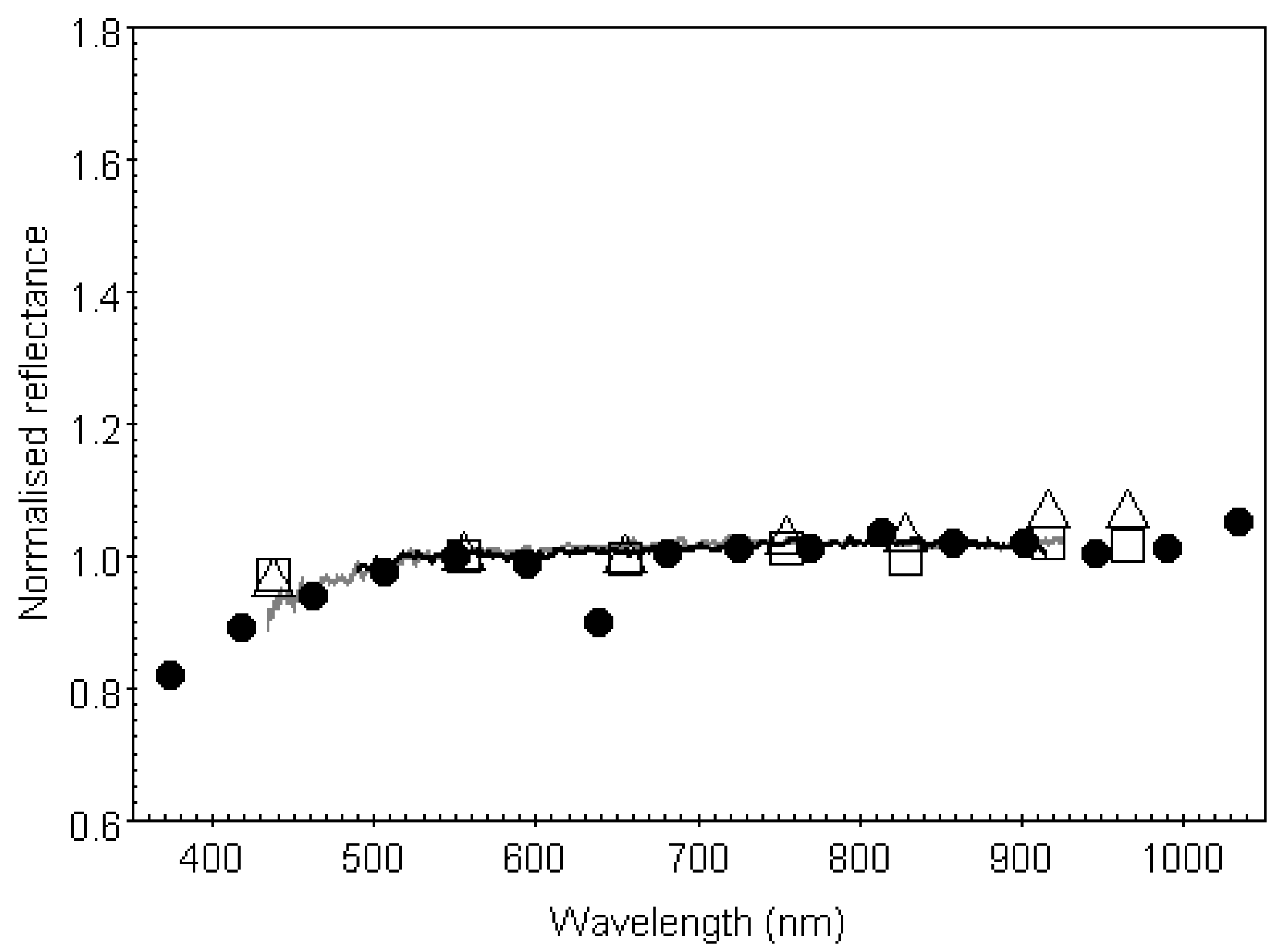}
   \caption{Gaia mean reflectance spectrum of the asteroid (1) Ceres, obtained at an average phase angle of 17.5\deg ,  plotted with solid circles. Literature ground-based spectra from \cite{Bus2002Icar..158..106B} and \cite{LAZZARO2004} are shown with a grey line (phase angle of $\sim$18.6\deg) and a black line (phase angle of 16\deg), respectively. Data obtained in space by the NASA Dawn mission \citep{2016ApJ...817L..22L} during RC1 are displayed with black open squares (phase angle between 17.2 and 17.6\deg), and during the RC2 with black open triangles (phase angle 42.7 to 45.3\deg).}
\label{Fig:ceres}%
\end{figure}

(1) Ceres is the largest asteroid and the only dwarf planet in the main belt. 
It is classified as C type due to its relatively flat reflectance spectrum, which presents a UV downturn typical of this spectral class (Fig.~\ref{Fig:ceres}).  
The NASA space mission Dawn \citep{2004P&SS...52..465R}, launched in 2007, began observations of Ceres in December 2014. The Dawn framing camera (FC) instruments observed this body for at least one full rotation in three separate periods in February and April 2015 during its approach phase (at three different phase angles, distances, and image resolutions). These three epochs are referred to as rotational characterisations (RC1, RC2, and RC3). \cite{2016ApJ...817L..22L} compared spectra measured within RC1 and RC2 with several ground-based spectra and observed an increase in the spectral slope with increasing phase angle. In Fig.~\ref{Fig:ceres}, we  plot the \gaia mean reflectance spectrum of (1) Ceres and compare it with literature ground-based spectra \citep{Bus2002Icar..158..106B,LAZZARO2004} and the two Dawn spectra presented in \citep{2016ApJ...817L..22L}.
The agreement between \gaia's reflectance spectrum and those of the literature from the ground and space is good, with the exception of a \gaia data point at wavelength 632~nm. The latter is very likely an artefact due to a problem in the overlapping region of the BP and the RP. It is also visible in some other spectra, in particular for very bright SSOs. The asteroid Ceres has been observed by \gaia at an average phase angle of 17.5~\deg, which is approximately the same value as the previous cited ones. One can observe a slight increase in the spectral slope of Dawn's RC2 spectrum, as demonstrated by \cite{2016ApJ...817L..22L}.

\begin{figure}[!ht]
   \centering
   \includegraphics[width=\columnwidth]{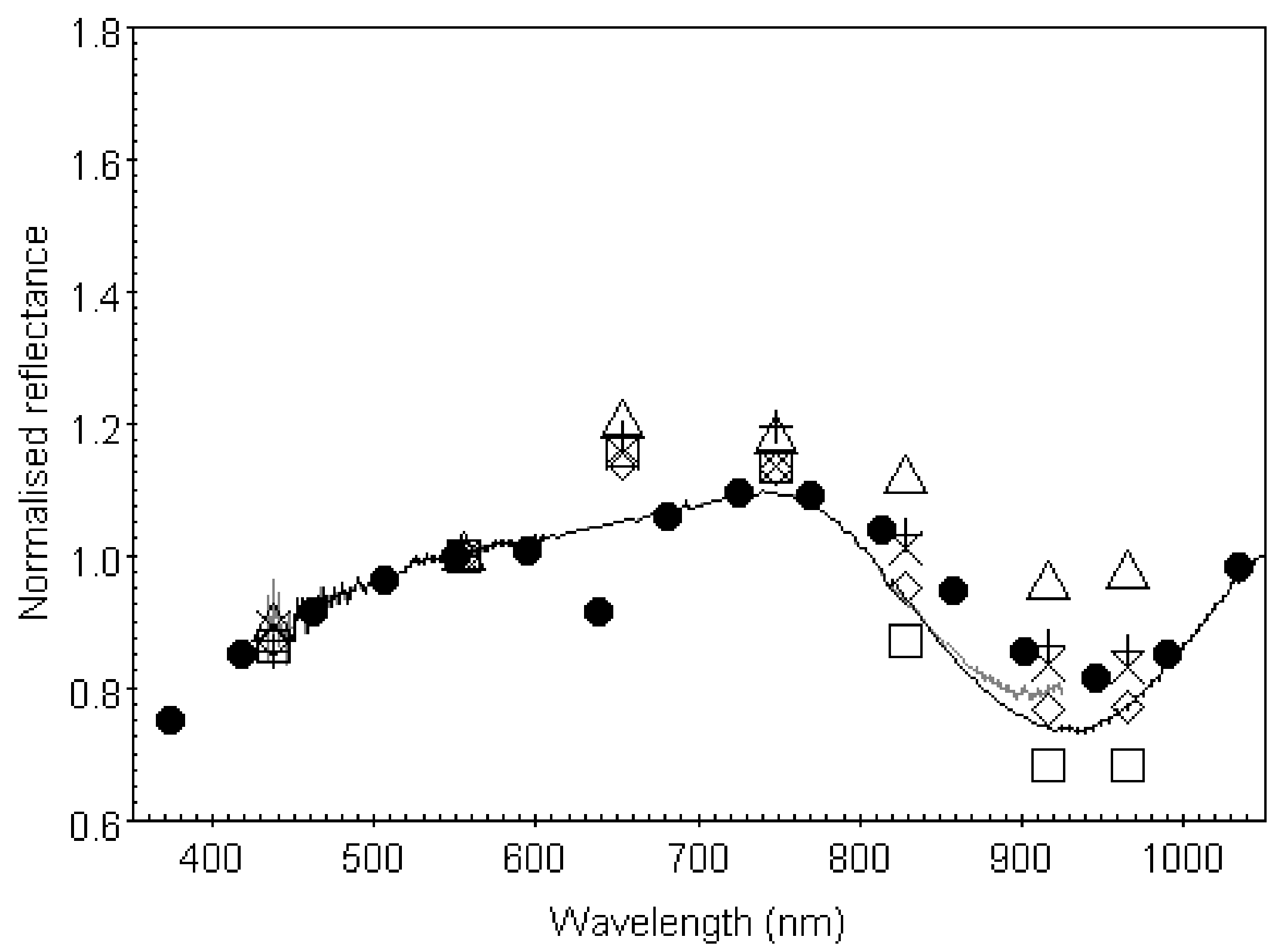}
   \caption{\gaia mean reflectance spectrum of the asteroid (4) Vesta, observed at an average phase angle of 21.3\deg, shown with black circles. Two ground-based spectra from \cite{Bus2002Icar..158..106B}, observed at a phase angle of 11.9\deg, and from MITHNEOS survey, observed at a phase angle of 23.5\deg, are shown with grey and black lines, respectively. Spectra from the NASA Dawn mission \citep{2012Sci...336..700R} for the bright, dark, grey, and orange terrains are plotted, respectively, with black open squares, black open upside down triangles, black crosses, and black plus symbols.  The red spectrum, plotted with black open diamonds, corresponds to the global average spectrum of Vesta. Vesta spectra presented in \citep{2012Sci...336..700R} were normalised at $\lambda = 750$ nm, and therefore we  re-normalised them at $\lambda = 550$ nm. These spectra were obtained at a phase angle of 30\deg.}
\label{Fig:vesta}%
\end{figure}

(4) Vesta is the second most massive asteroid in the main belt and is spectroscopically similar to basaltic achondrite meteorites, known as Howardites, Eucrites, Diogenites (HEDs), of which it is considered to be the parent body. Vesta's reflectance spectrum presents typically two strong absorption bands around $\sim$1 $\mu$m and $\sim$2 $\mu$m. The NASA Dawn mission has observed Vesta before continuing towards Ceres. Using images taken from the Dawn FC, \cite{2012Sci...336..700R} computed spectra from four different areas of Vesta, and classified these terrains as bright, corresponding to bright ejecta around the 11.2-km diameter fresh impact crater Canuleia, dark, in order to highlight the dark material on the crater wall and in the surroundings of the 30-km diameter Numisia crater, grey, relative to the grey ejecta blanket of the 58-km diameter Marcia crater, and orange, which is characteristic of the 34-km diameter impact crater Oppia. These authors produced one spectrum per area, plus a global average spectrum. \cite{2012Sci...336..700R} explained that fresh impact craters have higher reflectances than background surface and deeper absorption bands. In Fig.~\ref{Fig:vesta}, we plotted the \gaia reflectance spectrum against literature ground-based spectra \citep{Bus2002Icar..158..106B,Binzel2019Icar..324...41B} from the MITHNEOS survey \footnote{\url{http://smass.mit.edu/data/spex/sp86/}} and the five space-based Dawn spectra. The \gaia spectrum presents the same artefact already detected for the Ceres spectrum at $\lambda = 632~nm$. This point is affected by the overlapping of BP and RP. Otherwise, the \gaia spectrum is very similar to both the ground-based and the space-based spectra in its blue part. Its spectral slope is consistent with that of ground-based spectra. The \gaia spectrum appears less deep in the 1-$\mu$m absorption band than the ground-based ones but it is similar to the grey area spectrum. According to \cite{2012Sci...336..700R}, most of Vesta's surface is covered with grey material, which could the explain its match with the \gaia spectrum.

\begin{figure}[!ht]
   \centering
   \includegraphics[width=\columnwidth]{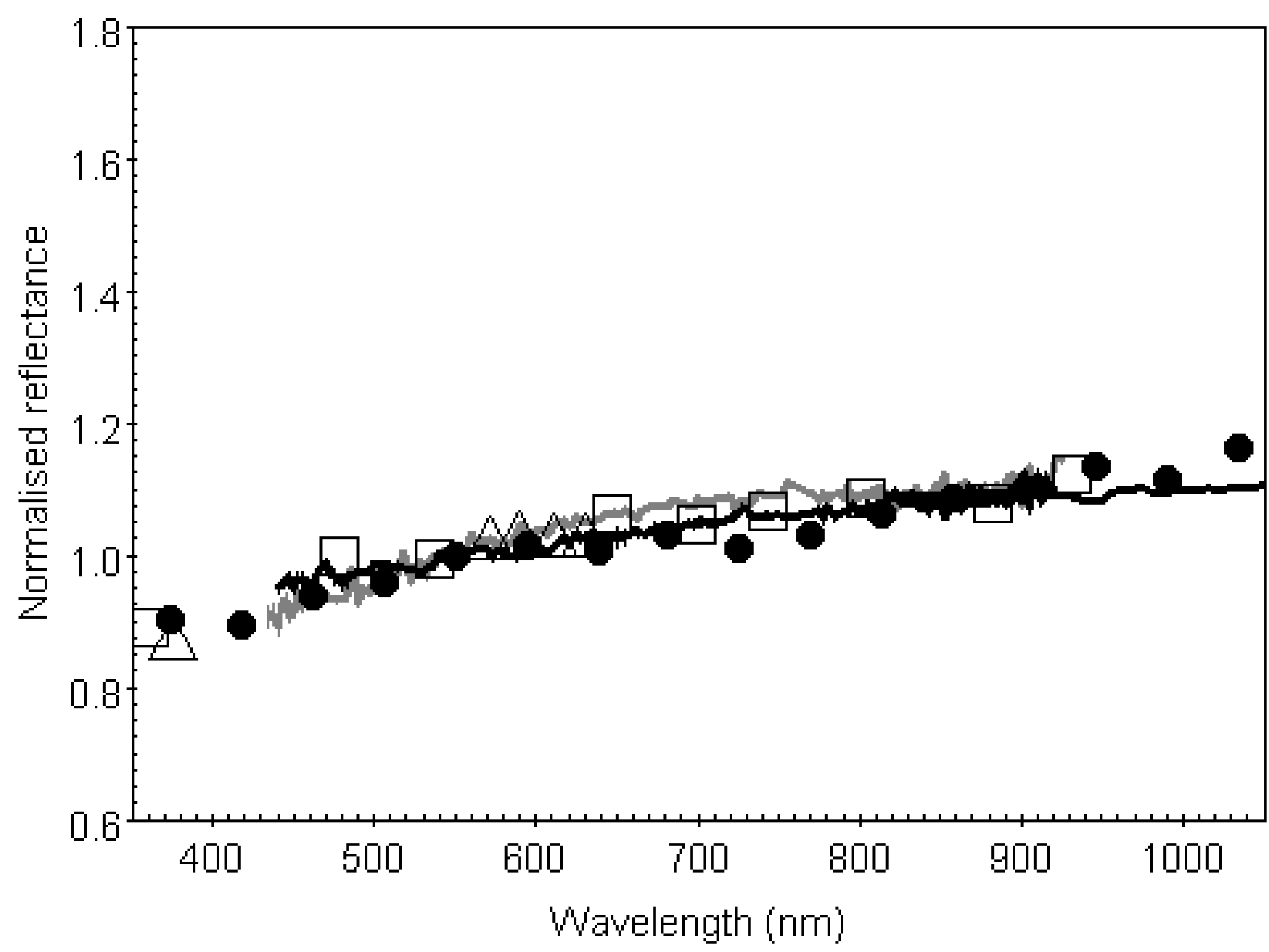}
   \caption{ \gaia mean reflectance spectrum of the asteroid (21) Lutetia is plotted, with black circles, together with literature ground-based spectra (grey line) and data obtained by the ESA Rosetta mission using OSIRIS NAC (black open square) and OSIRIS WAC  (black open  upside down triangle) at a phase angle of 7.74~\deg \citep{2011Sci...334..487S}.}
\label{Fig:lutetia}%
\end{figure}

(21) Lutetia is an M-type asteroid \citep{1989aste.conf.1139T}. It was visited in 2010 by the ESA mission Rosetta on its way to comet 67P/Churyumov-Gerasimenko. Rosetta observed Lutetia using the Optical, Spectroscopic, and Infrared Remote Imaging System (OSIRIS), which includes a wide-angle and a narrow-angle camera \citep[WAC and NAC, respectively;][]{2011Sci...334..487S}. In Fig.~\ref{Fig:lutetia}, we plot the \gaia mean reflectance spectrum together with literature ground-based spectra and the Rosetta spectra. There is an excellent match between all spectra. The slope of the \gaia reflectance spectrum is also consistent with those from the literature reflectance spectra.

\begin{figure}[!ht]
   \centering
   \includegraphics[width=\columnwidth]{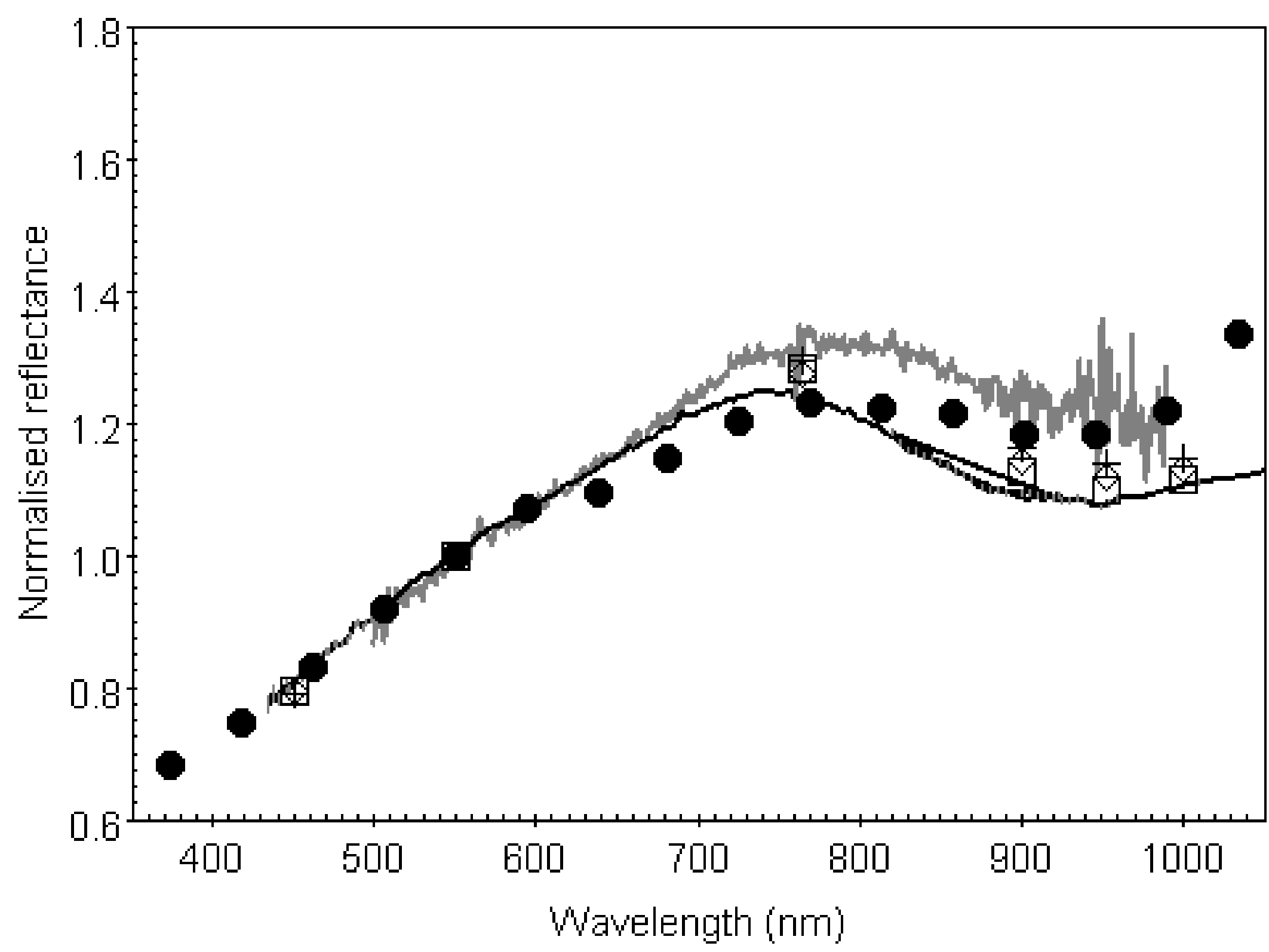}
   \caption{\gaia mean reflectance spectrum of the asteroid (433) Eros, shown with black circles, together with ground-based spectra from \cite{1992Icar..100...85V} and \cite{Binzel2019Icar..324...41B} represented with a grey and a black line, respectively. Data obtained in space by the NASA NEAR Shoemaker mission \citep{2000Sci...289.2088V} during the approach, at a phase angle of between 49 and 55\deg, and during the flyby at phase angles of 82\deg \ and 112\deg \ are shown with open diamonds, open squares, and black crosses, respectively.}
\label{Fig:eros}%
\end{figure}

(433) Eros is a near-Earth and Mars-crossing asteroid that has been visited by the NASA Near Earth Asteroid Rendezvous (NEAR) Shoemaker spacecraft in the early 2000s \citep{2000Sci...289.2088V}. Eros is spectroscopically consistent with a silicate-based composition and classified as S-type \citep{Bus2002Icar..158..106B}. Several spectra of Eros were measured by NEAR, one during the approach phase at large phase angles (between 49 and 55\deg) and two during the flyby phase at even larger phase angles (82 and 112\deg). In Fig.~\ref{Fig:eros}, we compare the \gaia reflectance spectrum with the ones obtained by the NEAR mission and two additional reflectance spectra acquired from ground-based telescopes \citep{1992Icar..100...85V, Binzel2019Icar..324...41B}. The slope of the \gaia spectrum is slightly less red than the other ones. The depth of the 1 $\mu$m absorption band is intermediate between those of the ground-based spectra, but slightly less intense than that measured by the NEAR mission. 

\begin{figure}[!ht]
   \centering
   \includegraphics[width=\columnwidth]{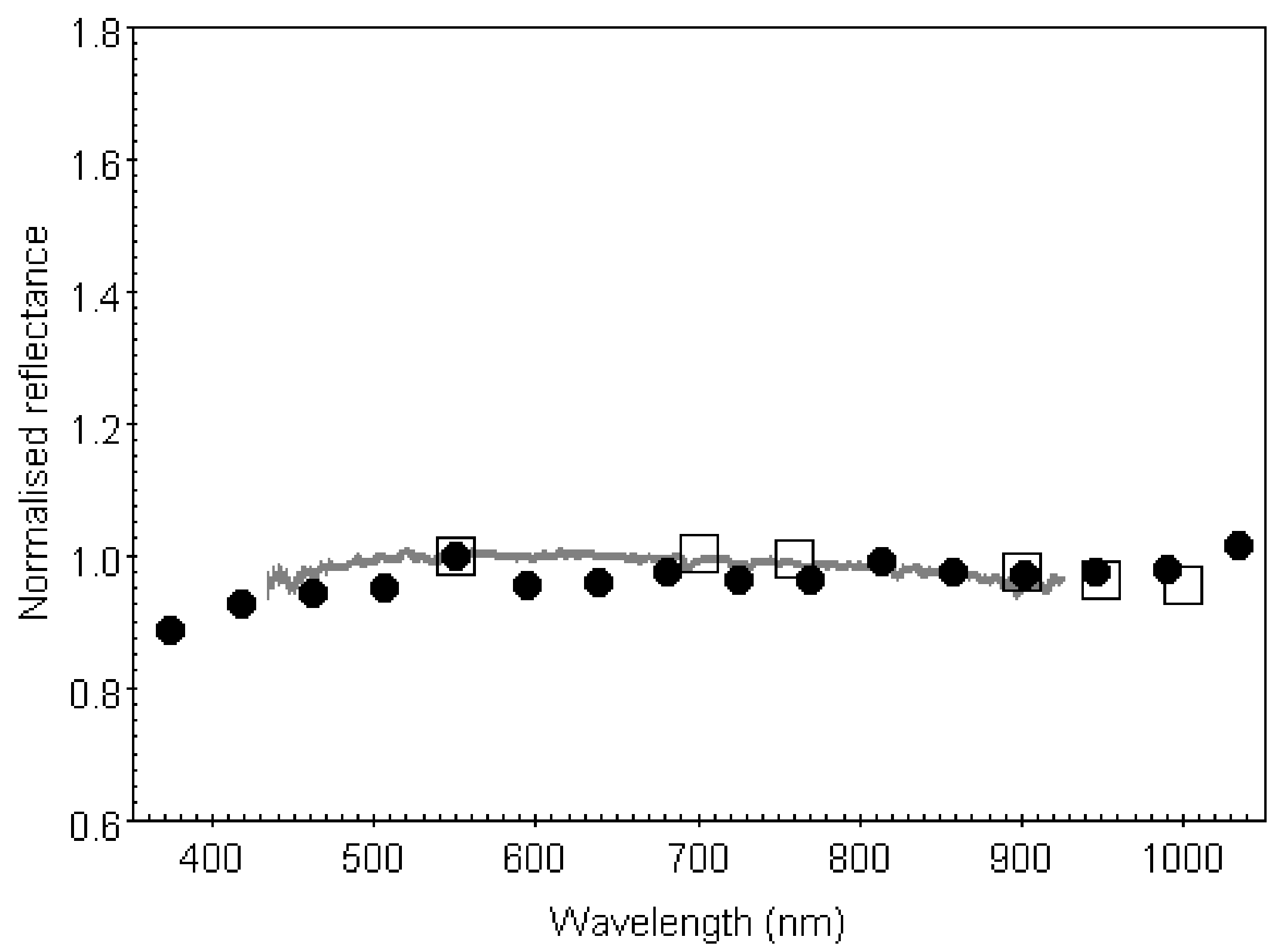}
   \caption{ \gaia mean reflectance spectrum of the asteroid (253) Mathilde, shown
with black circles, together with ground-based spectra 
from \cite{Bus2002Icar..158..106B} shown as a grey line, and data obtained in space by the NASA NEAR Shoemarker mission \citep{1999Icar..140...53C} with open squares (NEAR spectrum was derived from the images taken at an average solar phase angle of 41\deg).}
\label{Fig:mathilde}%
\end{figure}

(253) Mathilde is a main belt asteroid that the NASA Shoemaker mission flew by on its way to Eros. In Fig.~\ref{Fig:mathilde}, we plotted the \gaia reflectance spectrum compared to those obtained by the NEAR mission and also with spectra that were obtained from ground-based telescopes. The agreement between all spectra is quite excellent.

\begin{figure}[!ht]
   \centering
   \includegraphics[width=\columnwidth]{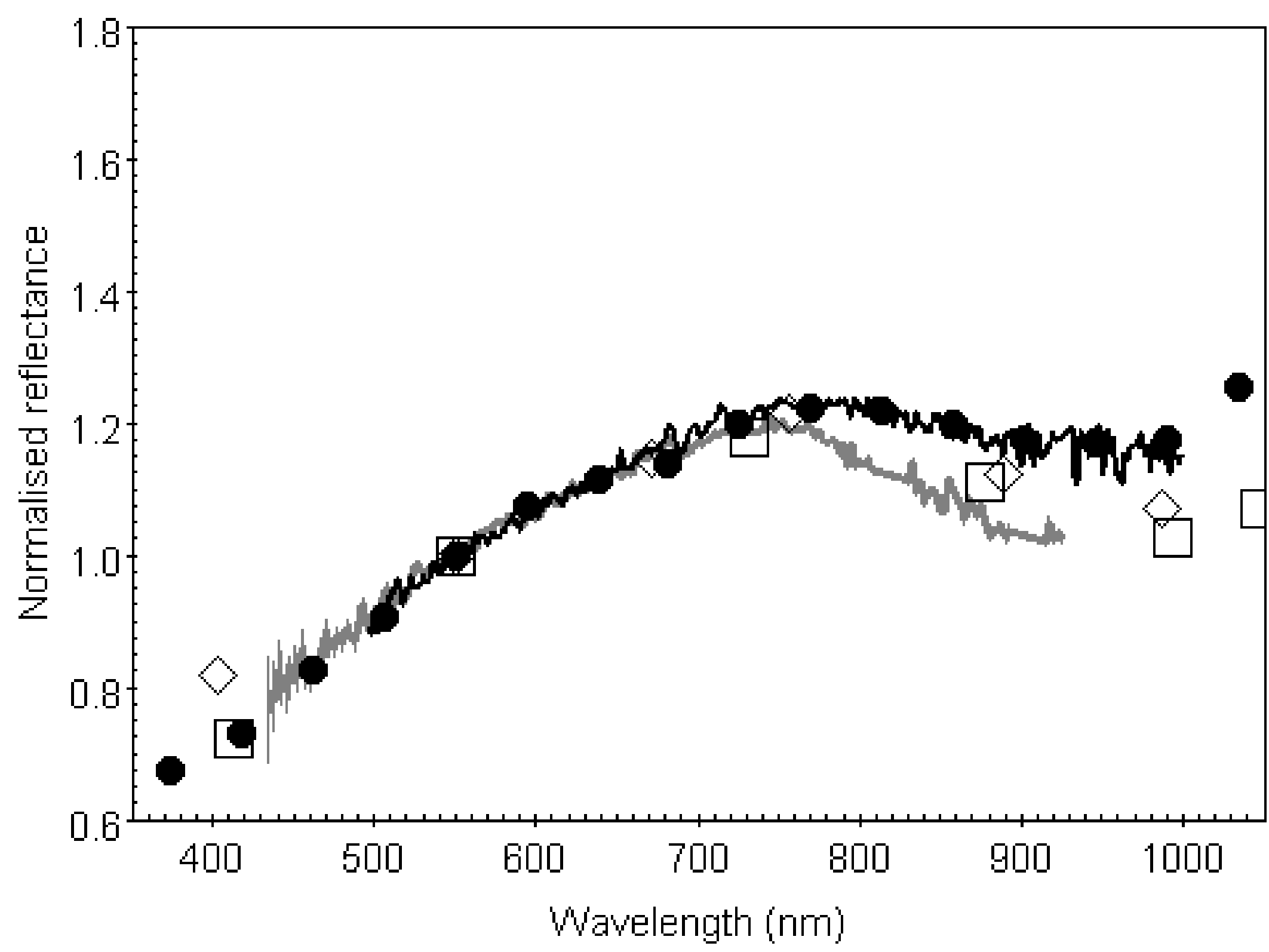}
   \caption{\gaia mean reflectance spectrum of the asteroid (951) Gaspra, shown with black circles, together with  literature ground-based spectra of \cite{Bus2002Icar..158..106B} and \cite{Xu1995Icar..115....1X} with grey and black lines, respectively. Data obtained in space by NASA Galileo mission \citep{1994LPI....25..453G} at phase angles of  51\deg and 31\deg are also shown with open squares and diamonds, respectively.} 
\label{Fig:gaspra}%
\end{figure}

(951) Gaspra is an S-type asteroid belonging to the Flora family \citep{Nesvorny2015aste.book..297N}. Gaspra was visited by the NASA Galileo spacecraft in October 1991. Figure~\ref{Fig:gaspra} reveals a very good match between the \gaia mean reflectance spectrum and the ground-based one from \cite{Xu1995Icar..115....1X}. Spectral slopes of all the compared spectra are also mutually consistent. The 1 $\mu$m-absorption is deeper in all the reflectance spectra obtained from the Galileo mission than in the one obtained by \gaia.

\begin{figure}[!ht]
   \centering
   \includegraphics[width=\columnwidth]{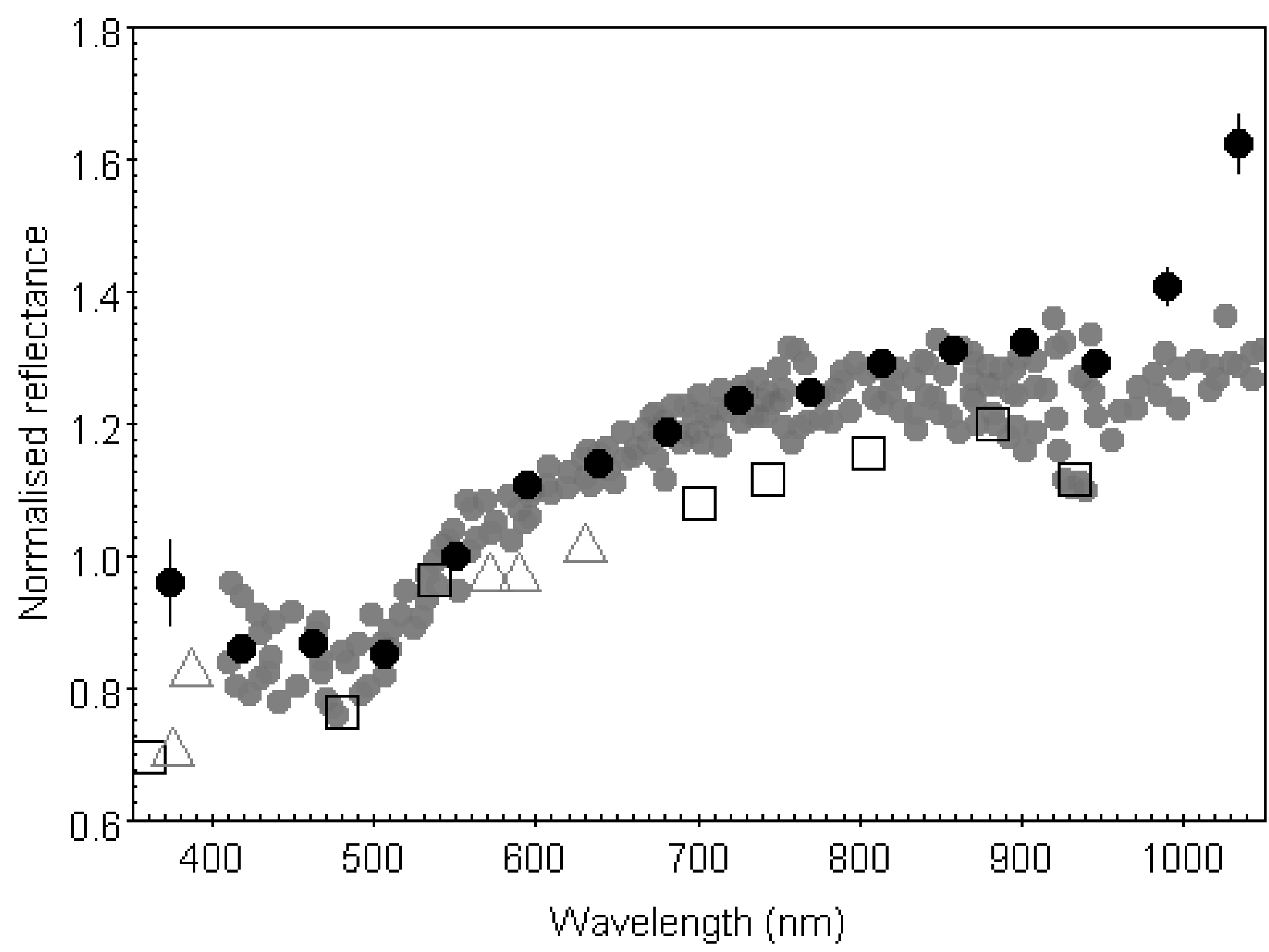}
   \caption{\gaia mean reflectance spectrum of the asteroid (2867) \u{S}teins, shown with black circles together with literature ground-based spectra from \cite{2005A&A...430..313B}, with grey circles, and data obtained in space by the ESA Rosetta mission using OSIRIS NAC, black open squares, and OSIRIS WAC, black open  upside-down triangle \citep[phase angles between 0 and 132\deg,][]{2010Sci...327..190K}.}
\label{Fig:steins}%
\end{figure}

(2867) \u{S}teins is a main-belt asteroid with a reflectance spectrum that is quite rare. It belongs to the E type spectral class. The ESA Rosetta mission flew by \u{S}teins in September 2008. Several spectra were taken using the OSIRIS cameras \citep{2010Sci...327..190K}. In Fig.~\ref{Fig:steins}, we compared the \gaia mean reflectance spectrum with a ground-based one from \cite{2005A&A...430..313B} and the ones derived from Rosetta data. We find good agreement with the \gaia spectrum and those presented by \cite{2005A&A...430..313B}. The slope of the Rosetta mission reflectance spectra appears less steep, but the overall comparison is also remarkable when taking into account the fact that the data from the Rosetta mission were acquired from a wide range of phase angles. The upturning of the \gaia data beyond 1000~nm is probably due to an artefact created by the method used for the calculation of the reflectance and by the `alien' photons problem (see Section~\ref{sec:discussion}).  

\subsection{\gaia view of space weathering in S-types}\label{ssec:spaceweatheringStype}
First of all, we selected asteroids with \gaia reflectance spectra compatible with spectral classes within the S-complex and belonging to collisional families \citep{Nesvorny2015aste.book..297N}. We did this by adopting the same boundaries in the spectral slope versus z-i colour space as in \citet{Demeo2013Icar..226..723D} for the S-complex. In particular, here, we used 6 $<$ spectral slope $<$ 25 \%/100~nm and -0.005 $<$ z-i  $<$ -0.265. We found that there are 21\,909 SSOs with \gaia spectral properties within this region, 9\,225 of which belong to known asteroid families according to the family identification of \citet{Nesvorny2015aste.book..297N}.  The age is known for several of these families, making it possible to study the variation of SSO spectral parameters as a function of age. 

However, because asteroids and asteroid families of the outer main belt tend to be observed with smaller phase angles than those of the inner main belt, it is important to correct the spectral slope and the z-i colour for potential dependence with the phase angle. To this aim, we fitted a straight line to the distribution of the spectral slope as a function of the average phase angle at which S-complex SSOs were observed by \gaia. We only used those SSOs that have an average reflectance spectrum with a S/N of higher than 50 and with an average phase angle of between 12 and 20 degrees. We found a value of the Pearson's correlation between phase angle and spectral slope of 0.30, which suggests a weak positive correlation between these two quantities. The fit of a straight line to the data resulted in an angular slope coefficient of 0.25 $\pm$ 0.04 \%/100~nm / degree (1$\sigma$). On the other hand, the z-i colour did not show a dependence with phase angle, as the Pearson correlation is -0.1; in addition, the angular coefficient of a straight line fit is only -0.04 $\pm$ 0.01 z-i mag / degree. 

\begin{figure*}[!ht]
   \centering
   \includegraphics[width=2\columnwidth]{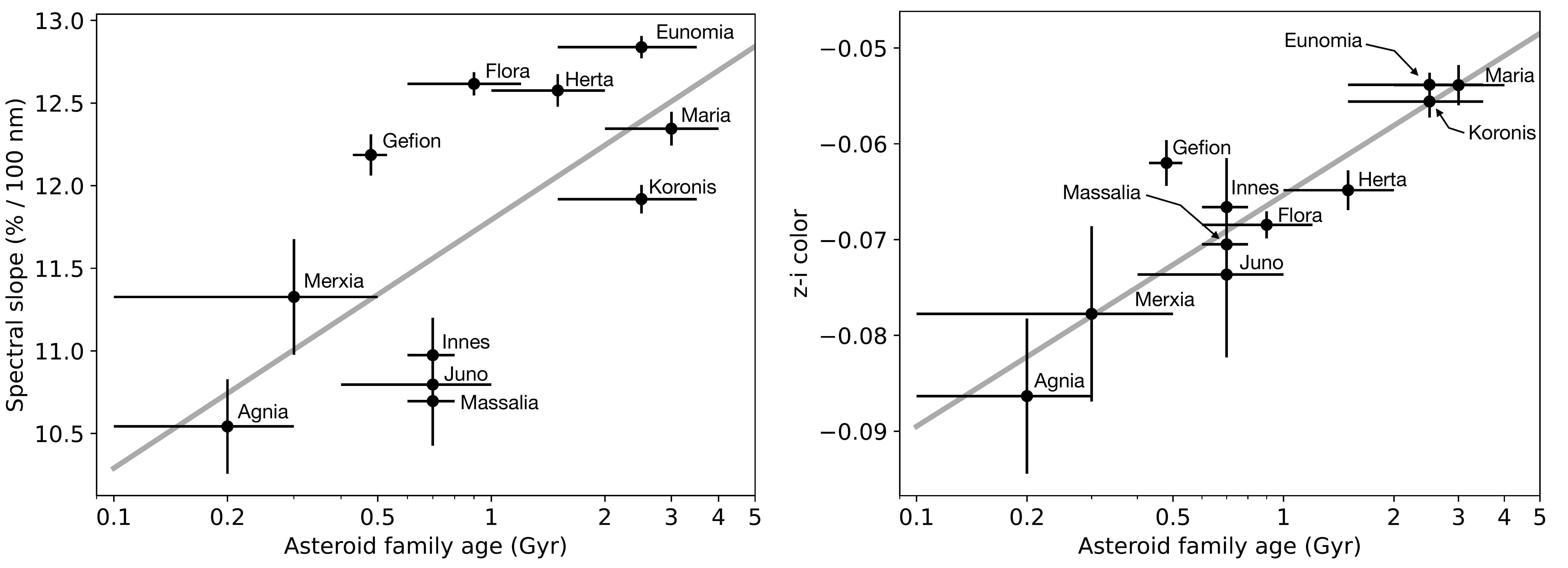}
   \caption{Left panel: \gdr{3} mean and standard deviation of the spectral slope of S-type families plotted against family age. Right panel: Mean and standard deviation of the z-i colour of S-type families plotted against family age. No correction for the composition was applied to either of the plots (see Section~\ref{sec:discussion}).}
\label{Fig:slope_ZI_S-families}%
\end{figure*}

For each member of the S-type asteroid families from the catalogue of \cite{Nesvorny2015aste.book..297N} that have known ages and a sufficient number ($>$10) of SSOs with \gdr{3} mean reflectance spectra (Table~\ref{Tab:StypeFamilies}), we calculated the spectral slope and z-i colour following the procedure described above. We then calculated the average phase angle of each family, $\alpha$, the mean spectral slope, $\xi$, and mean z-i, and the standard errors of these means. We calculated a corrected mean spectral slopes $\xi_C$ at a common phase angle of 20.6$^\circ$ by applying Eq.~\ref{Eq:phase_angle_corr}:
\begin{equation}
\xi_C = \xi -  0.25 (\alpha - 20.6),
\label{Eq:phase_angle_corr}
\end{equation}
where 0.25 is the angular slope coefficient determined before. This correction is needed to take into account  the spectral slope -- phase angle dependance of S-type asteroids. On the other hand, we did not correct the z-i colours for the phase angle. Results are shown in Fig.~\ref{Fig:slope_ZI_S-families}. The correlation between z-i colour and the logarithm of the family age is very robust, with a Pearson correlation coefficient of 0.91 (p value = 0.0001). The correlation between the spectral slope and the logarithm of the family age has a Pearson correlation coefficient of 0.67 (p value = 0.023).

\begin{table*}
\caption{S-type families for which we calculated an average spectral slope and z-i colour. Ages are from \cite{Broz2013A&A...551A.117B}. $N_{fam}$ and $N_{ref}$ are the number of family members and the number of asteroids used to calculate the average spectral slope and z-i colour.}
\label{Tab:StypeFamilies}
\begin{tabular}{l c c c c c c c c c}
\hline
Name & Age   & Uncertainty & Phase angle & $N_{fam}$ & $N_{ref}$ & Slope & Uncertainty & z-i & Uncertainty\\
     & (Gyr) & (Gyr) & ($^\circ$) & & & (\%/100 nm) & (\%/100 nm) & &  \\
\hline
\hline
Juno     & 0.7  & 0.3  & 20.8 &  1684 &   65 & 10.80 & 0.37 & -0.074 & 0.009\\
Flora    & 0.9  & 0.3  & 22.7 & 13786 & 1532 & 12.62 & 0.07 & -0.068 & 0.001\\
Eunomia  & 2.5  & 1.0  & 19.5 &  5670 & 1277 & 12.84 & 0.07 & -0.054 & 0.001\\
Massalia & 0.7  & 0.1  & 22.2 &  6424 &  113 & 10.70 & 0.23 & -0.071 & 0.006\\
Koronis  & 2.5  & 1.0  & 17.5 &  5949 &  596 & 11.92 & 0.09 & -0.055 & 0.002\\
Maria    & 3.0  & 1.0  & 19.5 &  2940 &  455 & 12.35 & 0.10 & -0.054 & 0.002\\
Merxia   & 0.3  & 0.2  & 18.9 &  1215 &   62 & 11.33 & 0.35 & -0.078 & 0.009\\
Herta    & 1.5  & 0.5  & 22.0 & 19073 &  927 & 12.58 & 0.10 & -0.065 & 0.002\\
Agnia    & 0.2  & 0.1  & 18.1 &  2125 &   55 & 10.54 & 0.29 & -0.086 & 0.008\\
Gefion   & 0.48 & 0.05 & 18.5 &  2547 &  381 & 12.19 & 0.13 & -0.062 & 0.002\\
Innes    & 0.7  & 0.1  & 20.1 &  1295 &  115 & 11.00 & 0.22 & -0.067 & 0.005\\
\hline
\end{tabular}
\end{table*}

\section{Discussion}\label{sec:discussion}
Compared to ground-based observations, the \gaia reflectance spectra extend, in general, to bluer wavelengths. For the brightest objects, the data point at 374~nm is typically usable. This is useful to distinguish the UV drop-off of otherwise featureless spectra, such as those of asteroids belonging to the B or F class in the Tholen taxonomy \citep[see][]{2020A&A...642A..80C}. However, for the faintest SSOs, which form the majority, the reflectance values at 374~nm (and in some cases also those at 418~nm) are affected by large errors and sometimes flagged as anomalous by our internal validation. This is due to the larger uncertainties in the calibrated BP spectra in the wavelength range $[350, 400]$ nm, possibly extending to 450 nm at this stage in the mission (Fig.~\ref{Fig:XP_Raw_Data}). A relatively small number of astronomical sources have significant flux in this wavelength range, making the calibration task more challenging.

The data points at the very red end of the spectrum, namely at wavelengths longer than $\sim$ 950~nm, are also to be used with care, in particular for the faintest SSOs. These points can often deviate from the expected reflectance more than their statistical error bars. This is in part due to the drop-off in the sensitivity of the RP at those wavelengths. 

A comparison between Gaia mean reflectances and those obtained from telescopes on the ground and in space for a selected sample of SSOs reveals a satisfactory match (sections 4.2 and 4.3), independently of the SSO spectral class and size. However, spectra can legitimately differ if an asteroid has spectral variability due, for example, to non-uniform composition and the comparison spectrum is not a rotational average. The publication of SSO epoch reflectance spectra in later releases will also allow asteroid spectral variability to be considered.

It is clear that  XP reflectances of SSOs are in general affected by larger relative uncertainties compared to astrometry and G-band photometry for the same objects. This is  due to a number of factors, such as (i) the dispersive nature of the spectroscopic data, which lowers the number of photo-electrons per pixel compared to data obtained from the astrometric field (the largest portion of the Gaia focal plane is covered with CCDs devoted to astrometric and unfiltered photometric measurements); (ii) the position of the BP and RP instruments at the end of the transit on the focal plane, which means that an object can move away from the window before reaching the BP and RP detectors; and (iii) the difficulty in applying the wavelength calibration for moving objects, the latter requiring a special procedure compared to Galactic stellar sources and extragalactic sources. However, the large number of SSOs for which \gaia produced reflectance spectra is invaluable for asteroid population studies, as showcased by this work. Given the distribution of the uncertainties in the spectral data (Fig.~\ref {fig:cu4sso_validationspectra_SNRdistribution}), we chose to limit the average spectral S/N to 13 for the \gdr{3}. More data are expected to be published with the \gdr{4} and later releases. 

The average difference between the z-i colours of the SMASS and \gaia is -0.08 (see Fig.~\ref{Fig:comp_slope_zi}). In order to understand if this difference could be due to the choice of solar analogues, we compared the solar analogue used for the production of the \gaia reflectances with those typically used in asteroid astronomical observations from the ground. Under the  assumption that SMASS and \gaia observed the same asteroid spectrum, it easy to demonstrate that 
\begin{equation}
(z-i_{SMASS}) - (z-i_{\gaia}) = \\ 
z-i\left( \frac{\text{Gaia solar analogue}} {\text{SMASS solar analogue}} \right).
\label{E:zidiff}
\end{equation}
We calculated the right-hand term of Eq.~\ref{E:zidiff} between the mean solar analogue spectrum used for the production of the \gaia reflectances and each mean spectrum of the following trusted solar analogues typically used in the literature: SA93-101, Hyades64, SA98-978, SA102-1081, BS4486, SA107-684, SA107-998, SA110-361, 16CygB, SA112-1333, and SA115-271 \citep{Marsset2020ApJS..247...73M}. We find that the z-i of these ratios has a mean value of -0.007 and standard deviation of 0.009. Given this evidence, it is therefore reasonable to assume that the difference in the z-i values between \gaia and the SMASS is unlikely to be due to the choice of solar analogues. Other effects that could explain this discrepancy are described below.

Raw BP and RP spectra are affected by energy redistribution arising from broad wings of the star image profiles. As a result, each sample in the spectrum, in addition to the \textit{local} photons, will contain \textit{alien} photons with different wavelengths. As the wings of stellar images in the focal plane are quite large, a considerable portion of the photons from the yellow-red part of the spectrum will fall into samples covering the UV and blue wavelength ranges and vice versa. The internal calibration of the BP and RP spectra does not remove this effect and therefore the spectra used to generate the SSO reflectance spectra in \textit{Gaia} DR3 will be affected by this contamination by \textit{alien} photons. This implies that when an SSO spectrum is divided by the spectrum of a solar analogue, the deviation from the theoretical reflectance spectrum is a function of the difference in shape between the SSO spectrum and that of the solar analogue. In Appendix~\ref{A:gaiaVSsmass}, we suggest formulas for correcting the \gaia spectral parameters (slope and z-i colours) using the SMASSII as a reference. The user should exercise particular caution when using those formulas, which only represent an average case, because of the large scatter in the data.

The average reflectances of SSOs with colours in the S-type range show a trend of increasing spectral slope with increasing phase angle (Fig.~\ref{fig:phaseSlopeStype}). This is expected, and has been noted from astronomical observations \citep{Binzel2019Icar..324...41B} and laboratory spectra of S-type analogue meteorites such as the ordinary chondrites \citep{2012Icar..220...36S,Brunetto2015aste.book..597B}. 
The value of 0.25 \%/100~nm / degree that we derive here appears to be larger than those obtained by previous studies: for instance, \citet{Nathues2010Icar..208..252N} found a value of 0.067\%/100~nm / degree with increasing phase angle in the range between 2 and 24 degrees, although this range is different from that observed by \gaia (Fig.~\ref{fig:phaseSlopeStype}). On the other hand, a laboratory study of ordinary chondrites found a minimal spectral slope increase with increasing phase angle when the latter is below 30 degrees \citep{2012Icar..220...36S}, which is the case for \gaia data in general. 
\cite{2018P&SS..157...82P} also conducted a study of the spectral variation with phase angle on near-Earth asteroids. They showed that, for low-albedo and spectroscopically featureless asteroids, such as those of the C-type and D-type, no phase reddening is visible. However, these authors also identified strong phase reddening (for very large phase angles) for olivine-dominated asteroids, such as A-types and Q-types. This last result is in accordance with the study of \citep{2012Icar..220...36S}, where it is shown that olivine-rich ordinary chondrites are those most influenced by phase reddening. 

For SSOs that have spectral slope and z-i colours consistent with the C-type asteroids \citep[$-5 <  slope < 6$ \% / 100 nm and $-0.20 < z-i < 0.17$ as in][]{Demeo2013Icar..226..723D}, we find that the distribution of the slopes as a function of the phase angle has a Pearson coefficient of 0.03, which indicates that there is no correlation between phase angle and spectral slope. This is consistent with other studies, such as that by \cite{2018Icar..302...10L}, who found no evidence of spectral reddening with increasing phase angle on B-, C-, and Ch-types.

Understanding the implications of the illumination and observation geometry, for example, the phase angle or the heliocentric  distance ---the latter affecting the surface temperature of the asteroids \citep{2012Icar..220...36S}---, is important for discriminating among the effects of asteroid ages and composition on the reflectance spectrum. The age determines the exposure of a surface to space weathering agents. It is expected that this exposure can affect the spectral parameters. Indeed, here we show that the slope of the average reflectance spectra for asteroids of the S-complex increases with increasing age (Fig.~\ref{Fig:slope_ZI_S-families}). This figure also shows an important dispersion of the slope versus age in S-type families with respect to a common trend. This dispersion is probably caused by the fact that we did not correct for the composition of the asteroid. Correction for the composition has been shown to be important in particular for the very young families \citep{Vernazza2008Natur.454..858V}. Nevertheless, even without correction for composition, the trend of increasing spectral slope with age is clearly visible in the \gaia data. In addition, the correlations between the age and the spectral slope and the age and the z-i colour for S-type families of Fig.~\ref{Fig:slope_ZI_S-families} are unlikely to be caused by composition differences between the families. This is because it has been shown that the closeness of the  match between spectra of S-complex families and those of ordinary chondrite meteorites of subtypes LL, L, and H is not correlated with the spectral slopes and z-i values of the families \citep{Vernazza2014ApJ...791..120V}.
The correlation between the z-i colour and the asteroid family age is even tighter and more clear in the \gaia data. Using the collisional family ages from the work of \cite{Spoto2015Icar..257..275S} increases the scatter in the data of Fig.~\ref{fig:phaseSlopeStype}, whereas the correlations between the spectral slope and the z-i colour with the family ages remains. \cite{Spoto2015Icar..257..275S}  assumed zero initial scatter in the semi-major axis of family members, which would be due to the post-collision velocity dispersion of the fragments. Their family ages are therefore driven only by subsequent evolution due to the Yarkovsky effect. Therefore, the provided ages should be considered as upper limits.  

\section{Conclusions}
The \gaia DR3 contains the largest space-based survey of reflectance spectra of Solar System small bodies observed in the visible range of the spectrum.

This article presents the procedures used to compute the mean reflectance spectrum of Solar System small bodies (SSOs) from calibrated epoch spectra and a model solar analogue spectrum. Some filtering based on the quality of the reflectance spectra was applied, leading to a catalogue of 60\,518 mean reflectances sampled in 16 bands. Each mean reflectance spectrum corresponds to one SSO. Hence, the catalogue contains 60\,518 different SSOs. An array of flags (one value per wavelength-band) for each spectrum is provided to indicate the quality of the measurement of the band.

In order to perform an external validation, we compared Gaia SSO reflectance spectra to literature spectra obtained by ground-based and space-borne telescopes. We show that this comparison reveals a good overall match. We conclude that the spectra presented in the DR3 are of good quality and show good consistency with the literature.  As expected, the \gdr{3} SSO spectra show a correlation between the spectral slope and the age of S-type asteroids belonging to collisional families. The \gdr{3} SSO spectra also reveal a correlation between the z-i colour and the age of S-type collisional families. 

The \gaia DR3 is a worthwhile survey that will help the community to better understand the SSO population. \gaia DR4 is expected to outperform the DR3 given that the former will contain a larger sample of asteroid reflectances and that these reflectances will be computed using twice as many data as were used for the DR3. The quality of the data will definitely increase and could allow us to refine the spectral bands selected for DR3. \gaia spectra will allow us to develop a \gaia spectral taxonomy that will be one of the products of the \gaia DR4.  This will make it possible to link asteroid mass and bulk density information (expected to be one of the products of the Gaia DR4) with the surface reflectance properties of asteroids. It will also allow us to derive separate relations from the magnitude--phase relations derived by Gaia for different classes of objects, with a possible dependence on the geometric albedo.




%
%
\bibliographystyle{aa} 
\bibliography{sso_bibliography,additional,refs,dpac}

\begin{thebibliography}{166}
\expandafter\ifx\csname natexlab\endcsname\relax\def\natexlab#1{#1}\fi

\bibitem[{{Abe} {et~al.}(2006){Abe}, {Takagi}, {Kitazato}, {Abe}, {Hiroi},
  {Vilas}, {Clark}, {Abell}, {Lederer}, {Jarvis}, {Nimura}, {Ueda}, \&
  {Fujiwara}}]{Abe2006Sci...312.1334A}
{Abe}, M., {Takagi}, Y., {Kitazato}, K., {et~al.} 2006, Science, 312, 1334

\bibitem[{{Ahn} {et~al.}(2012){Ahn}, {Alexandroff}, {Allende Prieto},
  {Anderson}, {Anderton}, {Andrews}, {Aubourg}, {Bailey}, {Balbinot}, {Barnes},
  \& et~al.}]{SDSS9}
{Ahn}, C.~P., {Alexandroff}, R., {Allende Prieto}, C., {et~al.} 2012, \apjs,
  203, 21

\bibitem[{{Albareti} {et~al.}(2017){Albareti}, {Allende Prieto}, {Almeida},
  {Anders}, {Anderson}, {Andrews}, {Arag{\'o}n-Salamanca},
  {Argudo-Fern{\'a}ndez}, {Armengaud}, {Aubourg}, {Avila-Reese}, {Badenes},
  {Bailey}, {Barbuy}, {Barger}, {Barrera-Ballesteros}, {Bartosz}, {Basu},
  {Bates}, {Battaglia}, {Baumgarten}, {Baur}, {Bautista}, {Beers}, {Belfiore},
  {Bershady}, {Bertran de Lis}, {Bird}, {Bizyaev}, {Blanc}, {Blanton},
  {Blomqvist}, {Bolton}, {Borissova}, {Bovy}, {Brand t}, {Brinkmann},
  {Brownstein}, {Bundy}, {Burtin}, {Busca}, {Orlando Camacho Chavez}, {Cano
  D{\'\i}az}, {Cappellari}, {Carrera}, {Chen}, {Cherinka}, {Cheung},
  {Chiappini}, {Chojnowski}, {Chuang}, {Chung}, {Cirolini}, {Clerc}, {Cohen},
  {Comerford}, {Comparat}, {Correa do Nascimento}, {Cousinou}, {Covey},
  {Crane}, {Croft}, {Cunha}, {Darling}, {Davidson}, {Dawson}, {Da Costa}, {Da
  Silva Ilha}, {Deconto Machado}, {Delubac}, {De Lee}, {De la Macorra}, {De la
  Torre}, {Diamond-Stanic}, {Donor}, {Downes}, {Drory}, {Du}, {Du Mas des
  Bourboux}, {Dwelly}, {Ebelke}, {Eigenbrot}, {Eisenstein}, {Elsworth},
  {Emsellem}, {Eracleous}, {Escoffier}, {Evans}, {Falc{\'o}n-Barroso}, {Fan},
  {Favole}, {Fernandez-Alvar}, {Fernand ez-Trincado}, {Feuillet}, {Fleming},
  {Font-Ribera}, {Freischlad}, {Frinchaboy}, {Fu}, {Gao}, {Garcia},
  {Garcia-Dias}, {Garcia-Hern{\'a}ndez}, {Garcia P{\'e}rez}, {Gaulme}, {Ge},
  {Geisler}, {Gillespie}, {Gil Marin}, {Girardi}, {Goddard}, {Gomez Maqueo
  Chew}, {Gonzalez-Perez}, {Grabowski}, {Green}, {Grier}, {Grier}, {Guo},
  {Guy}, {Hagen}, {Hall}, {Harding}, {Harley}, {Hasselquist}, {Hawley},
  {Hayes}, {Hearty}, {Hekker}, {Hernandez Toledo}, {Ho}, {Hogg},
  {Holley-Bockelmann}, {Holtzman}, {Holzer}, {Hu}, {Huber}, {Hutchinson},
  {Hwang}, {Ibarra-Medel}, {Ivans}, {Ivory}, {Jaehnig}, {Jensen}, {Johnson},
  {Jones}, {Jullo}, {Kallinger}, {Kinemuchi}, {Kirkby}, {Klaene}, {Kneib},
  {Kollmeier}, {Lacerna}, {Lane}, {Lang}, {Laurent}, {Law}, {Leauthaud}, {Le
  Goff}, {Li}, {Li}, {Li}, {Li}, {Liang}, {Liang}, {Lima}, {Lin}, {Lin}, {Lin},
  {Liu}, {Long}, {Lucatello}, {MacDonald}, {MacLeod}, {Mackereth}, {Mahadevan},
  {Maia}, {Maiolino}, {Majewski}, {Malanushenko}, {Malanushenko}, {Mallmann},
  {Manchado}, {Maraston}, {Marques-Chaves}, {Martinez Valpuesta}, {Masters},
  {Mathur}, {McGreer}, {Merloni}, {Merrifield}, {M{\'e}sz{\'a}ros}, {Meza},
  {Miglio}, {Minchev}, {Molaverdikhani}, {Montero-Dorta}, {Mosser}, {Muna},
  {Myers}, {Nair}, {Nandra}, {Ness}, {Newman}, {Nichol}, {Nidever},
  {Nitschelm}, {O'Connell}, {Oravetz}, {Oravetz}, {Pace}, {Padilla},
  {Palanque-Delabrouille}, {Pan}, {Parejko}, {Paris}, {Park}, {Peacock},
  {Peirani}, {Pellejero-Ibanez}, {Penny}, {Percival}, {Percival},
  {Perez-Fournon}, {Petitjean}, {Pieri}, {Pinsonneault}, {Pisani}, {Prada},
  {Prakash}, {Price-Jones}, {Raddick}, {Rahman}, {Raichoor}, {Barboza Rembold},
  {Reyna}, {Rich}, {Richstein}, {Ridl}, {Riffel}, {Riffel}, {Rix}, {Robin},
  {Rockosi}, {Rodr{\'\i}guez-Torres}, {Rodrigues}, {Roe}, {Roman Lopes},
  {Rom{\'a}n-Z{\'u}{\~n}iga}, {Ross}, {Rossi}, {Ruan}, {Ruggeri}, {Runnoe},
  {Salazar-Albornoz}, {Salvato}, {Sanchez}, {Sanchez}, {Sanchez-Gallego},
  {Santiago}, {Schiavon}, {Schimoia}, {Schlafly}, {Schlegel}, {Schneider},
  {Sch{\"o}nrich}, {Schultheis}, {Schwope}, {Seo}, {Serenelli}, {Sesar},
  {Shao}, {Shetrone}, {Shull}, {Silva Aguirre}, {Skrutskie}, {Slosar}, {Smith},
  {Smith}, {Sobeck}, {Somers}, {Souto}, {Stark}, {Stassun}, {Steinmetz},
  {Stello}, {Storchi Bergmann}, {Strauss}, {Streblyanska}, {Stringfellow},
  {Suarez}, {Sun}, {Taghizadeh-Popp}, {Tang}, {Tao}, {Tayar}, {Tembe},
  {Thomas}, {Tinker}, {Tojeiro}, {Tremonti}, {Troup}, {Trump}, {Unda-Sanzana},
  {Valenzuela}, {Van den Bosch}, {Vargas-Maga{\~n}a}, {Vazquez}, {Villanova},
  {Vivek}, {Vogt}, {Wake}, {Walterbos}, {Wang}, {Wang}, {Weaver}, {Weijmans},
  {Weinberg}, {Westfall}, {Whelan}, {Wilcots}, {Wild}, {Williams}, {Wilson},
  {Wood-Vasey}, {Wylezalek}, {Xiao}, {Yan}, {Yang}, {Ybarra}, {Yeche}, {Yuan},
  {Zakamska}, {Zamora}, {Zasowski}, {Zhang}, {Zhao}, {Zhao}, {Zheng}, {Zheng},
  {Zhou}, {Zhu}, {Zinn}, \& {Zou}}]{2017ApJS..233...25A}
{Albareti}, F.~D., {Allende Prieto}, C., {Almeida}, A., {et~al.} 2017, \apjs,
  233, 25

\bibitem[{{Astropy Collaboration} {et~al.}(2018){Astropy Collaboration},
  {Price-Whelan}, {Sip{\H o}cz}, {G{\"u}nther}, {Lim}, {Crawford}, {Conseil},
  {Shupe}, {Craig}, {Dencheva}, {Ginsburg}, {VanderPlas}, {Bradley},
  {P{\'e}rez-Su{\'a}rez}, {de Val-Borro}, {Aldcroft}, {Cruz}, {Robitaille},
  {Tollerud}, {Ardelean}, {Babej}, {Bach}, {Bachetti}, {Bakanov}, {Bamford},
  {Barentsen}, {Barmby}, {Baumbach}, {Berry}, {Biscani}, {Boquien}, {Bostroem},
  {Bouma}, {Brammer}, {Bray}, {Breytenbach}, {Buddelmeijer}, {Burke},
  {Calderone}, {Cano Rodr{\'{\i}}guez}, {Cara}, {Cardoso}, {Cheedella},
  {Copin}, {Corrales}, {Crichton}, {D'Avella}, {Deil}, {Depagne}, {Dietrich},
  {Donath}, {Droettboom}, {Earl}, {Erben}, {Fabbro}, {Ferreira}, {Finethy},
  {Fox}, {Garrison}, {Gibbons}, {Goldstein}, {Gommers}, {Greco}, {Greenfield},
  {Groener}, {Grollier}, {Hagen}, {Hirst}, {Homeier}, {Horton}, {Hosseinzadeh},
  {Hu}, {Hunkeler}, {Ivezi{\'c}}, {Jain}, {Jenness}, {Kanarek}, {Kendrew},
  {Kern}, {Kerzendorf}, {Khvalko}, {King}, {Kirkby}, {Kulkarni}, {Kumar},
  {Lee}, {Lenz}, {Littlefair}, {Ma}, {Macleod}, {Mastropietro}, {McCully},
  {Montagnac}, {Morris}, {Mueller}, {Mumford}, {Muna}, {Murphy}, {Nelson},
  {Nguyen}, {Ninan}, {N{\"o}the}, {Ogaz}, {Oh}, {Parejko}, {Parley}, {Pascual},
  {Patil}, {Patil}, {Plunkett}, {Prochaska}, {Rastogi}, {Reddy Janga},
  {Sabater}, {Sakurikar}, {Seifert}, {Sherbert}, {Sherwood-Taylor}, {Shih},
  {Sick}, {Silbiger}, {Singanamalla}, {Singer}, {Sladen}, {Sooley},
  {Sornarajah}, {Streicher}, {Teuben}, {Thomas}, {Tremblay}, {Turner},
  {Terr{\'o}n}, {van Kerkwijk}, {de la Vega}, {Watkins}, {Weaver}, {Whitmore},
  {Woillez}, {Zabalza}, \& {Astropy Contributors}}]{2018AJ....156..123A}
{Astropy Collaboration}, {Price-Whelan}, A., {Sip{\H o}cz}, B.~M., {et~al.}
  2018, \aj, 156, 123

\bibitem[{{Barucci} {et~al.}(2005){Barucci}, {Fulchignoni}, {Fornasier},
  {Dotto}, {Vernazza}, {Birlan}, {Binzel}, {Carvano}, {Merlin}, {Barbieri}, \&
  {Belskaya}}]{2005A&A...430..313B}
{Barucci}, M.~A., {Fulchignoni}, M., {Fornasier}, S., {et~al.} 2005, \aap, 430,
  313

\bibitem[{{Bell} {et~al.}(1988){Bell}, {Owensby}, {Hawke}, \&
  {Gaffey}}]{Bell1988LPI....19...57B}
{Bell}, J.~F., {Owensby}, P.~D., {Hawke}, B.~R., \& {Gaffey}, M.~J. 1988, in
  Lunar and Planetary Science Conference, Vol.~19, Lunar and Planetary Science
  Conference, 57

\bibitem[{{Belton} {et~al.}(1995){Belton}, {Chapman}, {Thomas}, {Davies},
  {Greenberg}, {Klaasen}, {Byrnes}, {D'Amario}, {Synnott}, {Johnson}, {McEwen},
  {Merline}, {Davis}, {Petit}, {Storrs}, {Veverka}, \&
  {Zellner}}]{Belton1995Natur.374..785B}
{Belton}, M.~J.~S., {Chapman}, C.~R., {Thomas}, P.~C., {et~al.} 1995, \nat,
  374, 785

\bibitem[{{Belton} {et~al.}(1992){Belton}, {Veverka}, {Thomas}, {Helfenstein},
  {Simonelli}, {Chapman}, {Davies}, {Greeley}, {Greenberg}, {Head}, {Murchie},
  {Klaasen}, {Johnson}, {McEwen}, {Morrison}, {Neukum}, {Fanale}, {Anger},
  {Carr}, \& {Pilcher}}]{Belton1992Sci...257.1647B}
{Belton}, M.~J.~S., {Veverka}, J., {Thomas}, P., {et~al.} 1992, Science, 257,
  1647

\bibitem[{{Binzel} {et~al.}(2019){Binzel}, {DeMeo}, {Turtelboom}, {Bus},
  {Tokunaga}, {Burbine}, {Lantz}, {Polishook}, {Carry}, {Morbidelli}, {Birlan},
  {Vernazza}, {Burt}, {Moskovitz}, {Slivan}, {Thomas}, {Rivkin}, {Hicks},
  {Dunn}, {Reddy}, {Sanchez}, {Granvik}, \&
  {Kohout}}]{Binzel2019Icar..324...41B}
{Binzel}, R.~P., {DeMeo}, F.~E., {Turtelboom}, E.~V., {et~al.} 2019, \icarus,
  324, 41

\bibitem[{{Boch} \& {Fernique}(2014)}]{2014ASPC..485..277B}
{Boch}, T. \& {Fernique}, P. 2014, in Astronomical Society of the Pacific
  Conference Series, Vol. 485, Astronomical Data Analysis Software and Systems
  XXIII, ed. N.~{Manset} \& P.~{Forshay}, 277

\bibitem[{{Bonnarel} {et~al.}(2000){Bonnarel}, {Fernique}, {Bienaym{\'e}},
  {Egret}, {Genova}, {Louys}, {Ochsenbein}, {Wenger}, \&
  {Bartlett}}]{2000A&AS..143...33B}
{Bonnarel}, F., {Fernique}, P., {Bienaym{\'e}}, O., {et~al.} 2000, \aaps, 143,
  33

\bibitem[{{Bottke} {et~al.}(2000){Bottke}, {Rubincam}, \&
  {Burns}}]{Bottke2000Icar..145..301B}
{Bottke}, William~F., J., {Rubincam}, D.~P., \& {Burns}, J.~A. 2000, \icarus,
  145, 301

\bibitem[{{Bottke} {et~al.}(2006){Bottke}, {Vokrouhlick{\'y}}, {Rubincam}, \&
  {Nesvorn{\'y}}}]{Bottke2006AREPS..34..157B}
{Bottke}, William~F., J., {Vokrouhlick{\'y}}, D., {Rubincam}, D.~P., \&
  {Nesvorn{\'y}}, D. 2006, Annual Review of Earth and Planetary Sciences, 34,
  157

\bibitem[{{Bottke} {et~al.}(2015){Bottke}, {Bro{\v{z}}}, {O'Brien}, {Campo
  Bagatin}, {Morbidelli}, \& {Marchi}}]{Bottke2015aste.book..701B}
{Bottke}, W.~F., {Bro{\v{z}}}, M., {O'Brien}, D.~P., {et~al.} 2015, in
  Asteroids IV, 701--724

\bibitem[{{Breddels} \& {Veljanoski}(2018)}]{2018A&A...618A..13B}
{Breddels}, M.~A. \& {Veljanoski}, J. 2018, \aap, 618, A13

\bibitem[{{Britt} {et~al.}(1992){Britt}, {Bell}, {Haack}, \&
  {Scott}}]{1992Metic..27Q.207B}
{Britt}, D.~T., {Bell}, J.~F., {Haack}, H., \& {Scott}, E.~R.~D. 1992,
  Meteoritics, 27, 207

\bibitem[{{Bro{\v{z}}} {et~al.}(2013){Bro{\v{z}}}, {Morbidelli}, {Bottke},
  {Rozehnal}, {Vokrouhlick{\'y}}, \& {Nesvorn{\'y}}}]{Broz2013A&A...551A.117B}
{Bro{\v{z}}}, M., {Morbidelli}, A., {Bottke}, W.~F., {et~al.} 2013, \aap, 551,
  A117

\bibitem[{Brunetto {et~al.}(2015)Brunetto, Loeffler, Nesvorn{\'{y}}, Sasaki, \&
  Strazzulla}]{Brunetto2015aste.book..597B}
Brunetto, R., Loeffler, M.~J., Nesvorn{\'{y}}, D., Sasaki, S., \& Strazzulla,
  G. 2015, in Asteroids IV (P. Michel, et al. eds.) University of Arizona
  Press, Tucson., 597

\bibitem[{{Brunetto} {et~al.}(2006){Brunetto}, {Vernazza}, {Marchi}, {Birlan},
  {Fulchignoni}, {Orofino}, \& {Strazzulla}}]{Brunetto2006Icar..184..327B}
{Brunetto}, R., {Vernazza}, P., {Marchi}, S., {et~al.} 2006, \icarus, 184, 327

\bibitem[{{Burbine} \& {Binzel}(2002)}]{Burbine2002Icar..159..468B}
{Burbine}, T.~H. \& {Binzel}, R.~P. 2002, \icarus, 159, 468

\bibitem[{Bus \& Binzel(2002{\natexlab{a}})}]{Bus2002Icar..158..146B}
Bus, S.~J. \& Binzel, R.~P. 2002{\natexlab{a}}, Icarus, 158, 146

\bibitem[{Bus \& Binzel(2002{\natexlab{b}})}]{Bus2002Icar..158..106B}
Bus, S.~J. \& Binzel, R.~P. 2002{\natexlab{b}}, Icarus, 158, 106

\bibitem[{Bus {et~al.}(2002)Bus, Vilas, \& Barucci}]{Bus2002aste.conf..169B}
Bus, S.~J., Vilas, F., \& Barucci, M.~A. 2002, Asteroids III, W. F. Bottke Jr.,
  A. Cellino, P. Paolicchi, and R. P. Binzel (eds), University of Arizona
  Press, Tucson., 169

\bibitem[{{Carrasco} {et~al.}(2021){Carrasco}, {Weiler}, {Jordi}, {Fabricius},
  {De Angeli}, {Evans}, {van Leeuwen}, {Riello}, \&
  {Montegriffo}}]{2021A&A...652A..86C}
{Carrasco}, J.~M., {Weiler}, M., {Jordi}, C., {et~al.} 2021, \aap, 652, A86

\bibitem[{{Carry}(2018)}]{Carry2018A&A...609A.113C}
{Carry}, B. 2018, \aap, 609, A113

\bibitem[{{Carvano} \& {Davalos}(2015)}]{2015A&A...580A..98C}
{Carvano}, J.~M. \& {Davalos}, J.~A.~G. 2015, \aap, 580, A98

\bibitem[{{Cellino} {et~al.}(2020){Cellino}, {Bendjoya}, {Delbo'}, {Galluccio},
  {Gayon-Markt}, {Tanga}, \& {Tedesco}}]{2020A&A...642A..80C}
{Cellino}, A., {Bendjoya}, P., {Delbo'}, M., {et~al.} 2020, \aap, 642, A80

\bibitem[{{Chambers} {et~al.}(2016){Chambers}, {Magnier}, {Metcalfe},
  {Flewelling}, {Huber}, {Waters}, {Denneau}, {Draper}, {Farrow}, {Finkbeiner},
  {Holmberg}, {Koppenhoefer}, {Price}, {Saglia}, {Schlafly}, {Smartt},
  {Sweeney}, {Wainscoat}, {Burgett}, {Grav}, {Heasley}, {Hodapp}, {Jedicke},
  {Kaiser}, {Kudritzki}, {Luppino}, {Lupton}, {Monet}, {Morgan}, {Onaka},
  {Stubbs}, {Tonry}, {Banados}, {Bell}, {Bender}, {Bernard}, {Botticella},
  {Casertano}, {Chastel}, {Chen}, {Chen}, {Cole}, {Deacon}, {Frenk},
  {Fitzsimmons}, {Gezari}, {Goessl}, {Goggia}, {Goldman}, {Grebel}, {Hambly},
  {Hasinger}, {Heavens}, {Heckman}, {Henderson}, {Henning}, {Holman}, {Hopp},
  {Ip}, {Isani}, {Keyes}, {Koekemoer}, {Kotak}, {Long}, {Lucey}, {Liu},
  {Martin}, {McLean}, {Morganson}, {Murphy}, {Nieto-Santisteban}, {Norberg},
  {Peacock}, {Pier}, {Postman}, {Primak}, {Rae}, {Rest}, {Riess}, {Riffeser},
  {Rix}, {Roser}, {Schilbach}, {Schultz}, {Scolnic}, {Szalay}, {Seitz},
  {Shiao}, {Small}, {Smith}, {Soderblom}, {Taylor}, {Thakar}, {Thiel},
  {Thilker}, {Urata}, {Valenti}, {Walter}, {Watters}, {Werner}, {White},
  {Wood-Vasey}, \& {Wyse}}]{panstarrs1}
{Chambers}, K.~C., {Magnier}, E.~A., {Metcalfe}, N., {et~al.} 2016, ArXiv
  e-prints [\eprint[arXiv]{1612.05560}]

\bibitem[{{Chapman} {et~al.}(2005){Chapman}, {Gaffey}, \&
  {McFadden}}]{Chapman2005PDSS...27.....C}
{Chapman}, C.~R., {Gaffey}, M., \& {McFadden}, L. 2005, NASA Planetary Data
  System, EAR

\bibitem[{{Clark} {et~al.}(1993){Clark}, {Bell}, {Fanale}, \&
  {Lucey}}]{Clark1993LPI....24..299C}
{Clark}, B.~E., {Bell}, J.~F., {Fanale}, F.~P., \& {Lucey}, P.~G. 1993, in
  Lunar and Planetary Science Conference, Lunar and Planetary Science
  Conference, 299

\bibitem[{{Clark} {et~al.}(2002){Clark}, {Helfenstein}, {Bell}, {Peterson},
  {Veverka}, {Izenberg}, {Domingue}, {Wellnitz}, \&
  {McFadden}}]{2002Icar..155..189C}
{Clark}, B.~E., {Helfenstein}, P., {Bell}, J.~F., {et~al.} 2002, \icarus, 155,
  189

\bibitem[{{Clark} {et~al.}(1999){Clark}, {Veverka}, {Helfenstein}, {Thomas},
  {Bell}, {Harch}, {Robinson}, {Murchie}, {McFadden}, \&
  {Chapman}}]{1999Icar..140...53C}
{Clark}, B.~E., {Veverka}, J., {Helfenstein}, P., {et~al.} 1999, \icarus, 140,
  53

\bibitem[{{Davis} {et~al.}(1979){Davis}, {Chapman}, {Greenberg},
  {Weidenschilling}, \& {Harris}}]{Davis1979aste.book..528D}
{Davis}, D.~R., {Chapman}, C.~R., {Greenberg}, R., {Weidenschilling}, S.~J., \&
  {Harris}, A.~W. 1979, in Asteroids, ed. T.~{Gehrels} \& M.~S. {Matthews},
  528--557

\bibitem[{{Davis} {et~al.}(1985){Davis}, {Chapman}, {Weidenschilling}, \&
  {Greenberg}}]{Davis1985Icar...62...30D}
{Davis}, D.~R., {Chapman}, C.~R., {Weidenschilling}, S.~J., \& {Greenberg}, R.
  1985, \icarus, 62, 30

\bibitem[{{Davis} {et~al.}(2002){Davis}, {Durda}, {Marzari}, {Campo Bagatin},
  \& {Gil-Hutton}}]{Davis2002aste.book..545D}
{Davis}, D.~R., {Durda}, D.~D., {Marzari}, F., {Campo Bagatin}, A., \&
  {Gil-Hutton}, R. 2002, in Asteroids III, 545--558

\bibitem[{{De Angeli et al.}(2022)}]{DPACP-118}
{De Angeli et al.} 2022, \aap\ in prep.

\bibitem[{{de Leon} {et~al.}(2018){de Leon}, {Pinilla-Alonso}, {Campins},
  {Licandro}, {Morate}, {Lorenzi}, {De Pr{\'a}}, \&
  {Rizos}}]{deLeon2018DPS....5031005D}
{de Leon}, J., {Pinilla-Alonso}, N., {Campins}, H., {et~al.} 2018, in
  AAS/Division for Planetary Sciences Meeting Abstracts, Vol.~50, AAS/Division
  for Planetary Sciences Meeting Abstracts \#50, 310.05

\bibitem[{Delbo {et~al.}(2014)Delbo, Libourel, Wilkerson, Murdoch, Michel,
  Ramesh, Ganino, Verati, \& Marchi}]{Delbo2014Natur.508..233D}
Delbo, M., Libourel, G., Wilkerson, J., {et~al.} 2014, Nature, 508, 233

\bibitem[{DeMeo {et~al.}(2015)DeMeo, Alexander, Walsh, Chapman, \&
  Binzel}]{DeMeo2015aste.book...13D}
DeMeo, F.~E., Alexander, C. M.~O., Walsh, K.~J., Chapman, C.~R., \& Binzel,
  R.~P. 2015, in Asteroids IV (P. Michel, et al. eds.) University of Arizona
  Press, Tucson., 13

\bibitem[{DeMeo {et~al.}(2009)DeMeo, Binzel, Slivan, \&
  Bus}]{DeMeo2009Icar..202..160D}
DeMeo, F.~E., Binzel, R.~P., Slivan, S.~M., \& Bus, S.~J. 2009, Icarus, 202,
  160

\bibitem[{{DeMeo} {et~al.}(2022){DeMeo}, {Burt}, {Marsset}, {Polishook},
  {Burbine}, {Carry}, {Binzel}, {Vernazza}, {Reddy}, {Tang}, {Thomas},
  {Rivkin}, {Moskovitz}, {Slivan}, \& {Bus}}]{DeMeo2022Icar..38014971D}
{DeMeo}, F.~E., {Burt}, B.~J., {Marsset}, M., {et~al.} 2022, \icarus, 380,
  114971

\bibitem[{DeMeo \& Carry(2013)}]{Demeo2013Icar..226..723D}
DeMeo, F.~E. \& Carry, B. 2013, Icarus, 226, 723

\bibitem[{DeMeo \& Carry(2014)}]{DeMeo2014Natur.505..629D}
DeMeo, F.~E. \& Carry, B. 2014, Nature, 505, 629

\bibitem[{{Devog{\`e}le} {et~al.}(2019){Devog{\`e}le}, {Moskovitz}, {Thirouin},
  {Gustaffson}, {Magnuson}, {Thomas}, {Willman}, {Christensen}, {Person},
  {Binzel}, {Polishook}, {DeMeo}, {Hinkle}, {Trilling}, {Mommert}, {Burt}, \&
  {Skiff}}]{Devogele2019AJ....158..196D}
{Devog{\`e}le}, M., {Moskovitz}, N., {Thirouin}, A., {et~al.} 2019, \aj, 158,
  196

\bibitem[{{Elkins-Tanton} {et~al.}(2016){Elkins-Tanton}, {Asphaug}, {Bell},
  {Bercovici}, {Bills}, {Binzel}, {Bottke}, {Goldsten}, {Jaumann}, {Jun},
  {Lawrence}, {Marchi}, {Oh}, {Park}, {Peplowski}, {Polanskey}, {Prettyman},
  {Raymond}, {Russell}, {Weiss}, {Wenkert}, {Wieczorek}, \&
  {Zuber}}]{Elkins-Tanton2016LPI....47.1631E}
{Elkins-Tanton}, L.~T., {Asphaug}, E., {Bell}, J., {et~al.} 2016, in 47th
  Annual Lunar and Planetary Science Conference, Lunar and Planetary Science
  Conference, 1631

\bibitem[{Erasmus {et~al.}(2019)Erasmus, McNeill, Mommert, Trilling,
  Sickafoose, \& Paterson}]{Erasmus2019ApJS..242...15E}
Erasmus, N., McNeill, A., Mommert, M., {et~al.} 2019, The Astrophysical Journal
  Supplement Series, 242, 15

\bibitem[{{Fabricius} {et~al.}(2002){Fabricius}, {H{\o}g}, {Makarov}, {Mason},
  {Wycoff}, \& {Urban}}]{2002A&A...384..180F}
{Fabricius}, C., {H{\o}g}, E., {Makarov}, V.~V., {et~al.} 2002, \aap, 384, 180

\bibitem[{{Farinella} {et~al.}(1992){Farinella}, {Davis}, {Paolicchi},
  {Cellino}, \& {Zappala}}]{Farinella1992A&A...253..604F}
{Farinella}, P., {Davis}, D.~R., {Paolicchi}, P., {Cellino}, A., \& {Zappala},
  V. 1992, \aap, 253, 604

\bibitem[{{Farinella} {et~al.}(1981){Farinella}, {Paolicchi}, {Tedesco}, \&
  {Zappala}}]{Farinella1981Icar...46..114F}
{Farinella}, P., {Paolicchi}, P., {Tedesco}, E.~F., \& {Zappala}, V. 1981,
  \icarus, 46, 114

\bibitem[{{Flewelling} {et~al.}(2020){Flewelling}, {Magnier}, {Chambers},
  {Heasley}, {Holmberg}, {Huber}, {Sweeney}, {Waters}, {Calamida}, {Casertano},
  {Chen}, {Farrow}, {Hasinger}, {Henderson}, {Long}, {Metcalfe}, {Narayan},
  {Nieto-Santisteban}, {Norberg}, {Rest}, {Saglia}, {Szalay}, {Thakar},
  {Tonry}, {Valenti}, {Werner}, {White}, {Denneau}, {Draper}, {Hodapp},
  {Jedicke}, {Kaiser}, {Kudritzki}, {Price}, {Wainscoat}, {Chastel}, {McLean},
  {Postman}, \& {Shiao}}]{panstarrs1f}
{Flewelling}, H.~A., {Magnier}, E.~A., {Chambers}, K.~C., {et~al.} 2020, \apjs,
  251, 7

\bibitem[{Fornasier {et~al.}(2007)Fornasier, Dotto, Hainaut, Marzari,
  Boehnhardt, {De Luise}, \& Barucci}]{FORNASIER2007}
Fornasier, S., Dotto, E., Hainaut, O., {et~al.} 2007, Icarus, 190, 622, deep
  Impact Mission to Comet 9P/Tempel 1, Part 2

\bibitem[{{Fornasier} {et~al.}(2020){Fornasier}, {Hasselmann}, {Deshapriya},
  {Barucci}, {Clark}, {Praet}, {Hamilton}, {Simon}, {Li}, {Cloutis}, {Merlin},
  {Zou}, \& {Lauretta}}]{2020A&A...644A.142F}
{Fornasier}, S., {Hasselmann}, P.~H., {Deshapriya}, J.~D.~P., {et~al.} 2020,
  \aap, 644, A142

\bibitem[{{Gaffey}(2010)}]{2010Icar..209..564G}
{Gaffey}, M.~J. 2010, \icarus, 209, 564

\bibitem[{{Gaia Collaboration} {et~al.}(2016){Gaia Collaboration}, {Prusti},
  {de Bruijne}, {Brown}, {Vallenari}, {Babusiaux}, {Bailer-Jones}, {Bastian},
  {Biermann}, {Evans}, {Eyer}, {Jansen}, {Jordi}, {Klioner}, {Lammers},
  {Lindegren}, {Luri}, {Mignard}, {Milligan}, {Panem}, {Poinsignon},
  {Pourbaix}, {Randich}, {Sarri}, {Sartoretti}, {Siddiqui}, {Soubiran},
  {Valette}, {van Leeuwen}, {Walton}, {Aerts}, {Arenou}, {Cropper}, {Drimmel},
  {H{\o}g}, {Katz}, {Lattanzi}, {O'Mullane}, {Grebel}, {Holland}, {Huc},
  {Passot}, {Bramante}, {Cacciari}, {Casta{\~n}eda}, {Chaoul}, {Cheek}, {De
  Angeli}, {Fabricius}, {Guerra}, {Hern{\'a}ndez}, {Jean-Antoine-Piccolo},
  {Masana}, {Messineo}, {Mowlavi}, {Nienartowicz}, {Ord{\'o}{\~n}ez-Blanco},
  {Panuzzo}, {Portell}, {Richards}, {Riello}, {Seabroke}, {Tanga},
  {Th{\'e}venin}, {Torra}, {Els}, {Gracia-Abril}, {Comoretto},
  {Garcia-Reinaldos}, {Lock}, {Mercier}, {Altmann}, {Andrae}, {Astraatmadja},
  {Bellas-Velidis}, {Benson}, {Berthier}, {Blomme}, {Busso}, {Carry},
  {Cellino}, {Clementini}, {Cowell}, {Creevey}, {Cuypers}, {Davidson}, {De
  Ridder}, {de Torres}, {Delchambre}, {Dell'Oro}, {Ducourant}, {Fr{\'e}mat},
  {Garc{\'\i}a-Torres}, {Gosset}, {Halbwachs}, {Hambly}, {Harrison}, {Hauser},
  {Hestroffer}, {Hodgkin}, {Huckle}, {Hutton}, {Jasniewicz}, {Jordan},
  {Kontizas}, {Korn}, {Lanzafame}, {Manteiga}, {Moitinho}, {Muinonen},
  {Osinde}, {Pancino}, {Pauwels}, {Petit}, {Recio-Blanco}, {Robin}, {Sarro},
  {Siopis}, {Smith}, {Smith}, {Sozzetti}, {Thuillot}, {van Reeven}, {Viala},
  {Abbas}, {Abreu Aramburu}, {Accart}, {Aguado}, {Allan}, {Allasia},
  {Altavilla}, {{\'A}lvarez}, {Alves}, {Anderson}, {Andrei}, {Anglada Varela},
  {Antiche}, {Antoja}, {Ant{\'o}n}, {Arcay}, {Atzei}, {Ayache}, {Bach},
  {Baker}, {Balaguer-N{\'u}{\~n}ez}, {Barache}, {Barata}, {Barbier}, {Barblan},
  {Baroni}, {Barrado y Navascu{\'e}s}, {Barros}, {Barstow}, {Becciani},
  {Bellazzini}, {Bellei}, {Bello Garc{\'\i}a}, {Belokurov}, {Bendjoya},
  {Berihuete}, {Bianchi}, {Bienaym{\'e}}, {Billebaud}, {Blagorodnova},
  {Blanco-Cuaresma}, {Boch}, {Bombrun}, {Borrachero}, {Bouquillon}, {Bourda},
  {Bouy}, {Bragaglia}, {Breddels}, {Brouillet}, {Br{\"u}semeister},
  {Bucciarelli}, {Budnik}, {Burgess}, {Burgon}, {Burlacu}, {Busonero}, {Buzzi},
  {Caffau}, {Cambras}, {Campbell}, {Cancelliere}, {Cantat-Gaudin}, {Carlucci},
  {Carrasco}, {Castellani}, {Charlot}, {Charnas}, {Charvet}, {Chassat},
  {Chiavassa}, {Clotet}, {Cocozza}, {Collins}, {Collins}, {Costigan}, {Crifo},
  {Cross}, {Crosta}, {Crowley}, {Dafonte}, {Damerdji}, {Dapergolas}, {David},
  {David}, {De Cat}, {de Felice}, {de Laverny}, {De Luise}, {De March}, {de
  Martino}, {de Souza}, {Debosscher}, {del Pozo}, {Delbo}, {Delgado},
  {Delgado}, {di Marco}, {Di Matteo}, {Diakite}, {Distefano}, {Dolding}, {Dos
  Anjos}, {Drazinos}, {Dur{\'a}n}, {Dzigan}, {Ecale}, {Edvardsson}, {Enke},
  {Erdmann}, {Escolar}, {Espina}, {Evans}, {Eynard Bontemps}, {Fabre},
  {Fabrizio}, {Faigler}, {Falc{\~a}o}, {Farr{\`a}s Casas}, {Faye}, {Federici},
  {Fedorets}, {Fern{\'a}ndez-Hern{\'a}ndez}, {Fernique}, {Fienga}, {Figueras},
  {Filippi}, {Findeisen}, {Fonti}, {Fouesneau}, {Fraile}, {Fraser}, {Fuchs},
  {Furnell}, {Gai}, {Galleti}, {Galluccio}, {Garabato}, {Garc{\'\i}a-Sedano},
  {Gar{\'e}}, {Garofalo}, {Garralda}, {Gavras}, {Gerssen}, {Geyer}, {Gilmore},
  {Girona}, {Giuffrida}, {Gomes}, {Gonz{\'a}lez-Marcos},
  {Gonz{\'a}lez-N{\'u}{\~n}ez}, {Gonz{\'a}lez-Vidal}, {Granvik}, {Guerrier},
  {Guillout}, {Guiraud}, {G{\'u}rpide}, {Guti{\'e}rrez-S{\'a}nchez}, {Guy},
  {Haigron}, {Hatzidimitriou}, {Haywood}, {Heiter}, {Helmi}, {Hobbs},
  {Hofmann}, {Holl}, {Holland}, {Hunt}, {Hypki}, {Icardi}, {Irwin}, {Jevardat
  de Fombelle}, {Jofr{\'e}}, {Jonker}, {Jorissen}, {Julbe}, {Karampelas},
  {Kochoska}, {Kohley}, {Kolenberg}, {Kontizas}, {Koposov}, {Kordopatis},
  {Koubsky}, {Kowalczyk}, {Krone-Martins}, {Kudryashova}, {Kull}, {Bachchan},
  {Lacoste-Seris}, {Lanza}, {Lavigne}, {Le Poncin-Lafitte}, {Lebreton},
  {Lebzelter}, {Leccia}, {Leclerc}, {Lecoeur-Taibi}, {Lemaitre}, {Lenhardt},
  {Leroux}, {Liao}, {Licata}, {Lindstr{\o}m}, {Lister}, {Livanou}, {Lobel},
  {L{\"o}ffler}, {L{\'o}pez}, {Lopez-Lozano}, {Lorenz}, {Loureiro},
  {MacDonald}, {Magalh{\~a}es Fernandes}, {Managau}, {Mann}, {Mantelet},
  {Marchal}, {Marchant}, {Marconi}, {Marie}, {Marinoni}, {Marrese},
  {Marschalk{\'o}}, {Marshall}, {Mart{\'\i}n-Fleitas}, {Martino}, {Mary},
  {Matijevi{\v{c}}}, {Mazeh}, {McMillan}, {Messina}, {Mestre}, {Michalik},
  {Millar}, {Miranda}, {Molina}, {Molinaro}, {Molinaro}, {Moln{\'a}r},
  {Moniez}, {Montegriffo}, {Monteiro}, {Mor}, {Mora}, {Morbidelli}, {Morel},
  {Morgenthaler}, {Morley}, {Morris}, {Mulone}, {Muraveva}, {Musella},
  {Narbonne}, {Nelemans}, {Nicastro}, {Noval}, {Ord{\'e}novic},
  {Ordieres-Mer{\'e}}, {Osborne}, {Pagani}, {Pagano}, {Pailler}, {Palacin},
  {Palaversa}, {Parsons}, {Paulsen}, {Pecoraro}, {Pedrosa}, {Pentik{\"a}inen},
  {Pereira}, {Pichon}, {Piersimoni}, {Pineau}, {Plachy}, {Plum}, {Poujoulet},
  {Pr{\v{s}}a}, {Pulone}, {Ragaini}, {Rago}, {Rambaux}, {Ramos-Lerate},
  {Ranalli}, {Rauw}, {Read}, {Regibo}, {Renk}, {Reyl{\'e}}, {Ribeiro},
  {Rimoldini}, {Ripepi}, {Riva}, {Rixon}, {Roelens}, {Romero-G{\'o}mez},
  {Rowell}, {Royer}, {Rudolph}, {Ruiz-Dern}, {Sadowski}, {Sagrist{\`a}
  Sell{\'e}s}, {Sahlmann}, {Salgado}, {Salguero}, {Sarasso}, {Savietto},
  {Schnorhk}, {Schultheis}, {Sciacca}, {Segol}, {Segovia}, {Segransan},
  {Serpell}, {Shih}, {Smareglia}, {Smart}, {Smith}, {Solano}, {Solitro},
  {Sordo}, {Soria Nieto}, {Souchay}, {Spagna}, {Spoto}, {Stampa}, {Steele},
  {Steidelm{\"u}ller}, {Stephenson}, {Stoev}, {Suess}, {S{\"u}veges}, {Surdej},
  {Szabados}, {Szegedi-Elek}, {Tapiador}, {Taris}, {Tauran}, {Taylor},
  {Teixeira}, {Terrett}, {Tingley}, {Trager}, {Turon}, {Ulla}, {Utrilla},
  {Valentini}, {van Elteren}, {Van Hemelryck}, {van Leeuwen}, {Varadi},
  {Vecchiato}, {Veljanoski}, {Via}, {Vicente}, {Vogt}, {Voss}, {Votruba},
  {Voutsinas}, {Walmsley}, {Weiler}, {Weingrill}, {Werner}, {Wevers},
  {Whitehead}, {Wyrzykowski}, {Yoldas}, {{\v{Z}}erjal}, {Zucker}, {Zurbach},
  {Zwitter}, {Alecu}, {Allen}, {Allende Prieto}, {Amorim},
  {Anglada-Escud{\'e}}, {Arsenijevic}, {Azaz}, {Balm}, {Beck}, {Bernstein},
  {Bigot}, {Bijaoui}, {Blasco}, {Bonfigli}, {Bono}, {Boudreault}, {Bressan},
  {Brown}, {Brunet}, {Bunclark}, {Buonanno}, {Butkevich}, {Carret}, {Carrion},
  {Chemin}, {Ch{\'e}reau}, {Corcione}, {Darmigny}, {de Boer}, {de Teodoro}, {de
  Zeeuw}, {Delle Luche}, {Domingues}, {Dubath}, {Fodor}, {Fr{\'e}zouls},
  {Fries}, {Fustes}, {Fyfe}, {Gallardo}, {Gallegos}, {Gardiol}, {Gebran},
  {Gomboc}, {G{\'o}mez}, {Grux}, {Gueguen}, {Heyrovsky}, {Hoar}, {Iannicola},
  {Isasi Parache}, {Janotto}, {Joliet}, {Jonckheere}, {Keil}, {Kim},
  {Klagyivik}, {Klar}, {Knude}, {Kochukhov}, {Kolka}, {Kos}, {Kutka}, {Lainey},
  {LeBouquin}, {Liu}, {Loreggia}, {Makarov}, {Marseille}, {Martayan},
  {Martinez-Rubi}, {Massart}, {Meynadier}, {Mignot}, {Munari}, {Nguyen},
  {Nordlander}, {Ocvirk}, {O'Flaherty}, {Olias Sanz}, {Ortiz}, {Osorio},
  {Oszkiewicz}, {Ouzounis}, {Palmer}, {Park}, {Pasquato}, {Peltzer}, {Peralta},
  {P{\'e}turaud}, {Pieniluoma}, {Pigozzi}, {Poels}, {Prat}, {Prod'homme},
  {Raison}, {Rebordao}, {Risquez}, {Rocca-Volmerange}, {Rosen}, {Ruiz-Fuertes},
  {Russo}, {Sembay}, {Serraller Vizcaino}, {Short}, {Siebert}, {Silva},
  {Sinachopoulos}, {Slezak}, {Soffel}, {Sosnowska}, {Strai{\v{z}}ys}, {ter
  Linden}, {Terrell}, {Theil}, {Tiede}, {Troisi}, {Tsalmantza}, {Tur},
  {Vaccari}, {Vachier}, {Valles}, {Van Hamme}, {Veltz}, {Virtanen}, {Wallut},
  {Wichmann}, {Wilkinson}, {Ziaeepour}, \& {Zschocke}}]{2016Prusti}
{Gaia Collaboration}, {Prusti}, T., {de Bruijne}, J.~H.~J., {et~al.} 2016,
  \aap, 595, A1

\bibitem[{{Gilmore} {et~al.}(2022){Gilmore}, {Randich}, {Worley}, {Hourihane},
  {Gonneau}, {Sacco}, \& {et al.}}]{GES_final_release_paper_1}
{Gilmore}, G., {Randich}, S., {Worley}, C.~C., {et~al.} 2022, \aap\ in press

\bibitem[{{Gomes} {et~al.}(2005){Gomes}, {Levison}, {Tsiganis}, \&
  {Morbidelli}}]{Gomes2005Natur.435..466G}
{Gomes}, R., {Levison}, H.~F., {Tsiganis}, K., \& {Morbidelli}, A. 2005, \nat,
  435, 466

\bibitem[{{Gradie} \& {Tedesco}(1982)}]{Gradie1982Sci...216.1405G}
{Gradie}, J. \& {Tedesco}, E. 1982, Science, 216, 1405

\bibitem[{{Gradie} {et~al.}(1989){Gradie}, {Chapman}, \&
  {Tedesco}}]{Gradie1989aste.conf..316G}
{Gradie}, J.~C., {Chapman}, C.~R., \& {Tedesco}, E.~F. 1989, in Asteroids II,
  ed. R.~P. {Binzel}, T.~{Gehrels}, \& M.~S. {Matthews}, 316--335

\bibitem[{{Granahan} {et~al.}(1994){Granahan}, {Fanale}, {Robinson}, {Carlson},
  {Kamp}, {Klaasen}, {Weissman}, {Belton}, {Cook}, {Edwards}, {McEwen},
  {Soderblom}, {Carcich}, {Helfenstein}, {Simonelli}, {Thomas}, \&
  {Veverka}}]{1994LPI....25..453G}
{Granahan}, J.~C., {Fanale}, F.~P., {Robinson}, M.~S., {et~al.} 1994, in Lunar
  and Planetary Science Conference, Lunar and Planetary Science Conference, 453

\bibitem[{Granvik {et~al.}(2018)Granvik, Morbidelli, Jedicke, Bolin, Bottke,
  Beshore, Vokrouhlick{\'{y}}, Nesvorn{\'{y}}, \&
  Michel}]{Granvik2018Icar..312..181G}
Granvik, M., Morbidelli, A., Jedicke, R., {et~al.} 2018, Icarus, 312, 181

\bibitem[{Granvik {et~al.}(2017)Granvik, Morbidelli, Vokrouhlick{\'{y}},
  Bottke, Nesvorn{\'{y}}, \& Jedicke}]{Granvik2017A&A...598A..52G}
Granvik, M., Morbidelli, A., Vokrouhlick{\'{y}}, D., {et~al.} 2017, Astronomy
  and Astrophysics, 598, A52

\bibitem[{{Hasegawa} {et~al.}(2019){Hasegawa}, {Hiroi}, {Ohtsuka}, {Ishiguro},
  {Kuroda}, {Ito}, \& {Sasaki}}]{Hasegawa2019PASJ...71..103H}
{Hasegawa}, S., {Hiroi}, T., {Ohtsuka}, K., {et~al.} 2019, \pasj, 71, 103

\bibitem[{{Hasegawa} {et~al.}(2021){Hasegawa}, {Marsset}, {DeMeo}, {Bus},
  {Geem}, {Ishiguro}, {Im}, {Kuroda}, \&
  {Vernazza}}]{Hasegawa2021ApJ...916L...6H}
{Hasegawa}, S., {Marsset}, M., {DeMeo}, F.~E., {et~al.} 2021, \apjl, 916, L6

\bibitem[{{Henden} {et~al.}(2016){Henden}, {Templeton}, {Terrell}, {Smith},
  {Levine}, \& {Welch}}]{apass9}
{Henden}, A.~A., {Templeton}, M., {Terrell}, D., {et~al.} 2016, VizieR Online
  Data Catalogue, 2336

\bibitem[{{H{\o}g} {et~al.}(2000){H{\o}g}, {Fabricius}, {Makarov}, {Urban},
  {Corbin}, {Wycoff}, {Bastian}, {Schwekendiek}, \&
  {Wicenec}}]{2000A&A...355L..27H}
{H{\o}g}, E., {Fabricius}, C., {Makarov}, V.~V., {et~al.} 2000, \aap, 355, L27

\bibitem[{{Huang} {et~al.}(2013){Huang}, {Ji}, {Ye}, {Wang}, {Yan}, {Meng},
  {Wang}, {Li}, {Li}, {Qiao}, {Zhao}, {Zhao}, {Zhang}, {Liu}, {Jiang}, {Rao},
  {Li}, {Huang}, {Ip}, {Hu}, {Zhu}, {Yu}, {Zou}, {Tang}, {Li}, {Zhao}, {Huang},
  {Jiang}, \& {Bai}}]{Huang2013NatSR...3E3411H}
{Huang}, J., {Ji}, J., {Ye}, P., {et~al.} 2013, Scientific Reports, 3, 3411

\bibitem[{{Huber} {et~al.}(2016){Huber}, {Bryson}, {Haas}, {Barclay},
  {Barentsen}, {Howell}, {Sharma}, {Stello}, \&
  {Thompson}}]{epic-2016ApJS..224....2H}
{Huber}, D., {Bryson}, S.~T., {Haas}, M.~R., {et~al.} 2016, \apjs, 224, 2

\bibitem[{Hunter(2007)}]{Hunter:2007}
Hunter, J.~D. 2007, Computing In Science \& Engineering, 9, 90

\bibitem[{Ivezi{\'c} {et~al.}(2019)Ivezi{\'c}, Kahn, Tyson, Abel, Acosta,
  Allsman, Alonso, AlSayyad, Anderson, Andrew, Angel, Angeli, Ansari,
  Antilogus, Araujo, Armstrong, Arndt, Astier, Aubourg, Auza, Axelrod, Bard,
  Barr, Barrau, Bartlett, Bauer, Bauman, Baumont, Bechtol, Bechtol, Becker,
  Becla, Beldica, Bellavia, Bianco, Biswas, Blanc, Blazek, Blandford, Bloom,
  Bogart, Bond, Booth, Borgland, Borne, Bosch, Boutigny, Brackett, Bradshaw,
  Brandt, Brown, Bullock, Burchat, Burke, Cagnoli, Calabrese, Callahan, Callen,
  Carlin, Carlson, Chandrasekharan, Charles-Emerson, Chesley, Cheu, Chiang,
  Chiang, Chirino, Chow, Ciardi, Claver, Cohen-Tanugi, Cockrum, Coles,
  Connolly, Cook, Cooray, Covey, Cribbs, Cui, Cutri, Daly, Daniel, Daruich,
  Daubard, Daues, Dawson, Delgado, Dellapenna, de~Peyster, de~Val-Borro, Digel,
  Doherty, Dubois, Dubois-Felsmann, {\v{D}}urech, Economou, Eifler, Eracleous,
  Emmons, Fausti~Neto, Ferguson, Figueroa, Fisher-Levine, Focke, Foss, Frank,
  Freemon, Gangler, Gawiser, Geary, Gee, Geha, Gessner, Gibson, Gilmore,
  Glanzman, Glick, Goldina, Goldstein, Goodenow, Graham, Gressler, Gris, Guy,
  Guyonnet, Haller, Harris, Hascall, Haupt, Hernandez, Herrmann, Hileman,
  Hoblitt, Hodgson, Hogan, Howard, Huang, Huffer, Ingraham, Innes, Jacoby,
  Jain, Jammes, Jee, Jenness, Jernigan, Jevremovi{\'c}, Johns, Johnson,
  Johnson, Jones, Juramy-Gilles, Juri{\'c}, Kalirai, Kallivayalil, Kalmbach,
  Kantor, Karst, Kasliwal, Kelly, Kessler, Kinnison, Kirkby, Knox, Kotov,
  Krabbendam, Krughoff, Kub{\'a}nek, Kuczewski, Kulkarni, Ku, Kurita, Lage,
  Lambert, Lange, Langton, Le~Guillou, Levine, Liang, Lim, Lintott, Long,
  Lopez, Lotz, Lupton, Lust, MacArthur, Mahabal, Mandelbaum, Markiewicz, Marsh,
  Marshall, Marshall, May, McKercher, McQueen, Meyers, Migliore, Miller, Mills,
  Miraval, Moeyens, Moolekamp, Monet, Moniez, Monkewitz, Montgomery, Morrison,
  Mueller, Muller, Mu{\~n}oz~Arancibia, Neill, Newbry, Nief, Nomerotski,
  Nordby, O{\textquoteright}Connor, Oliver, Olivier, Olsen,
  O{\textquoteright}Mullane, Ortiz, Osier, Owen, Pain, Palecek, Parejko,
  Parsons, Pease, Peterson, \& Peter...}]{Ivezic2019ApJ...873..111I}
Ivezi{\'c}, {\v Z}., Kahn, S.~M., Tyson, J.~A., {et~al.} 2019, The
  Astrophysical Journal, 873, 111

\bibitem[{{Ivezi{\'c}} {et~al.}(2019){Ivezi{\'c}}, {Kahn}, {Tyson}, {Abel},
  {Acosta}, {Allsman}, {Alonso}, {AlSayyad}, {Anderson}, {Andrew}, {Angel},
  {Angeli}, {Ansari}, {Antilogus}, {Araujo}, {Armstrong}, {Arndt}, {Astier},
  {Aubourg}, {Auza}, {Axelrod}, {Bard}, {Barr}, {Barrau}, {Bartlett}, {Bauer},
  {Bauman}, {Baumont}, {Bechtol}, {Bechtol}, {Becker}, {Becla}, {Beldica},
  {Bellavia}, {Bianco}, {Biswas}, {Blanc}, {Blazek}, {Blandford}, {Bloom},
  {Bogart}, {Bond}, {Booth}, {Borgland}, {Borne}, {Bosch}, {Boutigny},
  {Brackett}, {Bradshaw}, {Brandt}, {Brown}, {Bullock}, {Burchat}, {Burke},
  {Cagnoli}, {Calabrese}, {Callahan}, {Callen}, {Carlin}, {Carlson},
  {Chandrasekharan}, {Charles-Emerson}, {Chesley}, {Cheu}, {Chiang}, {Chiang},
  {Chirino}, {Chow}, {Ciardi}, {Claver}, {Cohen-Tanugi}, {Cockrum}, {Coles},
  {Connolly}, {Cook}, {Cooray}, {Covey}, {Cribbs}, {Cui}, {Cutri}, {Daly},
  {Daniel}, {Daruich}, {Daubard}, {Daues}, {Dawson}, {Delgado}, {Dellapenna},
  {de Peyster}, {de Val-Borro}, {Digel}, {Doherty}, {Dubois},
  {Dubois-Felsmann}, {Durech}, {Economou}, {Eifler}, {Eracleous}, {Emmons},
  {Fausti Neto}, {Ferguson}, {Figueroa}, {Fisher-Levine}, {Focke}, {Foss},
  {Frank}, {Freemon}, {Gangler}, {Gawiser}, {Geary}, {Gee}, {Geha}, {Gessner},
  {Gibson}, {Gilmore}, {Glanzman}, {Glick}, {Goldina}, {Goldstein}, {Goodenow},
  {Graham}, {Gressler}, {Gris}, {Guy}, {Guyonnet}, {Haller}, {Harris},
  {Hascall}, {Haupt}, {Hernandez}, {Herrmann}, {Hileman}, {Hoblitt}, {Hodgson},
  {Hogan}, {Howard}, {Huang}, {Huffer}, {Ingraham}, {Innes}, {Jacoby}, {Jain},
  {Jammes}, {Jee}, {Jenness}, {Jernigan}, {Jevremovi{\'c}}, {Johns}, {Johnson},
  {Johnson}, {Jones}, {Juramy-Gilles}, {Juri{\'c}}, {Kalirai}, {Kallivayalil},
  {Kalmbach}, {Kantor}, {Karst}, {Kasliwal}, {Kelly}, {Kessler}, {Kinnison},
  {Kirkby}, {Knox}, {Kotov}, {Krabbendam}, {Krughoff}, {Kub{\'a}nek},
  {Kuczewski}, {Kulkarni}, {Ku}, {Kurita}, {Lage}, {Lambert}, {Lange},
  {Langton}, {Le Guillou}, {Levine}, {Liang}, {Lim}, {Lintott}, {Long},
  {Lopez}, {Lotz}, {Lupton}, {Lust}, {MacArthur}, {Mahabal}, {Mandelbaum},
  {Markiewicz}, {Marsh}, {Marshall}, {Marshall}, {May}, {McKercher}, {McQueen},
  {Meyers}, {Migliore}, {Miller}, {Mills}, {Miraval}, {Moeyens}, {Moolekamp},
  {Monet}, {Moniez}, {Monkewitz}, {Montgomery}, {Morrison}, {Mueller},
  {Muller}, {Mu{\~n}oz Arancibia}, {Neill}, {Newbry}, {Nief}, {Nomerotski},
  {Nordby}, {O'Connor}, {Oliver}, {Olivier}, {Olsen}, {O'Mullane}, {Ortiz},
  {Osier}, {Owen}, {Pain}, {Palecek}, {Parejko}, {Parsons}, {Pease},
  {Peterson}, {Peterson}, {Petravick}, {Libby Petrick}, {Petry},
  {Pierfederici}, {Pietrowicz}, {Pike}, {Pinto}, {Plante}, {Plate}, {Plutchak},
  {Price}, {Prouza}, {Radeka}, {Rajagopal}, {Rasmussen}, {Regnault}, {Reil},
  {Reiss}, {Reuter}, {Ridgway}, {Riot}, {Ritz}, {Robinson}, {Roby}, {Roodman},
  {Rosing}, {Roucelle}, {Rumore}, {Russo}, {Saha}, {Sassolas}, {Schalk},
  {Schellart}, {Schindler}, {Schmidt}, {Schneider}, {Schneider}, {Schoening},
  {Schumacher}, {Schwamb}, {Sebag}, {Selvy}, {Sembroski}, {Seppala}, {Serio},
  {Serrano}, {Shaw}, {Shipsey}, {Sick}, {Silvestri}, {Slater}, {Smith},
  {Smith}, {Sobhani}, {Soldahl}, {Storrie-Lombardi}, {Stover}, {Strauss},
  {Street}, {Stubbs}, {Sullivan}, {Sweeney}, {Swinbank}, {Szalay}, {Takacs},
  {Tether}, {Thaler}, {Thayer}, {Thomas}, {Thornton}, {Thukral}, {Tice},
  {Trilling}, {Turri}, {Van Berg}, {Vanden Berk}, {Vetter}, {Virieux},
  {Vucina}, {Wahl}, {Walkowicz}, {Walsh}, {Walter}, {Wang}, {Wang}, {Warner},
  {Wiecha}, {Willman}, {Winters}, {Wittman}, {Wolff}, {Wood-Vasey}, {Wu},
  {Xin}, {Yoachim}, \& {Zhan}}]{lsst2019ApJ...873..111I}
{Ivezi{\'c}}, {\v{Z}}., {Kahn}, S.~M., {Tyson}, J.~A., {et~al.} 2019, \apj,
  873, 111

\bibitem[{{Jordi} {et~al.}(2010){Jordi}, {Gebran}, {Carrasco}, {de Bruijne},
  {Voss}, {Fabricius}, {Knude}, {Vallenari}, {Kohley}, \&
  {Mora}}]{Jordi2010A&A...523A..48J}
{Jordi}, C., {Gebran}, M., {Carrasco}, J.~M., {et~al.} 2010, \aap, 523, A48

\bibitem[{{Keller} {et~al.}(2010){Keller}, {Barbieri}, {Koschny}, {Lamy},
  {Rickman}, {Rodrigo}, {Sierks}, {A'Hearn}, {Angrilli}, {Barucci}, {Bertaux},
  {Cremonese}, {Da Deppo}, {Davidsson}, {De Cecco}, {Debei}, {Fornasier},
  {Fulle}, {Groussin}, {Gutierrez}, {Hviid}, {Ip}, {Jorda}, {Knollenberg},
  {Kramm}, {K{\"u}hrt}, {K{\"u}ppers}, {Lara}, {Lazzarin}, {Moreno}, {Marzari},
  {Michalik}, {Naletto}, {Sabau}, {Thomas}, {Wenzel}, {Bertini}, {Besse},
  {Ferri}, {Kaasalainen}, {Lowry}, {Marchi}, {Mottola}, {Sabolo},
  {Schr{\"o}der}, {Spjuth}, \& {Vernazza}}]{2010Sci...327..190K}
{Keller}, H.~U., {Barbieri}, C., {Koschny}, D., {et~al.} 2010, Science, 327,
  190

\bibitem[{{Lantz} {et~al.}(2018){Lantz}, {Binzel}, \&
  {DeMeo}}]{2018Icar..302...10L}
{Lantz}, C., {Binzel}, R.~P., \& {DeMeo}, F.~E. 2018, \icarus, 302, 10

\bibitem[{{Lasker} {et~al.}(2008){Lasker}, {Lattanzi}, {McLean}, {Bucciarelli},
  {Drimmel}, {Garcia}, {Greene}, {Guglielmetti}, {Hanley}, {Hawkins},
  {Laidler}, {Loomis}, {Meakes}, {Mignani}, {Morbidelli}, {Morrison},
  {Pannunzio}, {Rosenberg}, {Sarasso}, {Smart}, {Spagna}, {Sturch},
  {Volpicelli}, {White}, {Wolfe}, \& {Zacchei}}]{2008AJ....136..735L}
{Lasker}, B.~M., {Lattanzi}, M.~G., {McLean}, B.~J., {et~al.} 2008, \aj, 136,
  735

\bibitem[{Lauretta {et~al.}(2019)Lauretta, DellaGiustina, Bennett, Becker,
  Barnouin, Bottke, Drouet~d{\textquoteright}Aubigny, Dworkin, Emery, Hamilton,
  Hergenrother, Izawa, Rizk, Walsh, Wolner, Team, Connolly, Clark, Campins,
  Becker, Boynton, d{\textquoteright}Aubigny, Enos, Golish, Howell,
  Balram-Knutson, Nolan, Roper, Smith, Scheeres, \&
  Kaplan}]{Lauretta2019Natur.568...55L}
Lauretta, D.~S., DellaGiustina, D.~N., Bennett, C.~A., {et~al.} 2019, Nature,
  568, 55

\bibitem[{Lazzaro {et~al.}(2004{\natexlab{a}})Lazzaro, Angeli, Carvano,
  Mothé-Diniz, Duffard, \& Florczak}]{LAZZARO2004}
Lazzaro, D., Angeli, C., Carvano, J., {et~al.} 2004{\natexlab{a}}, Icarus, 172,
  179, special Issue: Cassini-Huygens at Jupiter

\bibitem[{Lazzaro {et~al.}(2004{\natexlab{b}})Lazzaro, Angeli, Carvano,
  Moth{\'e}-Diniz, Duffard, \& FLORCZAK}]{Lazzaro2004Icar..172..179L}
Lazzaro, D., Angeli, C.~A., Carvano, J.~M., {et~al.} 2004{\natexlab{b}},
  Icarus, 172, 179

\bibitem[{{Li} {et~al.}(2016){Li}, {Reddy}, {Nathues}, {Le Corre}, {Izawa},
  {Cloutis}, {Sykes}, {Carsenty}, {Castillo-Rogez}, {Hoffmann}, {Jaumann},
  {Krohn}, {Mottola}, {Prettyman}, {Schaefer}, {Schenk}, {Schr{\"o}der},
  {Williams}, {Smith}, {Zuber}, {Konopliv}, {Park}, {Raymond}, \&
  {Russell}}]{2016ApJ...817L..22L}
{Li}, J.-Y., {Reddy}, V., {Nathues}, A., {et~al.} 2016, \apjl, 817, L22

\bibitem[{Libourel {et~al.}(2017)Libourel, Michel, Delbo, Ganino, Recio-Blanco,
  de~Laverny, Zolensky, \& Krot}]{Libourel2017Icar..282..375L}
Libourel, G., Michel, P., Delbo, M., {et~al.} 2017, Icarus, 282, 375

\bibitem[{{Lindegren} {et~al.}(2012){Lindegren}, {Lammers}, {Hobbs},
  {O'Mullane}, {Bastian}, \& {Hern{\'a}ndez}}]{Lindegren2012A&A...538A..78L}
{Lindegren}, L., {Lammers}, U., {Hobbs}, D., {et~al.} 2012, \aap, 538, A78

\bibitem[{{LSST Science Collaboration} {et~al.}(2009){LSST Science
  Collaboration}, {Abell}, {Allison}, {Anderson}, {Andrew}, {Angel}, {Armus},
  {Arnett}, {Asztalos}, {Axelrod}, {Bailey}, {Ballantyne}, {Bankert},
  {Barkhouse}, {Barr}, {Barrientos}, {Barth}, {Bartlett}, {Becker}, {Becla},
  {Beers}, {Bernstein}, {Biswas}, {Blanton}, {Bloom}, {Bochanski}, {Boeshaar},
  {Borne}, {Bradac}, {Brandt}, {Bridge}, {Brown}, {Brunner}, {Bullock},
  {Burgasser}, {Burge}, {Burke}, {Cargile}, {Chandrasekharan}, {Chartas},
  {Chesley}, {Chu}, {Cinabro}, {Claire}, {Claver}, {Clowe}, {Connolly}, {Cook},
  {Cooke}, {Cooray}, {Covey}, {Culliton}, {de Jong}, {de Vries}, {Debattista},
  {Delgado}, {Dell'Antonio}, {Dhital}, {Di Stefano}, {Dickinson}, {Dilday},
  {Djorgovski}, {Dobler}, {Donalek}, {Dubois-Felsmann}, {Durech},
  {Eliasdottir}, {Eracleous}, {Eyer}, {Falco}, {Fan}, {Fassnacht}, {Ferguson},
  {Fernandez}, {Fields}, {Finkbeiner}, {Figueroa}, {Fox}, {Francke}, {Frank},
  {Frieman}, {Fromenteau}, {Furqan}, {Galaz}, {Gal-Yam}, {Garnavich},
  {Gawiser}, {Geary}, {Gee}, {Gibson}, {Gilmore}, {Grace}, {Green}, {Gressler},
  {Grillmair}, {Habib}, {Haggerty}, {Hamuy}, {Harris}, {Hawley}, {Heavens},
  {Hebb}, {Henry}, {Hileman}, {Hilton}, {Hoadley}, {Holberg}, {Holman},
  {Howell}, {Infante}, {Ivezic}, {Jacoby}, {Jain}, {R}, {Jedicke}, {Jee},
  {Garrett Jernigan}, {Jha}, {Johnston}, {Jones}, {Juric}, {Kaasalainen},
  {Styliani}, {Kafka}, {Kahn}, {Kaib}, {Kalirai}, {Kantor}, {Kasliwal},
  {Keeton}, {Kessler}, {Knezevic}, {Kowalski}, {Krabbendam}, {Krughoff},
  {Kulkarni}, {Kuhlman}, {Lacy}, {Lepine}, {Liang}, {Lien}, {Lira}, {Long},
  {Lorenz}, {Lotz}, {Lupton}, {Lutz}, {Macri}, {Mahabal}, {Mandelbaum},
  {Marshall}, {May}, {McGehee}, {Meadows}, {Meert}, {Milani}, {Miller},
  {Miller}, {Mills}, {Minniti}, {Monet}, {Mukadam}, {Nakar}, {Neill}, {Newman},
  {Nikolaev}, {Nordby}, {O'Connor}, {Oguri}, {Oliver}, {Olivier}, {Olsen},
  {Olsen}, {Olszewski}, {Oluseyi}, {Padilla}, {Parker}, {Pepper}, {Peterson},
  {Petry}, {Pinto}, {Pizagno}, {Popescu}, {Prsa}, {Radcka}, {Raddick},
  {Rasmussen}, {Rau}, {Rho}, {Rhoads}, {Richards}, {Ridgway}, {Robertson},
  {Roskar}, {Saha}, {Sarajedini}, {Scannapieco}, {Schalk}, {Schindler},
  {Schmidt}, {Schmidt}, {Schneider}, {Schumacher}, {Scranton}, {Sebag},
  {Seppala}, {Shemmer}, {Simon}, {Sivertz}, {Smith}, {Allyn Smith}, {Smith},
  {Spitz}, {Stanford}, {Stassun}, {Strader}, {Strauss}, {Stubbs}, {Sweeney},
  {Szalay}, {Szkody}, {Takada}, {Thorman}, {Trilling}, {Trimble}, {Tyson}, {Van
  Berg}, {Vanden Berk}, {VanderPlas}, {Verde}, {Vrsnak}, {Walkowicz},
  {Wandelt}, {Wang}, {Wang}, {Warner}, {Wechsler}, {West}, {Wiecha},
  {Williams}, {Willman}, {Wittman}, {Wolff}, {Wood-Vasey}, {Wozniak}, {Young},
  {Zentner}, \& {Zhan}}]{lsst2009arXiv0912.0201L}
{LSST Science Collaboration}, {Abell}, P.~A., {Allison}, J., {et~al.} 2009,
  arXiv e-prints, arXiv:0912.0201

\bibitem[{Lucas {et~al.}(2019)Lucas, Emery, MacLennan, Pinilla-Alonso,
  Cartwright, Lindsay, Reddy, Sanchez, Thomas, \& Lorenzi}]{LUCAS2019}
Lucas, M.~P., Emery, J.~P., MacLennan, E.~M., {et~al.} 2019, Icarus, 322, 227

\bibitem[{{Luo} {et~al.}(2015){Luo}, {Zhao}, {Zhao}, {Deng}, {Liu}, {Jing},
  {Wang}, {Zhang}, {Shi}, {Cui}, {Chu}, {Li}, {Bai}, {Wu}, {Cai}, {Cao}, {Cao},
  {Carlin}, {Chen}, {Chen}, {Chen}, {Chen}, {Chen}, {Chen}, {Chen},
  {Christlieb}, {Chu}, {Cui}, {Dong}, {Du}, {Fan}, {Feng}, {Fu}, {Gao}, {Gong},
  {Gu}, {Guo}, {Han}, {He}, {Hou}, {Hou}, {Hou}, {Hu}, {Hu}, {Hu}, {Huo},
  {Jia}, {Jiang}, {Jiang}, {Jiang}, {Jin}, {Kong}, {Kong}, {Lei}, {Li}, {Li},
  {Li}, {Li}, {Li}, {Li}, {Li}, {Li}, {Li}, {Li}, {Li}, {Li}, {Liang}, {Lin},
  {Liu}, {Liu}, {Liu}, {Liu}, {Lu}, {Luo}, {Mao}, {Newberg}, {Ni}, {Qi}, {Qi},
  {Shen}, {Shi}, {Song}, {Song}, {Su}, {Su}, {Tang}, {Tao}, {Tian}, {Wang},
  {Wang}, {Wang}, {Wang}, {Wang}, {Wang}, {Wang}, {Wang}, {Wang}, {Wang},
  {Wang}, {Wang}, {Wang}, {Wang}, {Wang}, {Wang}, {Wang}, {Wang}, {Wang},
  {Wang}, {Wei}, {Wei}, {Wu}, {Wu}, {Wu}, {Wu}, {Xing}, {Xu}, {Xu}, {Xu},
  {Yan}, {Yang}, {Yang}, {Yang}, {Yang}, {Yao}, {Yu}, {Yuan}, {Yuan}, {Yuan},
  {Yuan}, {Zhai}, {Zhang}, {Zhang}, {Zhang}, {Zhang}, {Zhang}, {Zhang},
  {Zhang}, {Zhang}, {Zhao}, {Zhou}, {Zhou}, {Zhu}, {Zhu}, {Zou}, \&
  {Zuo}}]{LamostDR1}
{Luo}, A.~L., {Zhao}, Y.-H., {Zhao}, G., {et~al.} 2015, Research in Astronomy
  and Astrophysics, 15, 1095

\bibitem[{{Magnier} {et~al.}(2020{\natexlab{a}}){Magnier}, {Chambers},
  {Flewelling}, {Hoblitt}, {Huber}, {Price}, {Sweeney}, {Waters}, {Denneau},
  {Draper}, {Hodapp}, {Jedicke}, {Kaiser}, {Kudritzki}, {Metcalfe}, {Stubbs},
  \& {Wainscoat}}]{panstarrs1b}
{Magnier}, E.~A., {Chambers}, K.~C., {Flewelling}, H.~A., {et~al.}
  2020{\natexlab{a}}, \apjs, 251, 3

\bibitem[{{Magnier} {et~al.}(2020{\natexlab{b}}){Magnier}, {Schlafly},
  {Finkbeiner}, {Tonry}, {Goldman}, {R{\"o}ser}, {Schilbach}, {Casertano},
  {Chambers}, {Flewelling}, {Huber}, {Price}, {Sweeney}, {Waters}, {Denneau},
  {Draper}, {Hodapp}, {Jedicke}, {Kaiser}, {Kudritzki}, {Metcalfe}, {Stubbs},
  \& {Wainscoat}}]{panstarrs1e}
{Magnier}, E.~A., {Schlafly}, E.~F., {Finkbeiner}, D.~P., {et~al.}
  2020{\natexlab{b}}, \apjs, 251, 6

\bibitem[{{Magnier} {et~al.}(2020{\natexlab{c}}){Magnier}, {Sweeney},
  {Chambers}, {Flewelling}, {Huber}, {Price}, {Waters}, {Denneau}, {Draper},
  {Farrow}, {Jedicke}, {Hodapp}, {Kaiser}, {Kudritzki}, {Metcalfe}, {Stubbs},
  \& {Wainscoat}}]{panstarrs1d}
{Magnier}, E.~A., {Sweeney}, W.~E., {Chambers}, K.~C., {et~al.}
  2020{\natexlab{c}}, \apjs, 251, 5

\bibitem[{Mainzer {et~al.}(2011)Mainzer, Grav, Bauer, Masiero, McMillan, Cutri,
  Walker, Wright, Eisenhardt, Tholen, Spahr, JEDICKE, Denneau, DeBaun, Elsbury,
  Gautier, Gomillion, Hand, Mo, Watkins, Wilkins, Bryngelson, Del Pino~Molina,
  Desai, G{\'o}mez~Camus, Hidalgo, Konstantopoulos, Larsen, Maleszewski,
  Malkan, Mauduit, Mullan, Olszewski, Pforr, Saro, Scotti, \&
  Wasserman}]{Mainzer2011ApJ...743..156M}
Mainzer, A., Grav, T., Bauer, J., {et~al.} 2011, The Astrophysical Journal,
  743, 156

\bibitem[{Mainzer {et~al.}(2015)Mainzer, Usui, \&
  Trilling}]{Mainzer2015aste.book...89M}
Mainzer, A., Usui, F., \& Trilling, D.~E. 2015, in Asteroids IV (P. Michel, et
  al. eds.) University of Arizona Press, Tucson., 89

\bibitem[{{Marsset} {et~al.}(2020){Marsset}, {DeMeo}, {Binzel}, {Bus},
  {Burbine}, {Burt}, {Moskovitz}, {Polishook}, {Rivkin}, {Slivan}, \&
  {Thomas}}]{Marsset2020ApJS..247...73M}
{Marsset}, M., {DeMeo}, F.~E., {Binzel}, R.~P., {et~al.} 2020, \apjs, 247, 73

\bibitem[{{Marsset} {et~al.}(2022){Marsset}, {DeMeo}, {Burt}, {Polishook},
  {Binzel}, {Granvik}, {Vernazza}, {Carry}, {Bus}, {Slivan}, {Thomas},
  {Moskovitz}, \& {Rivkin}}]{Marsset2022AJ....163..165M}
{Marsset}, M., {DeMeo}, F.~E., {Burt}, B., {et~al.} 2022, \aj, 163, 165

\bibitem[{{Michel}(2013)}]{Michel2013AcAau..90....6M}
{Michel}, P. 2013, Acta Astronautica, 90, 6

\bibitem[{{Michel} {et~al.}(2021){Michel}, {Kueppers}, {Fitzsimmons}, {Green},
  {Lazzarin}, {Ulamec}, {Carnelli}, \& {Martino}}]{Michel2021EPSC...15...71M}
{Michel}, P., {Kueppers}, M., {Fitzsimmons}, A., {et~al.} 2021, in European
  Planetary Science Congress, EPSC2021--71

\bibitem[{{Millis} {et~al.}(1976){Millis}, {Bowell}, \&
  {Thompson}}]{1976Icar...28...53M}
{Millis}, R.~L., {Bowell}, E., \& {Thompson}, D.~T. 1976, \icarus, 28, 53

\bibitem[{Minton \& Malhotra(2009)}]{Minton2009Natur.457.1109M}
Minton, D.~A. \& Malhotra, R. 2009, Nature, 457, 1109

\bibitem[{Molaro {et~al.}(2017)Molaro, Byrne, \&
  Le}]{Molaro2017Icar..294..247M}
Molaro, J.~L., Byrne, S., \& Le, J.~L. 2017, Icarus, 294, 247

\bibitem[{Molaro {et~al.}(2020)Molaro, Walsh, Jawin, Ballouz, Bennett,
  DellaGiustina, Golish, Drouet~d'Aubigny, Rizk, Schwartz, Hanna, Martel,
  Pajola, Campins, Ryan, Bottke, \& Lauretta}]{Molaro2020NatCo..11.2913M}
Molaro, J.~L., Walsh, K.~J., Jawin, E.~R., {et~al.} 2020, Nature
  Communications, 11, 2913

\bibitem[{{Morate} {et~al.}(2021){Morate}, {Marcio Carvano}, {Alvarez-Candal},
  {De Pr{\'a}}, {Licandro}, {Galarza}, {Mahlke}, {Solano-M{\'a}rquez},
  {Cenarro}, {Crist{\'o}bal-Hornillos}, {Hern{\'a}ndez-Monteagudo},
  {L{\'o}pez-Sanjuan}, {Mar{\'\i}n-Franch}, {Moles}, {Varela}, {V{\'a}zquez
  Rami{\'o}}, {Alcaniz}, {Dupke}, {Ederoclite}, {Sodr{\'e}}, {Angulo},
  {Jim{\'e}nez-Esteban}, {Siffert}, \& {J-PLUS
  Collaboration}}]{Morate2021A&A...655A..47M}
{Morate}, D., {Marcio Carvano}, J., {Alvarez-Candal}, A., {et~al.} 2021, \aap,
  655, A47

\bibitem[{Morbidelli {et~al.}(2009)Morbidelli, Bottke, Nesvorn{\'{y}}, \&
  Levison}]{Morbidelli2009Icar..204..558M}
Morbidelli, A., Bottke, W.~F., Nesvorn{\'{y}}, D., \& Levison, H.~F. 2009,
  Icarus, 204, 558

\bibitem[{{Morbidelli} {et~al.}(2020){Morbidelli}, {Delbo}, {Granvik},
  {Bottke}, {Jedicke}, {Bolin}, {Michel}, \&
  {Vokrouhlicky}}]{Morbidelli2020Icar..34013631M}
{Morbidelli}, A., {Delbo}, M., {Granvik}, M., {et~al.} 2020, \icarus, 340,
  113631

\bibitem[{Morbidelli \& Raymond(2016)}]{Morbidelli2016JGRE..121.1962M}
Morbidelli, A. \& Raymond, S.~N. 2016, Journal of Geophysical Research:
  Planets, 121, 1962

\bibitem[{Morbidelli \&
  Vokrouhlick{\'{y}}(2003)}]{Morbidelli2003Icar..163..120M}
Morbidelli, A. \& Vokrouhlick{\'{y}}, D. 2003, Icarus, 163, 120

\bibitem[{{Nathues}(2010)}]{Nathues2010Icar..208..252N}
{Nathues}, A. 2010, \icarus, 208, 252

\bibitem[{{Nesvorn{\'y}}(2018)}]{Nesvorny2018ARA&A..56..137N}
{Nesvorn{\'y}}, D. 2018, \araa, 56, 137

\bibitem[{Nesvorn{\'{y}} {et~al.}(2015)Nesvorn{\'{y}}, Bro{\v z}, \&
  Carruba}]{Nesvorny2015aste.book..297N}
Nesvorn{\'{y}}, D., Bro{\v z}, M., \& Carruba, V. 2015, in Asteroids IV (P.
  Michel, et al. eds.) University of Arizona Press, Tucson., 297

\bibitem[{Nesvorn{\'{y}} \& Morbidelli(2012)}]{Nesvorny2012AJ....144..117N}
Nesvorn{\'{y}}, D. \& Morbidelli, A. 2012, The Astronomical Journal, 144, 117

\bibitem[{{Novakovic} {et~al.}(2009){Novakovic}, {Balaz}, {Knezevic}, \&
  {Potocnik}}]{Novakovic2009SerAJ.179...75N}
{Novakovic}, B., {Balaz}, A., {Knezevic}, Z., \& {Potocnik}, M. 2009, Serbian
  Astronomical Journal, 179, 75

\bibitem[{{Ochsenbein} {et~al.}(2000){Ochsenbein}, {Bauer}, \&
  {Marcout}}]{2000A&AS..143...23O}
{Ochsenbein}, F., {Bauer}, P., \& {Marcout}, J. 2000, \aaps, 143, 23

\bibitem[{{Olkin} {et~al.}(2021){Olkin}, {Levison}, {Vincent}, {Noll},
  {Andrews}, {Gray}, {Good}, {Marchi}, {Christensen}, {Reuter}, {Weaver},
  {P{\"a}tzold}, {Bell}, {Hamilton}, {Dello Russo}, {Simon}, {Beasley},
  {Grundy}, {Howett}, {Spencer}, {Ravine}, \&
  {Caplinger}}]{Olkin2021PSJ.....2..172O}
{Olkin}, C.~B., {Levison}, H.~F., {Vincent}, M., {et~al.} 2021, The Planetary
  Science Journal, 2, 172

\bibitem[{{Onken} {et~al.}(2019){Onken}, {Wolf}, {Bessell}, {Chang}, {Da
  Costa}, {Luvaul}, {Mackey}, {Schmidt}, \& {Shao}}]{2019PASA...36...33O}
{Onken}, C.~A., {Wolf}, C., {Bessell}, M.~S., {et~al.} 2019, \pasa, 36, e033

\bibitem[{{Parker} {et~al.}(2008){Parker}, {Ivezi{\'c}}, {Juri{\'c}}, {Lupton},
  {Sekora}, \& {Kowalski}}]{Parker2008Icar..198..138P}
{Parker}, A., {Ivezi{\'c}}, {\v{Z}}., {Juri{\'c}}, M., {et~al.} 2008, \icarus,
  198, 138

\bibitem[{P\'erez \& Granger(2007)}]{PER-GRA:2007}
P\'erez, F. \& Granger, B.~E. 2007, Computing in Science and Engineering, 9, 21

\bibitem[{{Perna} {et~al.}(2018{\natexlab{a}}){Perna}, {Barucci},
  {Fulchignoni}, {Popescu}, {Belskaya}, {Fornasier}, {Doressoundiram}, {Lantz},
  \& {Merlin}}]{Perna2018P&SS..157...82P}
{Perna}, D., {Barucci}, M.~A., {Fulchignoni}, M., {et~al.} 2018{\natexlab{a}},
  \planss, 157, 82

\bibitem[{{Perna} {et~al.}(2018{\natexlab{b}}){Perna}, {Barucci},
  {Fulchignoni}, {Popescu}, {Belskaya}, {Fornasier}, {Doressoundiram}, {Lantz},
  \& {Merlin}}]{2018P&SS..157...82P}
{Perna}, D., {Barucci}, M.~A., {Fulchignoni}, M., {et~al.} 2018{\natexlab{b}},
  \planss, 157, 82

\bibitem[{{Pierens} {et~al.}(2014){Pierens}, {Raymond}, {Nesvorny}, \&
  {Morbidelli}}]{Pierens2014ApJ...795L..11P}
{Pierens}, A., {Raymond}, S.~N., {Nesvorny}, D., \& {Morbidelli}, A. 2014,
  \apjl, 795, L11

\bibitem[{Popescu {et~al.}(2016)Popescu, Licandro, Morate, de~Le{\'o}n,
  Nedelcu, Rebolo, McMahon, Gonzalez-Solares, \&
  Irwin}]{Popescu2016A&A...591A.115P}
Popescu, M., Licandro, J., Morate, D., {et~al.} 2016, Astronomy and
  Astrophysics, 591, A115

\bibitem[{{Popescu, M.} {et~al.}(2014){Popescu, M.}, {Birlan, M.}, {Nedelcu, D.
  A.}, {Vaubaillon, J.}, \& {Cristescu, C. P.}}]{Popescu2014}
{Popescu, M.}, {Birlan, M.}, {Nedelcu, D. A.}, {Vaubaillon, J.}, \& {Cristescu,
  C. P.} 2014, A\&A, 572, A106

\bibitem[{{Popescu, M.} {et~al.}(2019){Popescu, M.}, {Vaduvescu, O.}, {de
  Le\'on, J.}, {Gherase, R. M.}, {Licandro, J.}, {Boaca, I. L.}, {Sonka, A.
  B.}, {Ashley, R. P.}, {Mocnik, T.}, {Morate, D.}, {Predatu, M.}, {De Pr\'a,
  M.}, {Fari\~na, C.}, {Stoev, H.}, {D\'{\i}az Alfaro, M.},
  {Ordonez-Etxeberria, I.}, {L\'opez-Mart\'{\i}nez, F.}, \& {Errmann,
  R.}}]{Popescu2019}
{Popescu, M.}, {Vaduvescu, O.}, {de Le\'on, J.}, {et~al.} 2019, A\&A, 627, A124

\bibitem[{{R Core Team}(2013)}]{RManual}
{R Core Team}. 2013, {R: A Language and Environment for Statistical Computing},
  R Foundation for Statistical Computing, Vienna, Austria

\bibitem[{{Randich} {et~al.}(2022){Randich}, {Gilmore}, {Magrini}, {Sacco},
  {Jackson}, {Jeffries}, \& {et al.}}]{GES_final_release_paper_2}
{Randich}, S., {Gilmore}, G., {Magrini}, L., {et~al.} 2022, \aap\ in press

\bibitem[{Raymond \& Izidoro(2017)}]{Raymond2017SciA....3E1138R}
Raymond, S.~N. \& Izidoro, A. 2017, Science Advances, 3, e1701138

\bibitem[{{Raymond} {et~al.}(2020){Raymond}, {Izidoro}, \&
  {Morbidelli}}]{Raymond2020plas.book..287R}
{Raymond}, S.~N., {Izidoro}, A., \& {Morbidelli}, A. 2020, in Planetary
  Astrobiology, ed. V.~S. {Meadows}, G.~N. {Arney}, B.~E. {Schmidt}, \& D.~J.
  {Des Marais}, 287

\bibitem[{{Raymond} \& {Nesvorny}(2020)}]{Raymond2020arXiv201207932R}
{Raymond}, S.~N. \& {Nesvorny}, D. 2020, arXiv e-prints, arXiv:2012.07932

\bibitem[{{Reddy} {et~al.}(2015){Reddy}, {Dunn}, {Thomas}, {Moskovitz}, \&
  {Burbine}}]{2015aste.book...43R}
{Reddy}, V., {Dunn}, T.~L., {Thomas}, C.~A., {Moskovitz}, N.~A., \& {Burbine},
  T.~H. 2015, in Asteroids IV, 43--63

\bibitem[{{Reddy} {et~al.}(2012){Reddy}, {Nathues}, {Le Corre}, {Sierks}, {Li},
  {Gaskell}, {McCoy}, {Beck}, {Schr{\"o}der}, {Pieters}, {Becker}, {Buratti},
  {Denevi}, {Blewett}, {Christensen}, {Gaffey}, {Gutierrez-Marques}, {Hicks},
  {Keller}, {Maue}, {Mottola}, {McFadden}, {McSween}, {Mittlefehldt},
  {O'Brien}, {Raymond}, \& {Russell}}]{2012Sci...336..700R}
{Reddy}, V., {Nathues}, A., {Le Corre}, L., {et~al.} 2012, Science, 336, 700

\bibitem[{{Rivkin} {et~al.}(2021){Rivkin}, {Chabot}, {Stickle}, {Thomas},
  {Richardson}, {Barnouin}, {Fahnestock}, {Ernst}, {Cheng}, {Chesley}, {Naidu},
  {Statler}, {Barbee}, {Agrusa}, {Moskovitz}, {Terik Daly}, {Pravec},
  {Scheirich}, {Dotto}, {Della Corte}, {Michel}, {K{\"u}ppers}, {Atchison}, \&
  {Hirabayashi}}]{Rivkin2021PSJ.....2..173R}
{Rivkin}, A.~S., {Chabot}, N.~L., {Stickle}, A.~M., {et~al.} 2021, The
  Planetary Science Journal, 2, 173

\bibitem[{{Roeser} {et~al.}(2010){Roeser}, {Demleitner}, \&
  {Schilbach}}]{2010AJ....139.2440R}
{Roeser}, S., {Demleitner}, M., \& {Schilbach}, E. 2010, \aj, 139, 2440

\bibitem[{{Rubincam}(2000)}]{Rubincam2000Icar..148....2R}
{Rubincam}, D.~P. 2000, \icarus, 148, 2

\bibitem[{{Russell} {et~al.}(2004){Russell}, {Coradini}, {Christensen}, {De
  Sanctis}, {Feldman}, {Jaumann}, {Keller}, {Konopliv}, {McCord}, {McFadden},
  {McSween}, {Mottola}, {Neukum}, {Pieters}, {Prettyman}, {Raymond}, {Smith},
  {Sykes}, {Williams}, {Wise}, \& {Zuber}}]{2004P&SS...52..465R}
{Russell}, C.~T., {Coradini}, A., {Christensen}, U., {et~al.} 2004, \planss,
  52, 465

\bibitem[{{Sanchez} {et~al.}(2012){Sanchez}, {Reddy}, {Nathues}, {Cloutis},
  {Mann}, \& {Hiesinger}}]{2012Icar..220...36S}
{Sanchez}, J.~A., {Reddy}, V., {Nathues}, A., {et~al.} 2012, \icarus, 220, 36

\bibitem[{{Sergeyev} \& {Carry}(2021)}]{Sergeyev2021A&A...652A..59S}
{Sergeyev}, A.~V. \& {Carry}, B. 2021, \aap, 652, A59

\bibitem[{{Sergeyev} {et~al.}(2022){Sergeyev}, {Carry}, {Onken}, {Devillepoix},
  {Wolf}, \& {Chang}}]{Sergeyev2022A&A...658A.109S}
{Sergeyev}, A.~V., {Carry}, B., {Onken}, C.~A., {et~al.} 2022, \aap, 658, A109

\bibitem[{{Sierks} {et~al.}(2011){Sierks}, {Lamy}, {Barbieri}, {Koschny},
  {Rickman}, {Rodrigo}, {A'Hearn}, {Angrilli}, {Barucci}, {Bertaux}, {Bertini},
  {Besse}, {Carry}, {Cremonese}, {Da Deppo}, {Davidsson}, {Debei}, {De Cecco},
  {De Leon}, {Ferri}, {Fornasier}, {Fulle}, {Hviid}, {Gaskell}, {Groussin},
  {Gutierrez}, {Ip}, {Jorda}, {Kaasalainen}, {Keller}, {Knollenberg}, {Kramm},
  {K{\"u}hrt}, {K{\"u}ppers}, {Lara}, {Lazzarin}, {Leyrat}, {Moreno}, {Magrin},
  {Marchi}, {Marzari}, {Massironi}, {Michalik}, {Moissl}, {Naletto},
  {Preusker}, {Sabau}, {Sabolo}, {Scholten}, {Snodgrass}, {Thomas}, {Tubiana},
  {Vernazza}, {Vincent}, {Wenzel}, {Andert}, {P{\"a}tzold}, \&
  {Weiss}}]{2011Sci...334..487S}
{Sierks}, H., {Lamy}, P., {Barbieri}, C., {et~al.} 2011, Science, 334, 487

\bibitem[{{Skrutskie} {et~al.}(2006){Skrutskie}, {Cutri}, {Stiening},
  {Weinberg}, {Schneider}, {Carpenter}, {Beichman}, {Capps}, {Chester},
  {Elias}, {Huchra}, {Liebert}, {Lonsdale}, {Monet}, {Price}, {Seitzer},
  {Jarrett}, {Kirkpatrick}, {Gizis}, {Howard}, {Evans}, {Fowler}, {Fullmer},
  {Hurt}, {Light}, {Kopan}, {Marsh}, {McCallon}, {Tam}, {Van Dyk}, \&
  {Wheelock}}]{2006AJ....131.1163S}
{Skrutskie}, M.~F., {Cutri}, R.~M., {Stiening}, R., {et~al.} 2006, \aj, 131,
  1163

\bibitem[{{Soubiran, C.} \& {Triaud, A.}(2004)}]{Soubiran2004}
{Soubiran, C.} \& {Triaud, A.} 2004, A\&A, 418, 1089

\bibitem[{{Spoto} {et~al.}(2015){Spoto}, {Milani}, \&
  {Kne{\v{z}}evi{\'c}}}]{Spoto2015Icar..257..275S}
{Spoto}, F., {Milani}, A., \& {Kne{\v{z}}evi{\'c}}, Z. 2015, \icarus, 257, 275

\bibitem[{{Steinmetz} {et~al.}(2020{\natexlab{a}}){Steinmetz}, {Guiglion},
  {McMillan}, {Matijevi{\v{c}}}, {Enke}, {Kordopatis}, {Zwitter}, {Valentini},
  {Chiappini}, {Casagrande}, {Wojno}, {Anguiano}, {Bienaym{\'e}}, {Bijaoui},
  {Binney}, {Burton}, {Cass}, {de Laverny}, {Fiegert}, {Freeman}, {Fulbright},
  {Gibson}, {Gilmore}, {Grebel}, {Helmi}, {Kunder}, {Munari}, {Navarro},
  {Parker}, {Ruchti}, {Recio-Blanco}, {Reid}, {Seabroke}, {Siviero}, {Siebert},
  {Stupar}, {Watson}, {Williams}, {Wyse}, {Anders}, {Antoja}, {Birko},
  {Bland-Hawthorn}, {Bossini}, {Garc{\'\i}a}, {Carrillo}, {Chaplin},
  {Elsworth}, {Famaey}, {Gerhard}, {Jofre}, {Just}, {Mathur}, {Miglio},
  {Minchev}, {Monari}, {Mosser}, {Ritter}, {Rodrigues}, {Scholz}, {Sharma},
  {Sysoliatina}, \& {RAVE Collaboration}}]{2020AJ....160...83S}
{Steinmetz}, M., {Guiglion}, G., {McMillan}, P.~J., {et~al.}
  2020{\natexlab{a}}, \aj, 160, 83

\bibitem[{{Steinmetz} {et~al.}(2020{\natexlab{b}}){Steinmetz},
  {Matijevi{\v{c}}}, {Enke}, {Zwitter}, {Guiglion}, {McMillan}, {Kordopatis},
  {Valentini}, {Chiappini}, {Casagrande}, {Wojno}, {Anguiano}, {Bienaym{\'e}},
  {Bijaoui}, {Binney}, {Burton}, {Cass}, {de Laverny}, {Fiegert}, {Freeman},
  {Fulbright}, {Gibson}, {Gilmore}, {Grebel}, {Helmi}, {Kunder}, {Munari},
  {Navarro}, {Parker}, {Ruchti}, {Recio-Blanco}, {Reid}, {Seabroke}, {Siviero},
  {Siebert}, {Stupar}, {Watson}, {Williams}, {Wyse}, {Anders}, {Antoja},
  {Birko}, {Bland-Hawthorn}, {Bossini}, {Garc{\'\i}a}, {Carrillo}, {Chaplin},
  {Elsworth}, {Famaey}, {Gerhard}, {Jofre}, {Just}, {Mathur}, {Miglio},
  {Minchev}, {Monari}, {Mosser}, {Ritter}, {Rodrigues}, {Scholz}, {Sharma},
  {Sysoliatina}, \& {RAVE Collaboration}}]{rave6a}
{Steinmetz}, M., {Matijevi{\v{c}}}, G., {Enke}, H., {et~al.}
  2020{\natexlab{b}}, \aj, 160, 82

\bibitem[{Sugita {et~al.}(2019)Sugita, Honda, Morota, Kameda, Sawada, Tatsumi,
  Yamada, Honda, Yokota, Kouyama, Sakatani, Ogawa, Suzuki, Okada, Namiki,
  Tanaka, Iijima, Yoshioka, Hayakawa, Cho, Matsuoka, Hirata, Miyamoto,
  Domingue, Hirabayashi, Nakamura, Hiroi, Michikami, Michel, Ballouz, Barnouin,
  Ernst, Schr{\"o}der, Kikuchi, Hemmi, Komatsu, Fukuhara, Taguchi, Arai,
  Senshu, Demura, Ogawa, Shimaki, Sekiguchi, M{\"u}ller, Hagermann, Mizuno,
  Noda, Matsumoto, Yamada, Ishihara, Ikeda, Araki, Yamamoto, Abe, Yoshida,
  Higuchi, Sasaki, Oshigami, Tsuruta, Asari, Tazawa, Shizugami, Kimura, Otsubo,
  Yabuta, Hasegawa, Ishiguro, Tachibana, Palmer, Gaskell, Le~Corre, Jaumann,
  Otto, Schmitz, Abell, Barucci, Zolensky, Vilas, Thuillet, Sugimoto, Takaki,
  Suzuki, Kamiyoshihara, Okada, Nagata, Fujimoto, Yoshikawa, Yamamoto, Shirai,
  Noguchi, Ogawa, Terui, Kikuchi, Yamaguchi, Oki, Takao, Takeuchi, Ono, Mimasu,
  Yoshikawa, Takahashi, Takei, Fujii, Hirose, Nakazawa, Hosoda, Mori, Shimada,
  Soldini, Iwata, Abe, Yano, Tsukizaki, Ozaki, Nishiyama, Saiki, Watanabe, \&
  Tsuda}]{Sugita2019Sci...364..252S}
Sugita, S., Honda, R., Morota, T., {et~al.} 2019, Science, 364, 252

\bibitem[{{Tanga} \& {Mignard}(2012)}]{2012P&SS...73....5T}
{Tanga}, P. \& {Mignard}, F. 2012, \planss, 73, 5

\bibitem[{{Tanga}(2022)}]{DR3-DPACP-150}
{Tanga}, P. e.~a. 2022, \aap\ in prep.

\bibitem[{{Taylor}(2005{\natexlab{a}})}]{2005ASPC..347...29T}
{Taylor}, M.~B. 2005{\natexlab{a}}, in Astronomical Society of the Pacific
  Conference Series, Vol. 347, Astronomical Data Analysis Software and Systems
  XIV, ed. P.~{Shopbell}, M.~{Britton}, \& R.~{Ebert}, 29

\bibitem[{{Taylor}(2005{\natexlab{b}})}]{2005Taylor}
{Taylor}, M.~B. 2005{\natexlab{b}}, in Astronomical Society of the Pacific
  Conference Series, Vol. 347, Astronomical Data Analysis Software and Systems
  XIV, ed. P.~{Shopbell}, M.~{Britton}, \& R.~{Ebert}, 29

\bibitem[{{Taylor}(2006)}]{2006ASPC..351..666T}
{Taylor}, M.~B. 2006, in Astronomical Society of the Pacific Conference Series,
  Vol. 351, Astronomical Data Analysis Software and Systems XV, ed.
  C.~{Gabriel}, C.~{Arviset}, D.~{Ponz}, \& S.~{Enrique}, 666

\bibitem[{{Taylor} {et~al.}(1971){Taylor}, {Gehrels}, \&
  {Silvester}}]{1971AJ.....76..141T}
{Taylor}, R.~C., {Gehrels}, T., \& {Silvester}, A.~B. 1971, \aj, 76, 141

\bibitem[{{Tholen}(1989)}]{1989aste.conf.1139T}
{Tholen}, D.~J. 1989, in Asteroids II, ed. R.~P. {Binzel}, T.~{Gehrels}, \&
  M.~S. {Matthews}, 1139--1150

\bibitem[{Tsiganis {et~al.}(2005)Tsiganis, Gomes, Morbidelli, \&
  Levison}]{Tsiganis2005Natur.435..459T}
Tsiganis, K., Gomes, R., Morbidelli, A., \& Levison, H.~F. 2005, Nature, 435,
  459

\bibitem[{{van Leeuwen}(2007)}]{2007A&A...474..653V}
{van Leeuwen}, F. 2007, \aap, 474, 653

\bibitem[{{Vera C. Rubin Observatory LSST Solar System Science Collaboration}
  {et~al.}(2020){Vera C. Rubin Observatory LSST Solar System Science
  Collaboration}, {Jones}, {Bannister}, {Bolin}, {Chandler}, {Chesley}, {Eggl},
  {Greenstreet}, {Holt}, {Hsieh}, {Ivezi{\'c}}, {Juri{\'c}}, {Kelley},
  {Knight}, {Malhotra}, {Oldroyd}, {Sarid}, {Schwamb}, {Snodgrass}, {Solontoi},
  \& {Trilling}}]{lsst2020arXiv200907653V}
{Vera C. Rubin Observatory LSST Solar System Science Collaboration}, {Jones},
  R.~L., {Bannister}, M.~T., {et~al.} 2020, arXiv e-prints, arXiv:2009.07653

\bibitem[{Vernazza {et~al.}(2008)Vernazza, Binzel, Thomas, DeMeo, Bus, Rivkin,
  \& Tokunaga}]{Vernazza2008Natur.454..858V}
Vernazza, P., Binzel, R.~P., Thomas, C.~A., {et~al.} 2008, Nature, 454, 858

\bibitem[{{Vernazza} {et~al.}(2016){Vernazza}, {Marsset}, {Beck}, {Binzel},
  {Birlan}, {Cloutis}, {DeMeo}, {Dumas}, \&
  {Hiroi}}]{Vernazza2016AJ....152...54V}
{Vernazza}, P., {Marsset}, M., {Beck}, P., {et~al.} 2016, \aj, 152, 54

\bibitem[{{Vernazza} {et~al.}(2014){Vernazza}, {Zanda}, {Binzel}, {Hiroi},
  {DeMeo}, {Birlan}, {Hewins}, {Ricci}, {Barge}, \&
  {Lockhart}}]{Vernazza2014ApJ...791..120V}
{Vernazza}, P., {Zanda}, B., {Binzel}, R.~P., {et~al.} 2014, \apj, 791, 120

\bibitem[{Veverka {et~al.}(1996)Veverka, Helfenstein, Lee, Thomas, McEwen,
  Belton, Klaasen, Johnson, Granahan, Fanale, Geissler, \&
  Head}]{Veverka1996Icar..120...66V}
Veverka, J., Helfenstein, P., Lee, P., {et~al.} 1996, Icarus, 120, 66

\bibitem[{{Veverka} {et~al.}(2000){Veverka}, {Robinson}, {Thomas}, {Murchie},
  {Bell}, {Izenberg}, {Chapman}, {Harch}, {Bell}, {Carcich}, {Cheng}, {Clark},
  {Domingue}, {Dunham}, {Farquhar}, {Gaffey}, {Hawkins}, {Joseph}, {Kirk},
  {Li}, {Lucey}, {Malin}, {Martin}, {McFadden}, {Merline}, {Miller}, {Owen},
  {Peterson}, {Prockter}, {Warren}, {Wellnitz}, {Williams}, \&
  {Yeomans}}]{2000Sci...289.2088V}
{Veverka}, J., {Robinson}, M., {Thomas}, P., {et~al.} 2000, Science, 289, 2088

\bibitem[{{Vilas} {et~al.}(1993){Vilas}, {Larson}, {Hatch}, \&
  {Jarvis}}]{Vilas1993Icar..105...67V}
{Vilas}, F., {Larson}, S.~M., {Hatch}, E.~C., \& {Jarvis}, K.~S. 1993, \icarus,
  105, 67

\bibitem[{{Vilas} \& {McFadden}(1992)}]{1992Icar..100...85V}
{Vilas}, F. \& {McFadden}, L.~A. 1992, \icarus, 100, 85

\bibitem[{{Vokrouhlick{\'y}} \&
  {Farinella}(2000)}]{Vokrouhlicky2000Natur.407..606V}
{Vokrouhlick{\'y}}, D. \& {Farinella}, P. 2000, \nat, 407, 606

\bibitem[{Walsh {et~al.}(2011)Walsh, Morbidelli, Raymond, O'Brien, \&
  Mandell}]{Walsh2011Natur.475..206W}
Walsh, K.~J., Morbidelli, A., Raymond, S.~N., O'Brien, D.~P., \& Mandell, A.~M.
  2011, Nature, 475, 206

\bibitem[{{Walsh} {et~al.}(2012){Walsh}, {Morbidelli}, {Raymond}, {O'Brien}, \&
  {Mandell}}]{Walsh2012M&PS...47.1941W}
{Walsh}, K.~J., {Morbidelli}, A., {Raymond}, S.~N., {O'Brien}, D.~P., \&
  {Mandell}, A.~M. 2012, Meteoritics \& Planetary Science, 47, 1941

\bibitem[{{Waters} {et~al.}(2020){Waters}, {Magnier}, {Price}, {Chambers},
  {Burgett}, {Draper}, {Flewelling}, {Hodapp}, {Huber}, {Jedicke}, {Kaiser},
  {Kudritzki}, {Lupton}, {Metcalfe}, {Rest}, {Sweeney}, {Tonry}, {Wainscoat},
  \& {Wood-Vasey}}]{panstarrs1c}
{Waters}, C.~Z., {Magnier}, E.~A., {Price}, P.~A., {et~al.} 2020, \apjs, 251, 4

\bibitem[{{Wenger} {et~al.}(2000){Wenger}, {Ochsenbein}, {Egret}, {Dubois},
  {Bonnarel}, {Borde}, {Genova}, {Jasniewicz}, {Lalo{\"e}}, {Lesteven}, \&
  {Monier}}]{2000AAS..143....9W}
{Wenger}, M., {Ochsenbein}, F., {Egret}, D., {et~al.} 2000, \aaps, 143, 9

\bibitem[{{Winn} \& {Fabrycky}(2015)}]{Winn2015ARA&A..53..409W}
{Winn}, J.~N. \& {Fabrycky}, D.~C. 2015, \araa, 53, 409

\bibitem[{Xu {et~al.}(1995)Xu, Binzel, Burbine, \& Bus}]{Xu1995Icar..115....1X}
Xu, S., Binzel, R.~P., Burbine, T.~H., \& Bus, S.~J. 1995, Icarus, 115, 1

\bibitem[{{Zacharias} {et~al.}(2015){Zacharias}, {Finch}, {Subasavage},
  {Bredthauer}, {Crockett}, {Divittorio}, {Ferguson}, {Harris}, {Harris},
  {Henden}, {Kilian}, {Munn}, {Rafferty}, {Rhodes}, {Schultheiss}, {Tilleman},
  \& {Wieder}}]{urat1}
{Zacharias}, N., {Finch}, C., {Subasavage}, J., {et~al.} 2015, \aj, 150, 101

\bibitem[{{Zacharias} {et~al.}(2013){Zacharias}, {Finch}, {Girard}, {Henden},
  {Bartlett}, {Monet}, \& {Zacharias}}]{2013AJ....145...44Z}
{Zacharias}, N., {Finch}, C.~T., {Girard}, T.~M., {et~al.} 2013, \aj, 145, 44

\bibitem[{Zellner {et~al.}(1985)Zellner, Tholen, \&
  Tedesco}]{Zellner1985Icar...61..355Z}
Zellner, B., Tholen, D.~J., \& Tedesco, E.~F. 1985, Icarus, 61, 355

\bibitem[{{Zolensky} {et~al.}(2006){Zolensky}, {Bland}, {Brown}, \&
  {Halliday}}]{Zolensky2006mess.book..869Z}
{Zolensky}, M., {Bland}, P., {Brown}, P., \& {Halliday}, I. 2006, in Meteorites
  and the Early Solar System II, ed. D.~S. {Lauretta} \& H.~Y. {McSween}, 869

\end{thebibliography}
\begin{appendix} 
\section*{Acknowledgements\label{sec:acknowl}}

This work presents results from the European Space Agency (ESA) space mission \gaia. \gaia\ data are being processed by the \gaia\ Data Processing and Analysis Consortium (DPAC). Funding for the DPAC is provided by national institutions, in particular the institutions participating in the \gaia\ MultiLateral Agreement (MLA). The \gaia\ mission website is at \url{https://www.cosmos.esa.int/gaia}. The \gaia\ archive website is \url{https://archives.esac.esa.int/gaia}.

The \gaia\ mission and data processing have financially been supported by, in alphabetical order by country:
\begin{itemize}
\item the Algerian Centre de Recherche en Astronomie, Astrophysique et G\'{e}ophysique of Bouzareah Observatory;
\item the Austrian Fonds zur F\"{o}rderung der wissenschaftlichen Forschung (FWF) Hertha Firnberg Programme through grants T359, P20046, and P23737;
\item the BELgian federal Science Policy Office (BELSPO) through various PROgramme de D\'{e}veloppement d'Exp\'{e}riences scientifiques (PRODEX) grants and the Polish Academy of Sciences - Fonds Wetenschappelijk Onderzoek through grant VS.091.16N, and the Fonds de la Recherche Scientifique (FNRS), and the Research Council of Katholieke Universiteit (KU) Leuven through grant C16/18/005 (Pushing AsteRoseismology to the next level with TESS, GaiA, and the Sloan DIgital Sky SurvEy -- PARADISE);  
\item the Brazil-France exchange programmes Funda\c{c}\~{a}o de Amparo \`{a} Pesquisa do Estado de S\~{a}o Paulo (FAPESP) and Coordena\c{c}\~{a}o de Aperfeicoamento de Pessoal de N\'{\i}vel Superior (CAPES) - Comit\'{e} Fran\c{c}ais d'Evaluation de la Coop\'{e}ration Universitaire et Scientifique avec le Br\'{e}sil (COFECUB);
\item the Chilean Agencia Nacional de Investigaci\'{o}n y Desarrollo (ANID) through Fondo Nacional de Desarrollo Cient\'{\i}fico y Tecnol\'{o}gico (FONDECYT) Regular Project 1210992 (L.~Chemin);
\item the National Natural Science Foundation of China (NSFC) through grants 11573054, 11703065, and 12173069, the China Scholarship Council through grant 201806040200, and the Natural Science Foundation of Shanghai through grant 21ZR1474100;  
\item the Tenure Track Pilot Programme of the Croatian Science Foundation and the \'{E}cole Polytechnique F\'{e}d\'{e}rale de Lausanne and the project TTP-2018-07-1171 `Mining the Variable Sky', with the funds of the Croatian-Swiss Research Programme;
\item the Czech-Republic Ministry of Education, Youth, and Sports through grant LG 15010 and INTER-EXCELLENCE grant LTAUSA18093, and the Czech Space Office through ESA PECS contract 98058;
\item the Danish Ministry of Science;
\item the Estonian Ministry of Education and Research through grant IUT40-1;
\item the European Commission’s Sixth Framework Programme through the European Leadership in Space Astrometry (\href{https://www.cosmos.esa.int/web/gaia/elsa-rtn-programme}{ELSA}) Marie Curie Research Training Network (MRTN-CT-2006-033481), through Marie Curie project PIOF-GA-2009-255267 (Space AsteroSeismology \& RR Lyrae stars, SAS-RRL), and through a Marie Curie Transfer-of-Knowledge (ToK) fellowship (MTKD-CT-2004-014188); the European Commission's Seventh Framework Programme through grant FP7-606740 (FP7-SPACE-2013-1) for the \gaia\ European Network for Improved data User Services (\href{https://gaia.ub.edu/twiki/do/view/GENIUS/}{GENIUS}) and through grant 264895 for the \gaia\ Research for European Astronomy Training (\href{https://www.cosmos.esa.int/web/gaia/great-programme}{GREAT-ITN}) network;
\item the European Cooperation in Science and Technology (COST) through COST Action CA18104 `Revealing the Milky Way with \gaia (MW-Gaia)';
\item the European Research Council (ERC) through grants 320360, 647208, and 834148 and through the European Union’s Horizon 2020 research and innovation and excellent science programmes through Marie Sk{\l}odowska-Curie grant 745617 (Our Galaxy at full HD -- Gal-HD) and 895174 (The build-up and fate of self-gravitating systems in the Universe) as well as grants 687378 (Small Bodies: Near and Far), 682115 (Using the Magellanic Clouds to Understand the Interaction of Galaxies), 695099 (A sub-percent distance scale from binaries and Cepheids -- CepBin), 716155 (Structured ACCREtion Disks -- SACCRED), 951549 (Sub-percent calibration of the extragalactic distance scale in the era of big surveys -- UniverScale), and 101004214 (Innovative Scientific Data Exploration and Exploitation Applications for Space Sciences -- EXPLORE);
\item the European Science Foundation (ESF), in the framework of the \gaia\ Research for European Astronomy Training Research Network Programme (\href{https://www.cosmos.esa.int/web/gaia/great-programme}{GREAT-ESF});
\item the European Space Agency (ESA) in the framework of the \gaia\ project, through the Plan for European Cooperating States (PECS) programme through contracts C98090 and 4000106398/12/NL/KML for Hungary, through contract 4000115263/15/NL/IB for Germany, and through PROgramme de D\'{e}veloppement d'Exp\'{e}riences scientifiques (PRODEX) grant 4000127986 for Slovenia;  
\item the Academy of Finland through grants 299543, 307157, 325805, 328654, 336546, and 345115 and the Magnus Ehrnrooth Foundation;
\item the French Centre National d’\'{E}tudes Spatiales (CNES), the Agence Nationale de la Recherche (ANR) through grant ANR-10-IDEX-0001-02 for the `Investissements d'avenir' programme, through grant ANR-15-CE31-0007 for project `Modelling the Milky Way in the \gaia era’ (MOD4Gaia), through grant ANR-14-CE33-0014-01 for project `The Milky Way disc formation in the \gaia era’ (ARCHEOGAL), through grant ANR-15-CE31-0012-01 for project `Unlocking the potential of Cepheids as primary distance calibrators’ (UnlockCepheids), through grant ANR-19-CE31-0017 for project `Secular evolution of galxies' (SEGAL), and through grant ANR-18-CE31-0006 for project `Galactic Dark Matter' (GaDaMa), the Centre National de la Recherche Scientifique (CNRS) and its SNO \gaia of the Institut des Sciences de l’Univers (INSU), its Programmes Nationaux: Cosmologie et Galaxies (PNCG), Gravitation R\'{e}f\'{e}rences Astronomie M\'{e}trologie (PNGRAM), Plan\'{e}tologie (PNP), Physique et Chimie du Milieu Interstellaire (PCMI), and Physique Stellaire (PNPS), the `Action F\'{e}d\'{e}ratrice \gaia' of the Observatoire de Paris, the R\'{e}gion de Franche-Comt\'{e}, the Institut National Polytechnique (INP) and the Institut National de Physique nucl\'{e}aire et de Physique des Particules (IN2P3) co-funded by CNES;
\item the German Aerospace Agency (Deutsches Zentrum f\"{u}r Luft- und Raumfahrt e.V., DLR) through grants 50QG0501, 50QG0601, 50QG0602, 50QG0701, 50QG0901, 50QG1001, 50QG1101, 50\-QG1401, 50QG1402, 50QG1403, 50QG1404, 50QG1904, 50QG2101, 50QG2102, and 50QG2202, and the Centre for Information Services and High Performance Computing (ZIH) at the Technische Universit\"{a}t Dresden for generous allocations of computer time;
\item the Hungarian Academy of Sciences through the Lend\"{u}let Programme grants LP2014-17 and LP2018-7 and the Hungarian National Research, Development, and Innovation Office (NKFIH) through grant KKP-137523 (`SeismoLab');
\item the Science Foundation Ireland (SFI) through a Royal Society - SFI University Research Fellowship (M.~Fraser);
\item the Israel Ministry of Science and Technology through grant 3-18143 and the Tel Aviv University Center for Artificial Intelligence and Data Science (TAD) through a grant;
\item the Agenzia Spaziale Italiana (ASI) through contracts I/037/08/0, I/058/10/0, 2014-025-R.0, 2014-025-R.1.2015, and 2018-24-HH.0 to the Italian Istituto Nazionale di Astrofisica (INAF), contract 2014-049-R.0/1/2 to INAF for the Space Science Data Centre (SSDC, formerly known as the ASI Science Data Center, ASDC), contracts I/008/10/0, 2013/030/I.0, 2013-030-I.0.1-2015, and 2016-17-I.0 to the Aerospace Logistics Technology Engineering Company (ALTEC S.p.A.), INAF, and the Italian Ministry of Education, University, and Research (Ministero dell'Istruzione, dell'Universit\`{a} e della Ricerca) through the Premiale project `MIning The Cosmos Big Data and Innovative Italian Technology for Frontier Astrophysics and Cosmology' (MITiC);
\item the Netherlands Organisation for Scientific Research (NWO) through grant NWO-M-614.061.414, through a VICI grant (A.~Helmi), and through a Spinoza prize (A.~Helmi), and the Netherlands Research School for Astronomy (NOVA);
\item the Polish National Science Centre through HARMONIA grant 2018/30/M/ST9/00311 and DAINA grant 2017/27/L/ST9/03221 and the Ministry of Science and Higher Education (MNiSW) through grant DIR/WK/2018/12;
\item the Portuguese Funda\c{c}\~{a}o para a Ci\^{e}ncia e a Tecnologia (FCT) through national funds, grants SFRH/\-BD/128840/2017 and PTDC/FIS-AST/30389/2017, and work contract DL 57/2016/CP1364/CT0006, the Fundo Europeu de Desenvolvimento Regional (FEDER) through grant POCI-01-0145-FEDER-030389 and its Programa Operacional Competitividade e Internacionaliza\c{c}\~{a}o (COMPETE2020) through grants UIDB/04434/2020 and UIDP/04434/2020, and the Strategic Programme UIDB/\-00099/2020 for the Centro de Astrof\'{\i}sica e Gravita\c{c}\~{a}o (CENTRA);  
\item the Slovenian Research Agency through grant P1-0188;
\item the Spanish Ministry of Economy (MINECO/FEDER, UE), the Spanish Ministry of Science and Innovation (MICIN), the Spanish Ministry of Education, Culture, and Sports, and the Spanish Government through grants BES-2016-078499, BES-2017-083126, BES-C-2017-0085, ESP2016-80079-C2-1-R, ESP2016-80079-C2-2-R, FPU16/03827, PDC2021-121059-C22, RTI2018-095076-B-C22, and TIN2015-65316-P (`Computaci\'{o}n de Altas Prestaciones VII'), the Juan de la Cierva Incorporaci\'{o}n Programme (FJCI-2015-2671 and IJC2019-04862-I for F.~Anders), the Severo Ochoa Centre of Excellence Programme (SEV2015-0493), and MICIN/AEI/10.13039/501100011033 (and the European Union through European Regional Development Fund `A way of making Europe') through grant RTI2018-095076-B-C21, the Institute of Cosmos Sciences University of Barcelona (ICCUB, Unidad de Excelencia `Mar\'{\i}a de Maeztu’) through grant CEX2019-000918-M, the University of Barcelona's official doctoral programme for the development of an R+D+i project through an Ajuts de Personal Investigador en Formaci\'{o} (APIF) grant, the Spanish Virtual Observatory through project AyA2017-84089, the Galician Regional Government, Xunta de Galicia, through grants ED431B-2021/36, ED481A-2019/155, and ED481A-2021/296, the Centro de Investigaci\'{o}n en Tecnolog\'{\i}as de la Informaci\'{o}n y las Comunicaciones (CITIC), funded by the Xunta de Galicia and the European Union (European Regional Development Fund -- Galicia 2014-2020 Programme), through grant ED431G-2019/01, the Red Espa\~{n}ola de Supercomputaci\'{o}n (RES) computer resources at MareNostrum, the Barcelona Supercomputing Centre - Centro Nacional de Supercomputaci\'{o}n (BSC-CNS) through activities AECT-2017-2-0002, AECT-2017-3-0006, AECT-2018-1-0017, AECT-2018-2-0013, AECT-2018-3-0011, AECT-2019-1-0010, AECT-2019-2-0014, AECT-2019-3-0003, AECT-2020-1-0004, and DATA-2020-1-0010, the Departament d'Innovaci\'{o}, Universitats i Empresa de la Generalitat de Catalunya through grant 2014-SGR-1051 for project `Models de Programaci\'{o} i Entorns d'Execuci\'{o} Parallels' (MPEXPAR), and Ramon y Cajal Fellowship RYC2018-025968-I funded by MICIN/AEI/10.13039/501100011033 and the European Science Foundation (`Investing in your future');
\item the Swedish National Space Agency (SNSA/Rymdstyrelsen);
\item the Swiss State Secretariat for Education, Research, and Innovation through the Swiss Activit\'{e}s Nationales Compl\'{e}mentaires and the Swiss National Science Foundation through an Eccellenza Professorial Fellowship (award PCEFP2\_194638 for R.~Anderson);
\item the United Kingdom Particle Physics and Astronomy Research Council (PPARC), the United Kingdom Science and Technology Facilities Council (STFC), and the United Kingdom Space Agency (UKSA) through the following grants to the University of Bristol, the University of Cambridge, the University of Edinburgh, the University of Leicester, the Mullard Space Sciences Laboratory of University College London, and the United Kingdom Rutherford Appleton Laboratory (RAL): PP/D006511/1, PP/D006546/1, PP/D006570/1, ST/I000852/1, ST/J005045/1, ST/K00056X/1, ST/\-K000209/1, ST/K000756/1, ST/L006561/1, ST/N000595/1, ST/N000641/1, ST/N000978/1, ST/\-N001117/1, ST/S000089/1, ST/S000976/1, ST/S000984/1, ST/S001123/1, ST/S001948/1, ST/\-S001980/1, ST/S002103/1, ST/V000969/1, ST/W002469/1, ST/W002493/1, ST/W002671/1, ST/W002809/1, and EP/V520342/1.
\end{itemize}

The \gaia\ project and data processing have made use of:
\begin{itemize}
\item the Set of Identifications, Measurements, and Bibliography for Astronomical Data \citep[SIMBAD,][]{2000AAS..143....9W}, the `Aladin sky atlas' \citep{2000A&AS..143...33B,2014ASPC..485..277B}, and the VizieR catalogue access tool \citep{2000A&AS..143...23O}, all operated at the Centre de Donn\'{e}es astronomiques de Strasbourg (\href{http://cds.u-strasbg.fr/}{CDS});
\item the National Aeronautics and Space Administration (NASA) Astrophysics Data System (\href{http://adsabs.harvard.edu/abstract_service.html}{ADS});
\item the SPace ENVironment Information System (SPENVIS), initiated by the Space Environment and Effects Section (TEC-EES) of ESA and developed by the Belgian Institute for Space Aeronomy (BIRA-IASB) under ESA contract through ESA’s General Support Technologies Programme (GSTP), administered by the BELgian federal Science Policy Office (BELSPO);
\item the software products \href{http://www.starlink.ac.uk/topcat/}{TOPCAT}, \href{http://www.starlink.ac.uk/stil}{STIL}, and \href{http://www.starlink.ac.uk/stilts}{STILTS} \citep{2005ASPC..347...29T,2006ASPC..351..666T};
\item Matplotlib \citep{Hunter:2007};
\item IPython \citep{PER-GRA:2007};  
\item Astropy, a community-developed core Python package for Astronomy \citep{2018AJ....156..123A};
\item R \citep{RManual};
\item Vaex \citep{2018A&A...618A..13B};
\item the Hipparcos-2 catalogue \citep{2007A&A...474..653V}. The Hipparcos and Tycho catalogues were constructed under the responsibility of large scientific teams collaborating with ESA. The Consortia Leaders were Lennart Lindegren (Lund, Sweden: NDAC) and Jean Kovalevsky (Grasse, France: FAST), together responsible for the Hipparcos Catalogue; Erik H{\o}g (Copenhagen, Denmark: TDAC) responsible for the Tycho Catalogue; and Catherine Turon (Meudon, France: INCA) responsible for the Hipparcos Input Catalogue (HIC);  
\item the Tycho-2 catalogue \citep{2000A&A...355L..27H}, the construction of which was supported by the Velux Foundation of 1981 and the Danish Space Board;
\item The Tycho double star catalogue \citep[TDSC,][]{2002A&A...384..180F}, based on observations made with the ESA Hipparcos astrometry satellite, as supported by the Danish Space Board and the United States Naval Observatory through their double-star programme;
\item data products from the Two Micron All Sky Survey \citep[2MASS,][]{2006AJ....131.1163S}, which is a joint project of the University of Massachusetts and the Infrared Processing and Analysis Center (IPAC) / California Institute of Technology, funded by the National Aeronautics and Space Administration (NASA) and the National Science Foundation (NSF) of the USA;
\item the ninth data release of the AAVSO Photometric All-Sky Survey (\href{https://www.aavso.org/apass}{APASS}, \citealt{apass9}), funded by the Robert Martin Ayers Sciences Fund;
\item the first data release of the Pan-STARRS survey \citep{panstarrs1,panstarrs1b,panstarrs1c,panstarrs1d,panstarrs1e,panstarrs1f}. The Pan-STARRS1 Surveys (PS1) and the PS1 public science archive have been made possible through contributions by the Institute for Astronomy, the University of Hawaii, the Pan-STARRS Project Office, the Max-Planck Society and its participating institutes, the Max Planck Institute for Astronomy, Heidelberg and the Max Planck Institute for Extraterrestrial Physics, Garching, The Johns Hopkins University, Durham University, the University of Edinburgh, the Queen's University Belfast, the Harvard-Smithsonian Center for Astrophysics, the Las Cumbres Observatory Global Telescope Network Incorporated, the National Central University of Taiwan, the Space Telescope Science Institute, the National Aeronautics and Space Administration (NASA) through grant NNX08AR22G issued through the Planetary Science Division of the NASA Science Mission Directorate, the National Science Foundation through grant AST-1238877, the University of Maryland, Eotvos Lorand University (ELTE), the Los Alamos National Laboratory, and the Gordon and Betty Moore Foundation;
\item the second release of the Guide Star Catalogue \citep[GSC2.3,][]{2008AJ....136..735L}. The Guide Star Catalogue II is a joint project of the Space Telescope Science Institute (STScI) and the Osservatorio Astrofisico di Torino (OATo). STScI is operated by the Association of Universities for Research in Astronomy (AURA), for the National Aeronautics and Space Administration (NASA) under contract NAS5-26555. OATo is operated by the Italian National Institute for Astrophysics (INAF). Additional support was provided by the European Southern Observatory (ESO), the Space Telescope European Coordinating Facility (STECF), the International GEMINI project, and the European Space Agency (ESA) Astrophysics Division (nowadays SCI-S);
\item the eXtended, Large (XL) version of the catalogue of Positions and Proper Motions \citep[PPM-XL,][]{2010AJ....139.2440R};
\item data products from the Wide-field Infrared Survey Explorer (WISE), which is a joint project of the University of California, Los Angeles, and the Jet Propulsion Laboratory/California Institute of Technology, and NEOWISE, which is a project of the Jet Propulsion Laboratory/California Institute of Technology. WISE and NEOWISE are funded by the National Aeronautics and Space Administration (NASA);
\item the first data release of the United States Naval Observatory (USNO) Robotic Astrometric Telescope \citep[URAT-1,][]{urat1};
\item the fourth data release of the United States Naval Observatory (USNO) CCD Astrograph Catalogue \citep[UCAC-4,][]{2013AJ....145...44Z};
\item the sixth and final data release of the Radial Velocity Experiment \citep[RAVE DR6,][]{2020AJ....160...83S,rave6a}. Funding for RAVE has been provided by the Leibniz Institute for Astrophysics Potsdam (AIP), the Australian Astronomical Observatory, the Australian National University, the Australian Research Council, the French National Research Agency, the German Research Foundation (SPP 1177 and SFB 881), the European Research Council (ERC-StG 240271 Galactica), the Istituto Nazionale di Astrofisica at Padova, the Johns Hopkins University, the National Science Foundation of the USA (AST-0908326), the W.M.\ Keck foundation, the Macquarie University, the Netherlands Research School for Astronomy, the Natural Sciences and Engineering Research Council of Canada, the Slovenian Research Agency, the Swiss National Science Foundation, the Science \& Technology Facilities Council of the UK, Opticon, Strasbourg Observatory, and the Universities of Basel, Groningen, Heidelberg, and Sydney. The RAVE website is at \url{https://www.rave-survey.org/};
\item the first data release of the Large sky Area Multi-Object Fibre Spectroscopic Telescope \citep[LAMOST DR1,][]{LamostDR1};
\item the K2 Ecliptic Plane Input Catalogue \citep[EPIC,][]{epic-2016ApJS..224....2H};
\item the ninth data release of the Sloan Digitial Sky Survey \citep[SDSS DR9,][]{SDSS9}. Funding for SDSS-III has been provided by the Alfred P. Sloan Foundation, the Participating Institutions, the National Science Foundation, and the United States Department of Energy Office of Science. The SDSS-III website is \url{http://www.sdss3.org/}. SDSS-III is managed by the Astrophysical Research Consortium for the Participating Institutions of the SDSS-III Collaboration including the University of Arizona, the Brazilian Participation Group, Brookhaven National Laboratory, Carnegie Mellon University, University of Florida, the French Participation Group, the German Participation Group, Harvard University, the Instituto de Astrof\'{\i}sica de Canarias, the Michigan State/Notre Dame/JINA Participation Group, Johns Hopkins University, Lawrence Berkeley National Laboratory, Max Planck Institute for Astrophysics, Max Planck Institute for Extraterrestrial Physics, New Mexico State University, New York University, Ohio State University, Pennsylvania State University, University of Portsmouth, Princeton University, the Spanish Participation Group, University of Tokyo, University of Utah, Vanderbilt University, University of Virginia, University of Washington, and Yale University;
\item the thirteenth release of the Sloan Digital Sky Survey \citep[SDSS DR13,][]{2017ApJS..233...25A}. Funding for SDSS-IV has been provided by the Alfred P. Sloan Foundation, the United States Department of Energy Office of Science, and the Participating Institutions. SDSS-IV acknowledges support and resources from the Center for High-Performance Computing at the University of Utah. The SDSS web site is \url{https://www.sdss.org/}. SDSS-IV is managed by the Astrophysical Research Consortium for the Participating Institutions of the SDSS Collaboration including the Brazilian Participation Group, the Carnegie Institution for Science, Carnegie Mellon University, the Chilean Participation Group, the French Participation Group, Harvard-Smithsonian Center for Astrophysics, Instituto de Astrof\'isica de Canarias, The Johns Hopkins University, Kavli Institute for the Physics and Mathematics of the Universe (IPMU) / University of Tokyo, the Korean Participation Group, Lawrence Berkeley National Laboratory, Leibniz Institut f\"ur Astrophysik Potsdam (AIP),  Max-Planck-Institut f\"ur Astronomie (MPIA Heidelberg), Max-Planck-Institut f\"ur Astrophysik (MPA Garching), Max-Planck-Institut f\"ur Extraterrestrische Physik (MPE), National Astronomical Observatories of China, New Mexico State University, New York University, University of Notre Dame, Observat\'ario Nacional / MCTI, The Ohio State University, Pennsylvania State University, Shanghai Astronomical Observatory, United Kingdom Participation Group, Universidad Nacional Aut\'onoma de M\'{e}xico, University of Arizona, University of Colorado Boulder, University of Oxford, University of Portsmouth, University of Utah, University of Virginia, University of Washington, University of Wisconsin, Vanderbilt University, and Yale University;
\item the second release of the SkyMapper catalogue \citep[SkyMapper DR2,][Digital Object Identifier 10.25914/5ce60d31ce759]{2019PASA...36...33O}. The national facility capability for SkyMapper has been funded through grant LE130100104 from the Australian Research Council (ARC) Linkage Infrastructure, Equipment, and Facilities (LIEF) programme, awarded to the University of Sydney, the Australian National University, Swinburne University of Technology, the University of Queensland, the University of Western Australia, the University of Melbourne, Curtin University of Technology, Monash University, and the Australian Astronomical Observatory. SkyMapper is owned and operated by The Australian National University's Research School of Astronomy and Astrophysics. The survey data were processed and provided by the SkyMapper Team at the the Australian National University. The SkyMapper node of the All-Sky Virtual Observatory (ASVO) is hosted at the National Computational Infrastructure (NCI). Development and support the SkyMapper node of the ASVO has been funded in part by Astronomy Australia Limited (AAL) and the Australian Government through the Commonwealth's Education Investment Fund (EIF) and National Collaborative Research Infrastructure Strategy (NCRIS), particularly the National eResearch Collaboration Tools and Resources (NeCTAR) and the Australian National Data Service Projects (ANDS);
\item the \gaia-ESO Public Spectroscopic Survey \citep[GES,][]{GES_final_release_paper_1,GES_final_release_paper_2}. The \gaia-ESO Survey is based on data products from observations made with ESO Telescopes at the La Silla Paranal Observatory under programme ID 188.B-3002. Public data releases are available through the \href{https://www.gaia-eso.eu/data-products/public-data-releases}{ESO Science Portal}. The project has received funding from the Leverhulme Trust (project RPG-2012-541), the European Research Council (project ERC-2012-AdG 320360-Gaia-ESO-MW), and the Istituto Nazionale di Astrofisica, INAF (2012: CRA 1.05.01.09.16; 2013: CRA 1.05.06.02.07).
\end{itemize}

The GBOT programme  uses observations collected at (i) the European Organisation for Astronomical Research in the Southern Hemisphere (ESO) with the VLT Survey Telescope (VST), under ESO programmes
092.B-0165,
093.B-0236,
094.B-0181,
095.B-0046,
096.B-0162,
097.B-0304,
098.B-0030,
099.B-0034,
0100.B-0131,
0101.B-0156,
0102.B-0174, and
0103.B-0165;
%
%
and (ii) the Liverpool Telescope, which is operated on the island of La Palma by Liverpool John Moores University in the Spanish Observatorio del Roque de los Muchachos of the Instituto de Astrof\'{\i}sica de Canarias with financial support from the United Kingdom Science and Technology Facilities Council, and (iii) telescopes of the Las Cumbres Observatory Global Telescope Network.

In case of errors or omissions, please contact the \href{https://www.cosmos.esa.int/web/gaia/gaia-helpdesk}{\gaia\ Helpdesk}.

This work was eased by the use of the data handling and visualisation software TOPCAT \citep{2005Taylor}, gnuplot,  python, and astropy. Auxiliary data are provided by the Minor Planet Physical Properties Catalogue \footnote{\url{mp3c.oca.eu}} of the Observatoire de la Côte d'Azur. M. Delbo wish to thank F. DeMeo, S. Raymond, C. Avdellidou for helpful discussions and M. Galinier (DPAC / CNES / OCA) for an editorial revision of the manuscript. The authors would like to thank editorial handling by T. Forveille and J. Neve, as well as comments and constructive criticisms from an anonymous reviewer.

\section{Dispersion of the photometric instrument}\label{appendix:xpresolution}
\begin{figure}[!ht]
    \centering
    \includegraphics[width=\columnwidth]{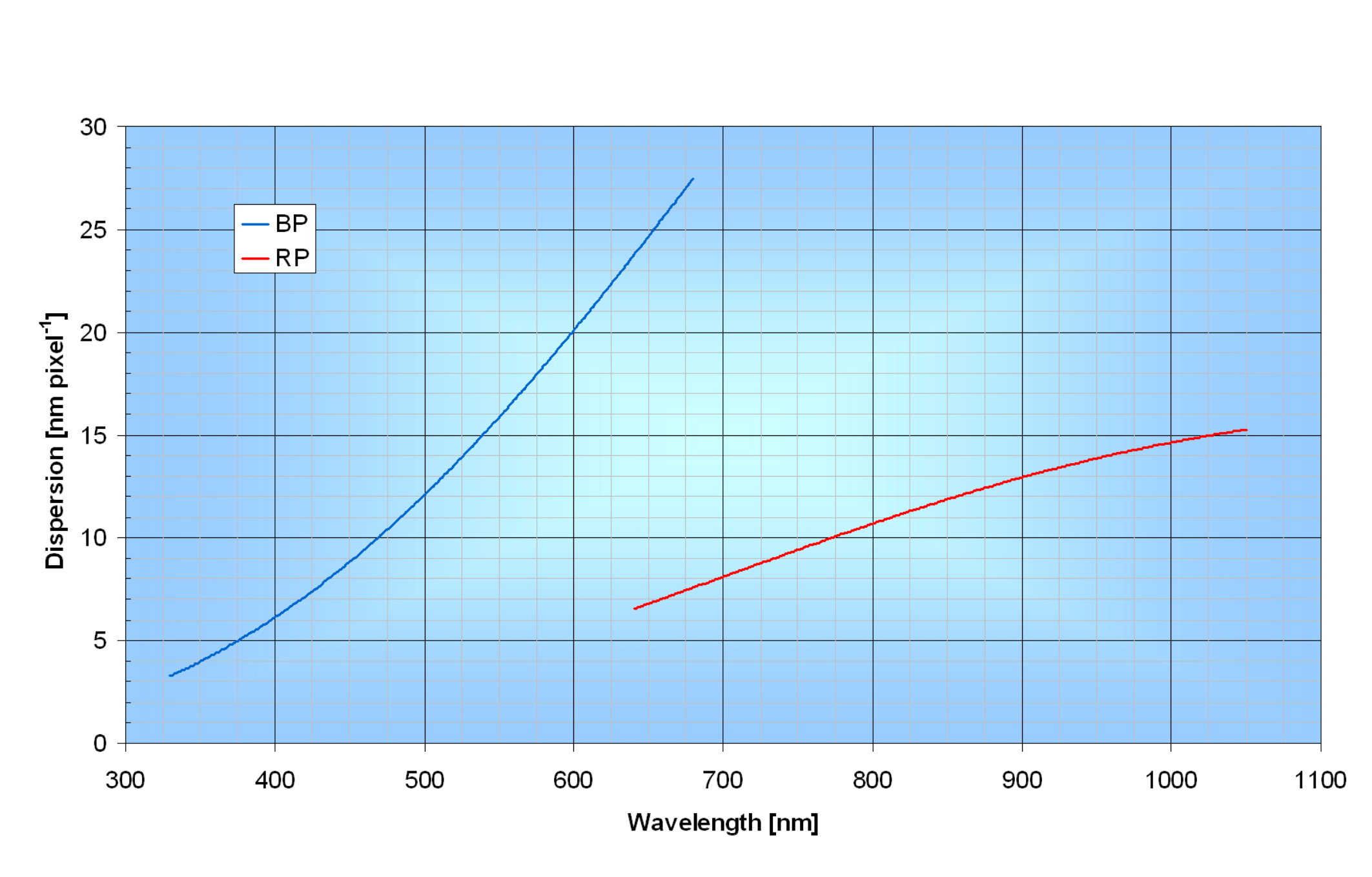}  
    \caption{Nominal dispersion curves of the BP/RP instrument.}
    \label{fig:xpresolution}
\end{figure}

The figure \ref{fig:xpresolution} of the nominal dispersion function of the BP and the RP is taken from the ESA website \footnote{\url{https://www.cosmos.esa.int/web/gaia/resolution}}.

\section{Astronomical Data Query Language (ADQL) queries}\label{sec:adql_queries}
\subsection{Request of a single spectrum of an asteroid}\label{sec:single_spectrum_request}

The following query returns the wavelengths, reflectances, and reflectance errors for the asteroid (21) Lutetia.

{\small
\begin{verbatim}
SELECT wavelength, reflectance_spectrum,
reflectance_spectrum_err 
FROM gaidr3.sso_reflectance_spectrum
WHERE number_mp = 21
\end{verbatim}
}

The same query can be written using the designation this time. The reader should be aware that no capital letters are taken into account.

{\small
\begin{verbatim}
SELECT wavelength, reflectance_spectrum,
reflectance_spectrum_err 
FROM gaidr3.sso_reflectance_spectrum
WHERE designation = 'lutetia'
\end{verbatim}
}



\section{Solar analogue}\label{a:solar_analog}
As shown by \cite{2020A&A...642A..80C}, the choice of the solar analogue used to compute the asteroid average reflectance spectrum is crucial. Depending on the spectrum of the taken solar analogue, one can observe effects on the slope or the absorption band of the resulting mean reflectance spectrum of the asteroid.  

The approach for computing the average solar analogue spectrum is described below: 
\begin{itemize}
    \item We created a list of known solar analogue stars used for asteroid spectroscopy.
    \item We computed an average spectrum from all the spectra of the list. Namely, firstly we normalised each spectrum by dividing point to point by the sum of the fluxes of BP and RP; next we calculated the average spectra of all normalised spectra and its uncertainty.
    \item From visual inspection of the internally calibrated XP spectra of the sources in the list, we noted that some of them are discrepant from the general spectrum of the set. 
\end{itemize}


From the ground, a series of trusted solar analogues have been used over the years. 
\gaia has observed these stars multiple times. We analysed their mean spectra and found little variation across the sample. In Table~\ref{tab:cu4sso_solar_analogue_stars}, the list of solar analogue stars is detailed with information such as the magnitude and the spectral type of the corresponding stars. 

\begin{table*}[h]
  \caption[Solar analogue stars selected to compute the mean spectra]{Solar analogue stars selected to compute the average spectrum.}
  \label{tab:cu4sso_solar_analogue_stars}
  \centering
\renewcommand{\arraystretch}{1.5} 
\begin{tabular}{l r r r r c l}
   \hline\hline 

   Denomination & \multicolumn{1}{c}{\gdr{3} sourceId} & \multicolumn{1}{c}{RA [\deg]} & \multicolumn{1}{c}{DEC [\deg]} & \multicolumn{1}{c}{\gmag} & \multicolumn{1}{c}{Spectral Type}  & \multicolumn{1}{l}{Reference} \\
   \hline
HD6400 & 2352876238295485440 &  16.178114       &-20.987929 &   9.30 &  G2V & a
\\
HD182081 & 4295820654469438848 & 290.679459 & 7.729991 & 9.39 & G2V & b
\\
HD146233 & 4345775217221821312 & 243.906303 &   -8.371572 &     5.30 &  G2V & c
\\
HD20926 & 5103333467421759616 & 50.470306 &     -19.395128       & 9.64 & G2V & a
\\
HD220764 & 2409392888309495424 & 351.641401 &   -12.538598 & 9.47 &     G2V & d
\\
SAO140573 & 4416093315142832128 &       232.604610      & -1.318933 &   8.99 &       G3V & e
\\
16CygB & 2135550755683405952  & 295.452970       & 50.524376 &  5.80 &  G1.5V & f
\\
SA110-361 & 4272476270261273216 &       280.687525      & 0.134614 &    12.27 &        & g    
\\
HD220022 & 2385351065840541184 &        350.185850 & -22.308895         & 9.54 & G3V &  e
\\
HD202282 & 6883947988319565568 &        318.798564 & -15.741823 & 8.82 & G3V & a
\\
HD16640 & 2502734072523738496 & 40.036149       & 2.918174 &    8.61 &  G2V & a
\\
SAO185145 & 4109488347412656640 &       258.130251 &    -25.227387 &    8.16 &       G2V & e
\\
SA107-684 & 4416641352970215936 &       234.325877 &    -0.163956 &     8.26 &       G2/3V & g
\\
SAO63954 & 1478734360725538944 &        212.205186       & 32.949572 &  9.01 &       G0V & e
\\
HD100044 & 3560743225161573120 &        172.653749 &    -15.103249 &    9.36 &       G2V & b
\\
HD154424 & 5965123126448738816 &        256.856628      & -43.837913    & 9.04 &  G2V& a
\\
SAO41869 & 924370390524853120 & 113.817066 &    40.506394 &     7.50 &  G0E & e
\\
HD144585 & 4341501106288171008 &        241.762879      & -14.071255  & 6.14 &       G2V & h
\\
SA98-978 & 3113329094598313472 &        102.890626 &    -0.192218 &     10.43    & F8E & a
\\[4pt]
   \hline
\multicolumn{7}{l}{a. \cite{2018P&SS..157...82P}, b. \cite{Popescu2019}, c. \cite{Soubiran2004}, 
d. \cite{Popescu2014},}\\
\multicolumn{7}{l}{e. \cite{LUCAS2019}, f. \cite{Bus2002Icar..158..106B}, g. \cite{FORNASIER2007},
h. \cite{LAZZARO2004}.}
\\[4pt]
   \hline
  \end{tabular}
\end{table*}

\begin{figure}[!ht]
   \centering
   \includegraphics[width=\columnwidth]{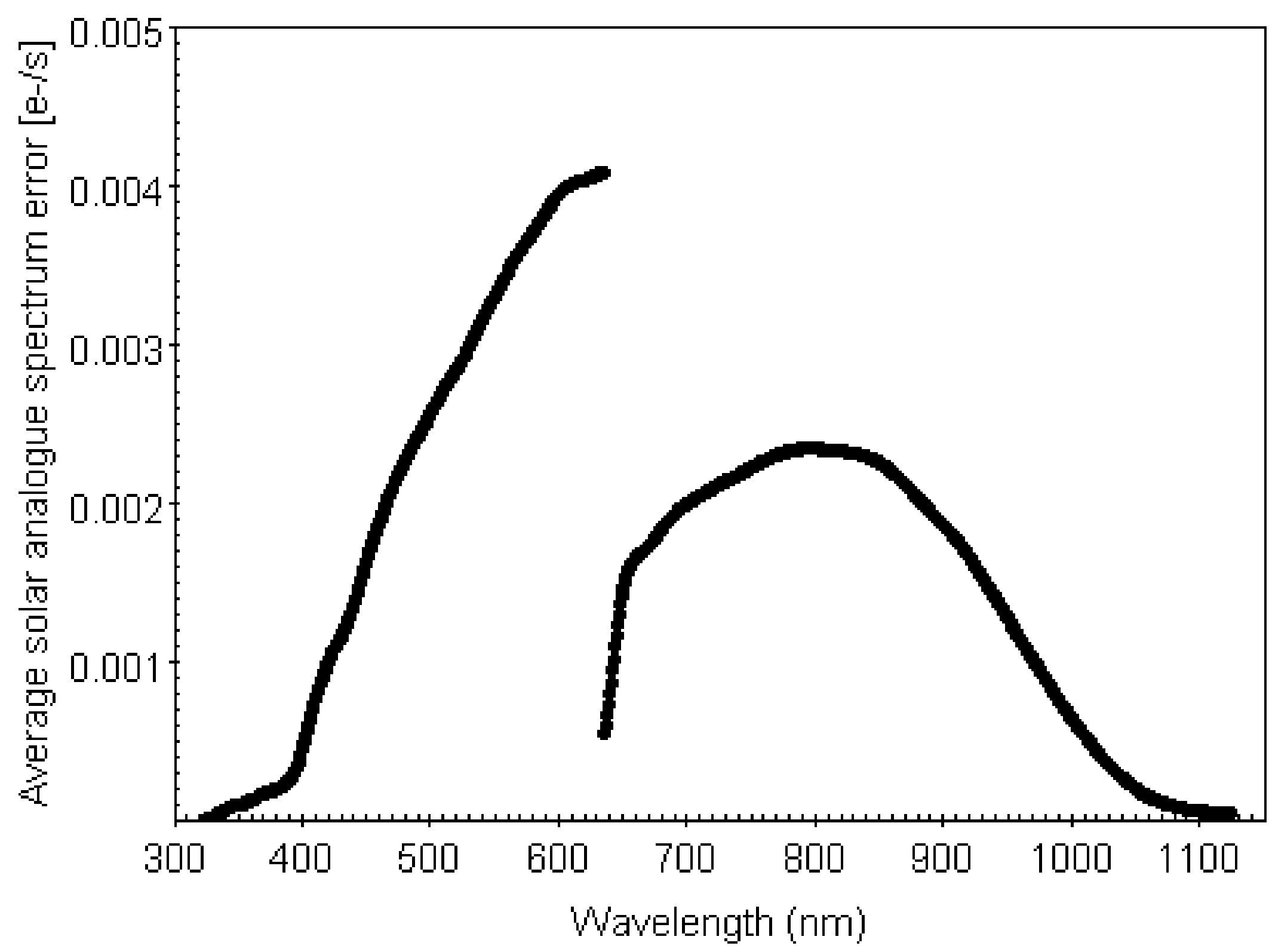}
    \includegraphics[width=\columnwidth]{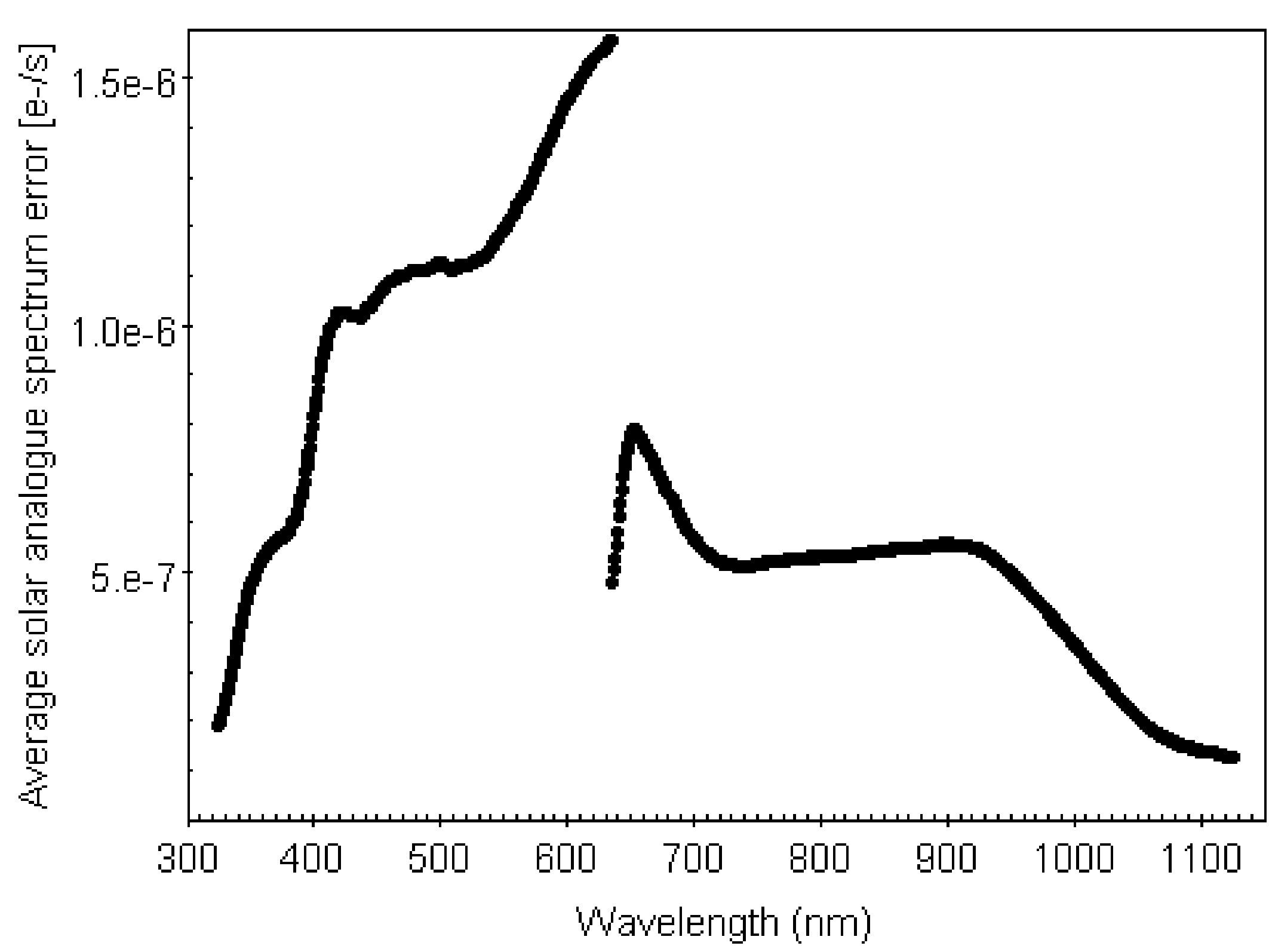}
   \caption{Average solar analogue spectrum computed for \gaia (top figure) and its error (bottom figure) are plotted. The curves with support in the wavelength range between 320 and 640 nm and between 640 and  1130 nm correspond to the BP and the RP, respectively.}
   \label{Fig:avg_solar_analog_spec}%
\end{figure}

The information relative to the solar analogues used in the context of asteroid spectroscopy is depicted in ESA Gaia DR3 auxiliary data webpage 
\footnote{\url{https://www.cosmos.esa.int/web/gaia/dr3-solar-analogue-spectrum}}.

\newpage 
\clearpage
\section{Colour palette code}\label{sec:appendix:spec2rgb}
Python example code:\\

The file specParam.dat contains the ssoId, the computed S/N, the spectral slope, and the z-i computed as explained in the corresponding session. The reader can easily recompute these data using the explanation provided in this paper.
\begin{lstlisting}[language=Python]
import numpy as np
import matplotlib.pyplot as plt

def spec2C(slope,zi):
    b=-1./25*(slope-20)
    b[b>1]=1.
    b[b<0]=0.

    r=1./24*(slope+5)
    r[r>1]=1.
    r[r<0]=0.

    g=-1./.3*zi;
    g[g>1]=1.
    g[g<0]=0.

    return np.transpose(np.array([r,g,b]))
    
#### LOAD DATA FILES
gid,snr,slope,zi = np.loadtxt('specParam.dat',
        usecols=(0,1,2,3), unpack=True)
a,e,sini,fam = np.loadtxt('orb_elem.dat', 
        usecols=(4,5,6,13), unpack=True)

### CALCULATE COLOURS
C=spec2C(slope,zi)
### SORT by SSO NUMBER
j=np.argsort(gid)[::-1]

### CLEAR PLOTS
plt.cla()
plt.clf()
### EXAMPLE OF PLOT
plt.scatter(a[j],sini[j],c=C[j],alpha=0.5,
    s=1.25, marker='o',edgecolors='none')

\end{lstlisting}
\newpage
\section{Phase-angle dependence of spectral parameters}\label{A:phaseAngleVsSlope}

\begin{figure}[!h]
    \centering
    \includegraphics[width=\columnwidth]{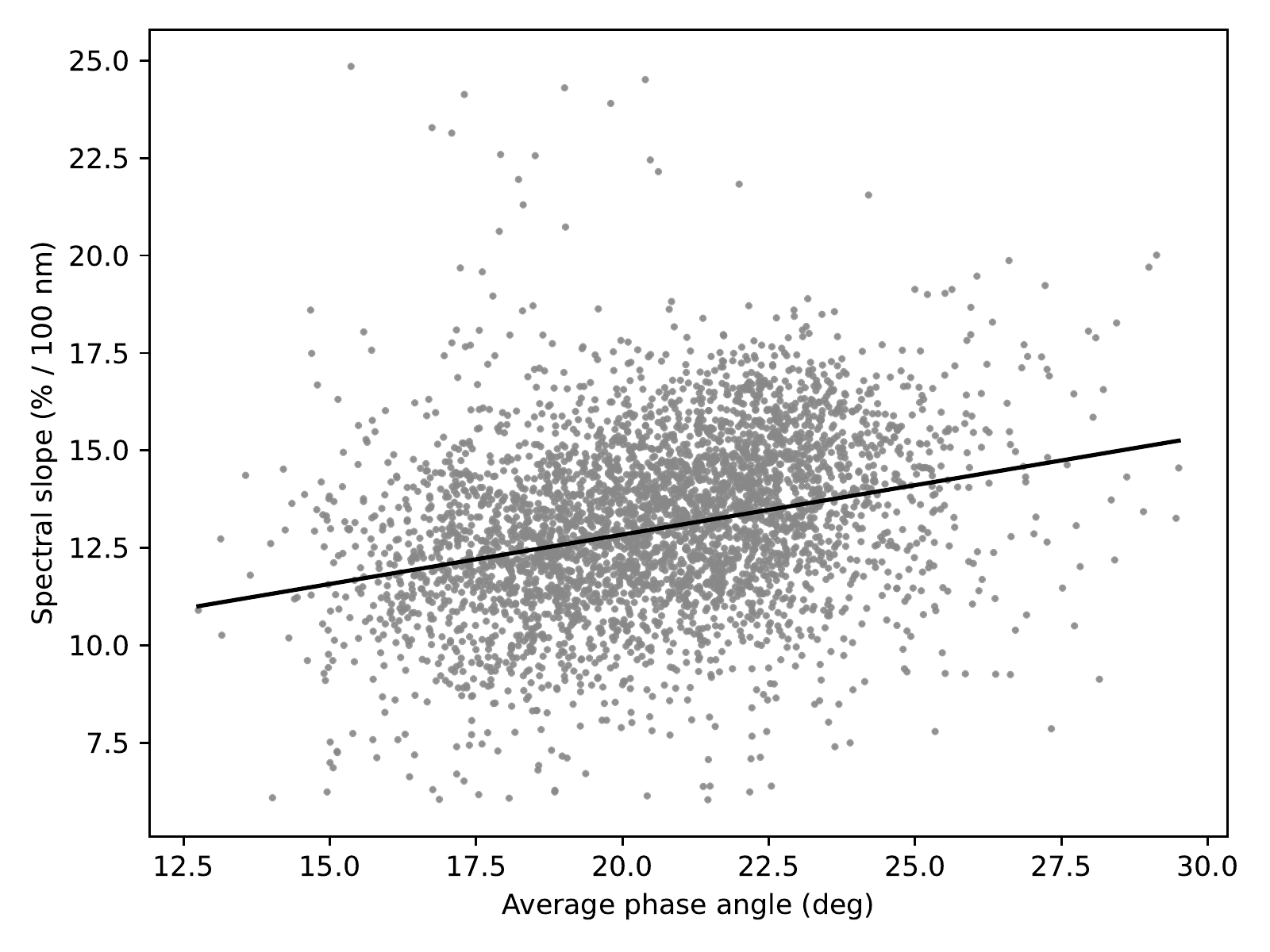}
    \caption{Distribution of the spectral slope as a function of the mean phase angle for asteroids with $\text{S/N}>50$ and spectral parameters within the region of the S-complex according to \citet{Demeo2013Icar..226..723D}. The straight line is fit to the underlining distribution. See Section~\ref{ssec:spaceweatheringStype} for further information.}
    \label{fig:phaseSlopeStype}
\end{figure}
\newpage 
\clearpage

\section{\gaia vs. SMASSII spectral parameters}\label{A:gaiaVSsmass}
\begin{figure}[h!]
    \centering
    \includegraphics[width=\columnwidth]{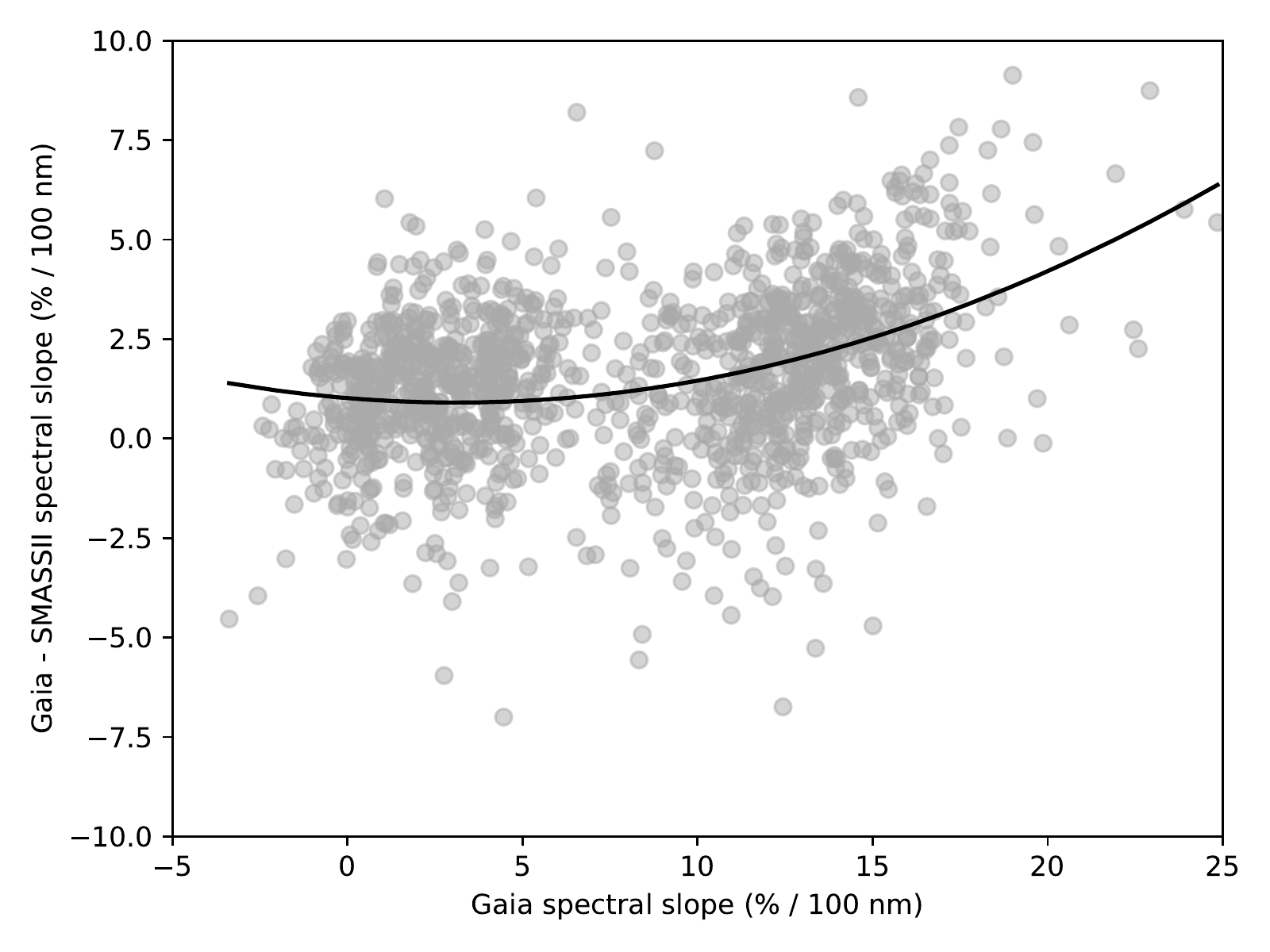}
    \includegraphics[width=\columnwidth]{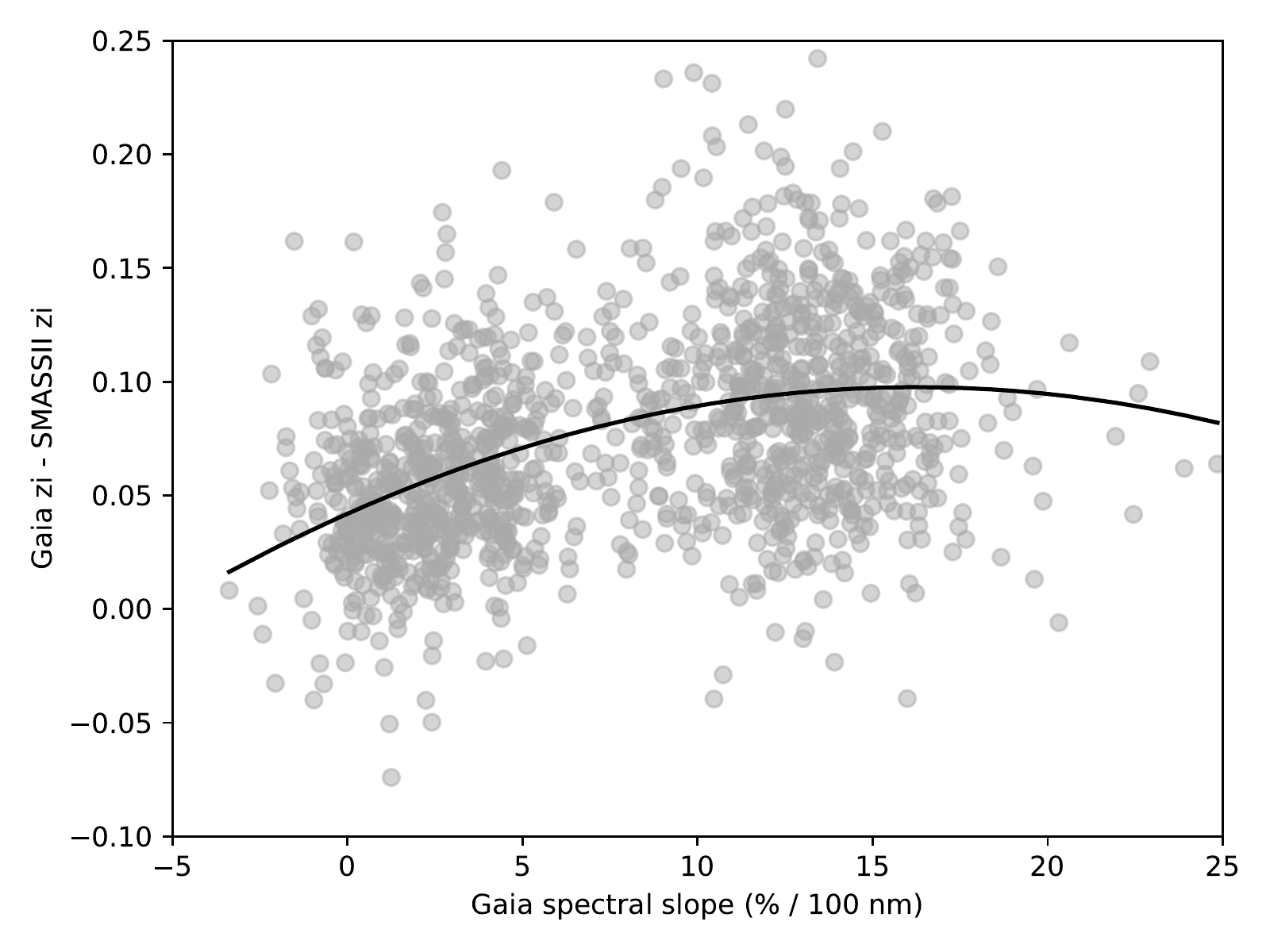}
    \caption{Spectral parameter difference between \gdr{3} and SMASSII. Top panel: Spectral slope difference. Bottom panel: z-i colour difference. Grey circles are the data, whereas the solid lines represent the second-order polynomial fit to the data (see Section~\ref{ssec:gbcomparison}).}
    \label{fig:SlopeZIGaiaSmass}
\end{figure}

We calculated spectral slope and z-i colours for the \gdr{3} asteroids and those of the SMASSII \footnote{\url{http://smass.mit.edu}} and selected only those objects common to both surveys. Figure~\ref{fig:SlopeZIGaiaSmass} shows that the spectral slope and z-i colour differences  between the two surveys, indicated by $\Delta S$ and $\Delta C$, respectively, is a weak function of \gaia's spectral slope $S_\text{gaia}$. Next, we fit a second-order polynomial to the spectral slope and z-i differences as a function of the \gaia spectral slope. We find that 
$\Delta S = 1.01532733 -0.07180434 S + 0.01156685 S^2 $, and $\Delta C = 0.04193216  + 0.00685427 S -0.000211 S^2 $.

\newpage

\section{Plots comparing \gaia spectra to ground-based  literature}\label{appendix:spectraPlots}
\begin{figure*}[!h]
   \centering
    \includegraphics[width=2\columnwidth]{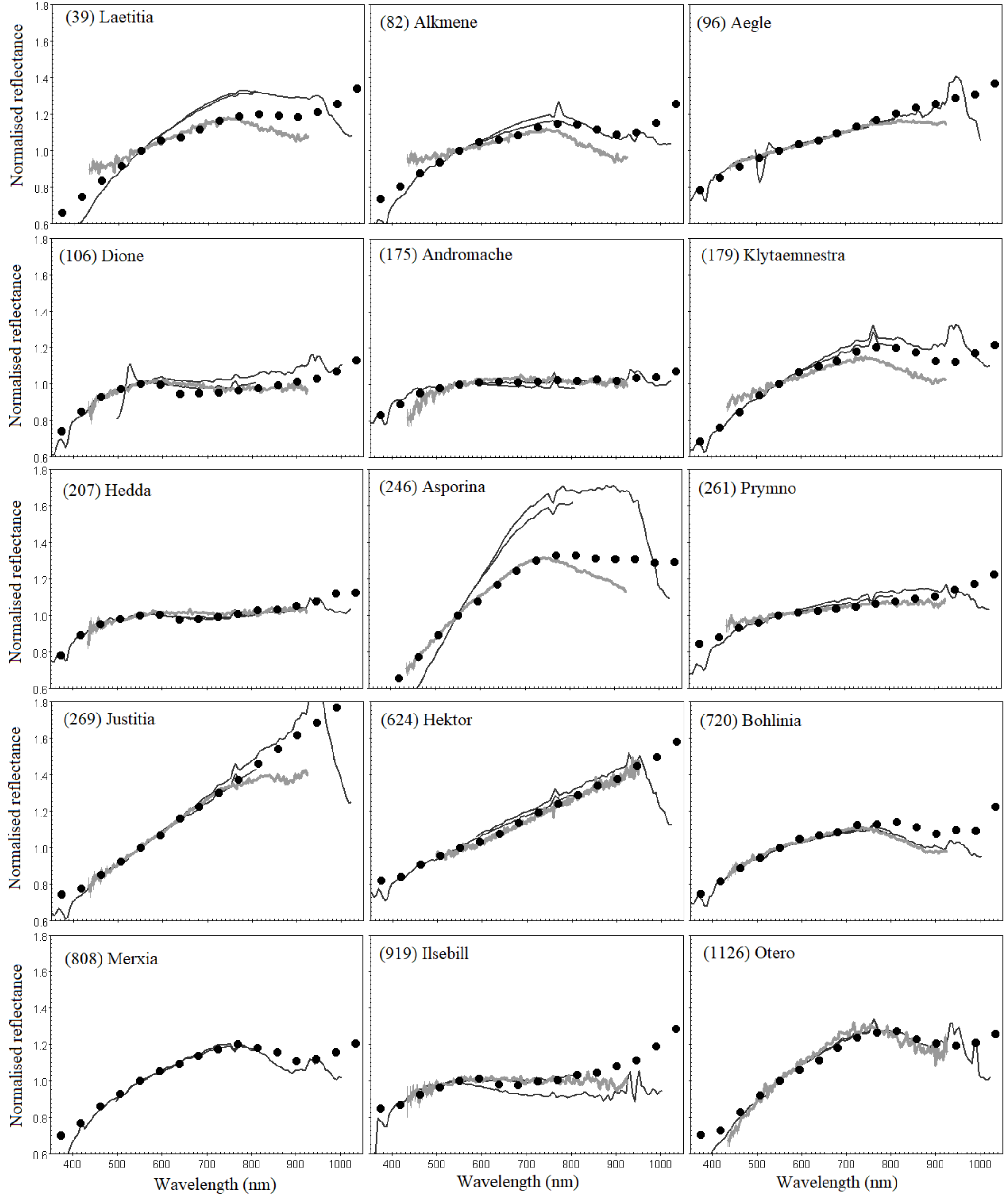}
 \caption{\gaia's mean reflectance spectra (black circles)  against ground-based observations;\cite{2020A&A...642A..80C} spectra in dark grey lines and in light grey lines for asteroids (624) \citep{Vilas1993Icar..105...67V} and (39), (82), (96), (106), (207), (246), (261), (269), (720), (919), (1126) \citep{Bus2002Icar..158..106B}.}
   \label{Fig:refl_comp_Cellino1}%
\end{figure*}

\begin{figure*}[!h]
   \centering
    \includegraphics[width=2\columnwidth]{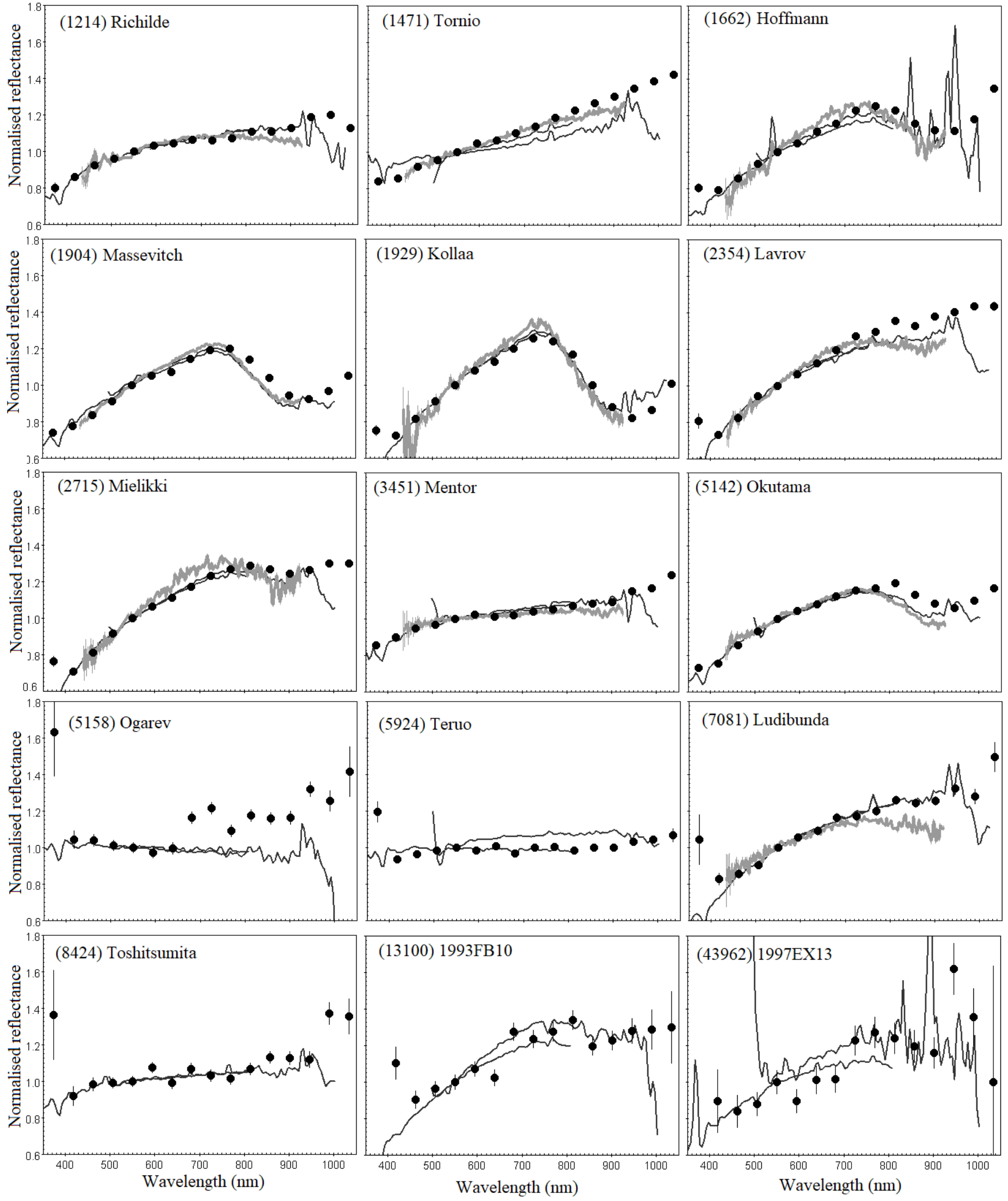}
 \caption{\gaia's mean reflectance spectra (black circles) against ground-based observations;  \cite{2020A&A...642A..80C} spectra in dark grey lines and those of \citep{Bus2002Icar..158..106B} in light grey lines for asteroids (1214), (1471), (1662), (1904), (2354), (2715), (3451), (5142), (7081).}
   \label{Fig:refl_comp_Cellino2}%
\end{figure*}

In this Appendix we add all the comparative plots (see Figures~\ref{Fig:refl_comp_Cellino1} and \ref{Fig:refl_comp_Cellino2}) of \gaia mean reflectance spectra and the literature ground-based spectra discussed in Section~\ref{ssec:gbcomparison}.

\end{appendix}
\end{document}